\documentclass[a4paper,11pt]{article}

\usepackage{jheppub} 
\usepackage[T1]{fontenc} 
\usepackage{amsmath,amssymb,xfrac,framed,verbatim,amsthm}
\usepackage{mathrsfs,accents,bbm,bbold} 
\usepackage{simplewick}
\usepackage{slashed,mathtools} 
\usepackage{sectsty} 
\usepackage{graphicx,enumerate,bm} 
\graphicspath{{figures/}}
\usepackage{verbatim}
\usepackage{comment}

\newcommand{\mc}[1]{\mathcal{#1}}

\newcommand{\tr}{\text{Tr}}

\newcommand{\ud}{\text{d}}
\newcommand{\tl}{\tilde}
\newcommand{\sgn}{\text{sgn}}
\renewcommand{\i}{\text{i}}

\newcommand{\g}{\gamma}

\renewcommand{\l}{\ell}

\preprint{TIFR/TH/20-38} 
\title{\boldmath On thermal correlators and bosonization duality in Chern-Simons theories with massive fundamental matter}

\author{Amiya Mishra}
\emailAdd{amiya.mishra@tifr.res.in}

\affiliation{Department of Theoretical Physics,\\Tata Institute of Fundamental Research, Homi Bhabha Road, Mumbai 400005, India}

\usepackage{feynmp}

\usepackage{tikz}
\usepackage[compat=1.1.0]{tikz-feynman}

\abstract{We consider Chern-Simons theory coupled to massive fundamental matter in three spacetime dimensions at finite temperature, in the large $N$ limit. We compute several thermal correlators in this theory for both fermionic and bosonic matter separately. The results are computed in the large $N$ 't Hooft limit but for arbitrary values of the 't Hooft coupling.
	Furthermore, we generalize the computations of the four-point function of fundamental scalars in the bosonic theory to finite temperature. As a consistency check, we see that the results obtained here agree with the existing previous results in different limiting cases. Moreover, we check that the results are consistent with the conjectured bosonization duality, providing an additional evidence of it.}

\begin{document}
\maketitle
\tableofcontents

\section{Introduction}

There is now considerable evidence that bosonic matter coupled to Chern-Simons (CS) gauge theories and fermionic matter coupled to - roughly speaking - `level-rank dual' CS gauge theories are dual to each other in three spacetime dimensions 
	\cite{Giombi:2011kc, Aharony:2011jz, Maldacena:2011jn, Maldacena:2012sf,
	Chang:2012kt, Jain:2012qi, Aharony:2012nh, Yokoyama:2012fa,
	GurAri:2012is, Aharony:2012ns, Jain:2013py, Takimi:2013zca,
	Jain:2013gza, Frishman:2013dvg, Yokoyama:2013pxa, Bardeen:2014paa, Jain:2014nza,
	Bardeen:2014qua, Gurucharan:2014cva, Dandekar:2014era,
	Frishman:2014cma, Moshe:2014bja, Aharony:2015pla, Inbasekar:2015tsa,
	Bedhotiya:2015uga, Gur-Ari:2015pca, Minwalla:2015sca,
	Radicevic:2015yla, Geracie:2015drf, Aharony:2015mjs,
	Yokoyama:2016sbx, Gur-Ari:2016xff, Karch:2016sxi, Murugan:2016zal,
	Seiberg:2016gmd, Giombi:2016ejx, Hsin:2016blu, Radicevic:2016wqn,
	Karch:2016aux, Giombi:2016zwa, Wadia:2016zpd, Aharony:2016jvv,
	Giombi:2017rhm, Benini:2017dus, Sezgin:2017jgm, Nosaka:2017ohr,
	Komargodski:2017keh, Giombi:2017txg, Gaiotto:2017tne,
	Jensen:2017dso, Jensen:2017xbs, Gomis:2017ixy, Inbasekar:2017ieo,
	Inbasekar:2017sqp, Cordova:2017vab, Charan:2017jyc, Benini:2017aed,
	Aitken:2017nfd, Jensen:2017bjo, Chattopadhyay:2018wkp,
	Turiaci:2018nua, Choudhury:2018iwf, Karch:2018mer, Aharony:2018npf,
	Yacoby:2018yvy, Aitken:2018cvh, Aharony:2018pjn, Dey:2018ykx, Skvortsov:2018uru, 
	Chattopadhyay:2019lpr, Dey:2019ihe, Halder:2019foo, Aharony:2019mbc,
	Li:2019twz, Jain:2019fja, Inbasekar:2019wdw, Inbasekar:2019azv,
	Jensen:2019mga, Kalloor:2019xjb, Ghosh:2019sqf, Inbasekar:2020hla, Jain:2020rmw, Minwalla:2020ysu, Jain:2020puw, amiyata1} \footnote{There are two pair of conjecturally dual theories in non-supersymmetric matter coupled Chern-Simons theories. One is regular fermion and critical boson which are together called quasi fermionic theories. The other pair is regular boson and critical fermions which are together called quasi bosonic theories. For details about these see e.g. in \cite{Maldacena:2011jn, Maldacena:2012sf, Aharony:2012ns, Jain:2013py, Jain:2014nza, Minwalla:2015sca, Choudhury:2018iwf, Dey:2018ykx, Aharony:2018pjn, Minwalla:2020ysu}.}. 
More specifically, $SU(N_B)$ CS gauge fields coupled to bosonic matter is level-rank dual to $U(N_F)$ CS gauge fields coupled to the fermionic matter (in the strict large $N$ limit - in which we will be working in - the difference between $SU(N)$ and $U(N)$ effectively disappears) \footnote{In the dimensional regulation scheme, the Chern-Simons levels are renormalized. In the large $N$ limit, the renormalized levels and ranks of the two gauge groups are related as $\kappa_F=-\kappa_B$ and $N_F=|\kappa_B|-N_B$ (see for details e.g. in \cite{Jain:2014nza, Minwalla:2015sca, Choudhury:2018iwf, Aharony:2018pjn, Dey:2018ykx, Minwalla:2020ysu}). For a precise form of conjectured duality see appendix A of \cite{Minwalla:2020ysu}.}. This is an example of strong-weak coupling duality. One of the exciting and interesting fact about these theories is that, in the 't Hooft large $N$ limit, exact computations of observables can be performed on both sides of the dual pair of theories and the duality can be explicitly checked in this limit. 

Previously there have been numerous computations and checks of this duality in the case of both the massless and massive matter theories at zero temperature. Even the computations have already been performed at finite temperature to compute the thermal free energy of these theories and it has already been shown that the thermal free energies in both the fermionic theories and the bosonic theories map to each other under duality \cite{Aharony:2012ns, Jain:2013py, Choudhury:2018iwf, Dey:2018ykx, Minwalla:2020ysu}. These theories also admit an infinite set of higher spin currents (spin $\geq$ 1) of single trace operators \cite{Giombi:2011kc, Maldacena:2011jn,Maldacena:2012sf, Frishman:2013dvg}. These theories have a global $U(1)$ symmetry. The spin one operator is identified as the corresponding $U(1)$ conserved current. The spin two operator is interpreted as the stress tensor of the corresponding theories. Apart from these, these theories also contain a single trace, gauge invariant scalar `current' operator which we refer to as spin-zero operator. Previously, there have been several computations of various correlation functions of these operators in different limiting cases. For example, the correlation functions of single trace operators of various spins were computed in the case of massless fundamental matter (for both fermions and bosons) coupled to Chern-Simons theory at zero temperature (see, e.g. in \cite{Aharony:2012nh, GurAri:2012is, Yacoby:2018yvy, Kalloor:2019xjb}).  In \cite{Geracie:2015drf, Gur-Ari:2016xff}, the two-point functions of spin-one current is computed in the massive fermionic matter theory coupled to Chern-Simons gauge fields at zero temperature. At finite temperature, the effects of holonomy of gauge fields - which is basically the zero-mode of the gauge field along the thermal circle - become important and one has to take into account that as well \cite{Aharony:2012ns, Jain:2013py}. In the large-$N$ limit, the effect of holonomy can be described by a continuous distribution function. There have been few computations of the current two-point functions in the massless fundamental matter theories at non-zero temperature with a specific holonomy distribution. For example, in \cite{Gur-Ari:2016xff}, the authors first computed the two-point function of $U(1)$ conserved current in the massless fermionic matter theories at finite temperature in the context of studying the Hall conductivity. In their analysis, however, they considered only a particular holonomy distribution of gauge field, i.e., the universal table-top distribution and the final results of that paper was in terms of some integral form. In a recent paper \cite{Ghosh:2019sqf} also, the authors have computed two-point functions of several current operators in the massless fermionic and massless bosonic matter theories at non-zero temperature. However, they had chosen a particular (table-top) holonomy distribution which is appropriate for infinite volume limit. In this work, we generalize the computations of various current correlators to the general case of the massive matter at finite temperature considering an arbitrary holonomy distribution for the gauge holonomy, in Chern-Simons coupled to fundamental matter theories \footnote{We have been able to solve the resultant Schwinger-Dyson equations explicitly by `effectively performing' the loop momenta integrals. From this point view, this work can also be thought of as the study of solving an interesting `mathematical problem'.}. We consider massive fermionic and bosonic matter theories separately at non-zero temeperature. There are two types of these theories: one is the regular matter theories and the other is the critical matter theories. The regular matter theories on one side of the duality is conjectured to be dual to the critical matter on the other side and vice versa. It was discussed in \cite{Aharony:2012nh, GurAri:2012is} and also pointed out in \cite{Ghosh:2019sqf} that the two-point correlators of various spin $s$ operators -  for spin $s\geq 1$ - are basically the same whether we consider the regular or critical matter theories, except the difference being implicitly through the exact masses of the fundamental excitations of fermion/boson. The two-point function of spin-zero, single trace scalar operator is different whether one considers the regular matter or critical matter; and it was discussed in \cite{Aharony:2012nh, GurAri:2012is} that they are related to each other in a particular way, at least in the massless theory. To be specific, we consider here the massive regular fermionic matter theory coupled to $U(N_F)$ Chern-Simons gauge fields. For convenience of the computation, we also consider the massive regular bosonic matter theory coupled to $SU(N_B)$ Chern-Simons gauge fields. We choose to work in the dimensional regulation scheme in which the Chern-Simons level is given by the renormalized parameter $\kappa$. We work in the 't Hooft large-$N$ limit, which is by taking $N,\kappa\to \infty$ but keeping $\lambda=\frac{N}{\kappa}$ fixed, and the modulus of the 't Hooft coupling $\lambda$ is less than unity. One of our goals in this paper is to check the conjectured bosonization duality. To do that we take the critical limit of the regular bosonic theory \cite{Aharony:2012nh, GurAri:2012is, Jain:2014nza} and check the duality. 
For the computation of bosonic correlators, we take a slightly different route than \cite{Ghosh:2019sqf}. Following \cite{Aharony:2012nh}, in the case of massive bosonic theories at finite temperature, we first compute the offshell thermal four-point functions of fundamental scalars; we do so by generalizing the results of \cite{Aharony:2012nh, Jain:2014nza} to include the finite temperature effects. The offshell four-point function of scalars is then used to compute the various current correlators. In the massive case, even at zero temperature, there are two phases of bosonic matter theories: one is unhiggsed phases of scalars and the other is the higgsed phases of $W$ and $Z$ bosons (for details on the Higgsed phases of bosonic matter coupled to Chern-Simons theories, see e.g. in \cite{Choudhury:2018iwf, Dey:2018ykx}). In the critical boson theory with a bare mass parameter $m_B^{\text{cri}}$, there are two possible signs. The unhiggsed phase of bosonic scalar corresponds to the case $m_B^{\text{cri}} > 0$ which under bose-fermi duality gets mapped to the regular fermionic matter theory with $\sgn(m_F\lambda_F)=+1$ \footnote{At finite temperature, this condition is replaced by a more general condition $\sgn(h_F\lambda_F)=+1$, where, $\sgn(h_F)$ is defined around \eqref{masseqnforcF}. The regular fermionic theory at nonzero temperature with $\sgn(h_F\lambda_F)=+1$ is dual to the unhiggsed phases of critical boson theory. The other case, i.e., the case $\sgn(h_F\lambda_F)=-1$ is dual to the higgsed phases of critical boson theory. At zero temperature, $\sgn(h_F)$ is replaced by $\sgn(m_F)$. For details about this see e.g., in \cite{Choudhury:2018iwf}, where $\sgn(h_F)$ is labelled by a different symbol $\sgn(X_F)$.}, where, $m_F$ is the bare mass parameter of the regular fermionic theory, and $\lambda_F=\frac{N_F}{\kappa_F}$ is 't Hooft coupling. We have explicitly checked the bose-fermi duality between the current correlators in these phases. However, for the fermionic theories, the results are valid for both possible signs of $m_F$. The other possible case, i.e., the case $\sgn(m_F\lambda_F)=-1$ in the fermionic theory is dual to the higgsed phases of the bosonic theory (which corresponds to the case $m_B^{\text{cri}} < 0$). We use the results of current correlators in the fermionic theory to predict the corresponding results in the Higgsed phases of bosonic matter.  We leave the excercise of explicit check of duality between the current correlators by exact computations in this phase for future work.

The organization of this paper is as follows. In section \ref{holonomy}, we briefly review and discuss about the effect of holonomy in Chern-Simons matter theories at finite temperature. In section \ref{fermth}, we study the fermionic matter coupled Chern-Simons theory and present the computations of thermal two-point functions of spin-zero and spin-one current operators. In section \ref{bosonth}, we study the bosonic matter coupled Chern-Simons theory and present the results of the offshell four-point function of fundamental scalars in the bosonic scalar theory at finite temperature (the details of which is presented in the appendix \ref{4ptscalar}) and also present the computations of thermal two-point functions of spin-zero and spin-one current operators. In section \ref{anlzresult}, we analyze the main results of this paper in various special limits and see that it agrees with the existing results in the limiting cases. In section \ref{dualitycheck}, we explicitly check that the results obtained in this paper are consistent with the conjectured bose-fermi duality. In section \ref{discussion}, we draw conclusions about our results, and discuss various outlook and interesting future directions. 


\section{A note on holonomy and related conventions}\label{holonomy}
In an Euclidean thermal field theory at some temperature $T$ in three dimensions, $x^3$ direction is put along a circle $\mathbf{S}^1$ of radius $\beta=T^{-1}$ \footnote{In other words, the identification $x^3 \sim x^3+\beta$ is used; here, $x^3$ is the Euclidean time coordinate.}. 
The position space integrals that we encounter here, are given by
\begin{equation}
\int {d}^3x \ f(x) = \int d^2\vec{x} \int_{0}^{\beta}  dx^3 \  f(\vec{x},x^3) \ .
\end{equation}
In Chern-Simons matter theory at finite temperature, holonomy, i.e., the zero mode of the gauge field along the thermal circle, becomes important and crucially effects the physical observables \cite{Aharony:2012ns, Jain:2013py}. Holonomy of the gauge field in a $U(N)$ gauge theory is completely specified by its eigenvalues, $e^{\i \alpha_j}$ where $j=1,\dots, N$ and $\alpha_j\in (-\pi,\pi]$ \cite{Choudhury:2018iwf, Minwalla:2020ysu, Aharony:2012ns}. In the large-$N$ limit, the location of eigenvalues on unit circle can be specified by a continuous distribution function $\rho(\alpha)$ defined by
\begin{equation}
\rho(\alpha)=\lim_{N\to \infty} \frac{1}{N}\sum_{j=1}^{N}\delta(\alpha-\alpha_j) \ .
\end{equation} 
It follows that the holonomy distribution function is normalized, i.e., $\int_{-\pi}^{\pi} \rho(\alpha)  ~d\alpha=1$. We choose the holonomy distribution $\rho(\alpha)$ to be arbitrary, throughout the paper, unless otherwise mentioned \footnote{Throughout, we assume the holonomy distribution $\rho(\alpha)$ to be an even function of $\alpha$, i.e., $\rho(-\alpha)=\rho(\alpha)$.}. The correlators that we compute in this paper, by summing Feynman diagrams and performing loop integrals, are in momentum space. Technically, the effect of non-trivial holonomy is to shift the loop momenta in the direction of $k_3$ for a generic loop momentum $k\equiv (k_1,k_2,k_3)$ \cite{Aharony:2012ns, Gur-Ari:2016xff, Choudhury:2018iwf, Ghosh:2019sqf}. Let us briefly explain this point. As $x^3$-direction is compactified in the thermal theory, the conjugate momenta along the third direction is quantized. We will not, however, explicitly write every time the quantized version of $k_3$ but will implicitly assume that is the case. The momentum space integrals are given by 
\begin{equation}\label{momint}
\int \frac{\mathcal{D}^3k}{(2\pi)^3} f(k)=\int \frac{d^2\vec{k}}{(2\pi)^2} \int \frac{\mathcal{D}k_3}{2\pi} f(\vec{k},k_3) \ ,
\end{equation}
where, $d^2\vec{k}$ is the usual integration measure $dk_1dk_2$ for the spatial momenta, and the momentum integration measure $\mc{D}k_3$ is defined by 
\begin{equation}\label{mom3int}
\int \frac{\mathcal{D}k_3}{2\pi} f(\vec{k},k_3) = \frac{1}{\beta} \int_{-\pi}^{\pi} \rho(\alpha) d\alpha \ \sum_{k_3} f\Big( \vec{k},k_3+\frac{\alpha}{\beta} \Big) \ ,
\end{equation}
Here, as mentioned earlier, the effect of holonomy is to shift the $k_3$ momenta by $\beta^{-1}\alpha$. Exact meaning of `integral over $k_3$' which is really a sum \cite{Choudhury:2018iwf}, is slightly different for bosons and fermions depending upon the periodic/antiperiodic boundary conditions on bosonic and fermionic fields along $\mathbf{S}^1$, the precise meaning of which is explained below in \eqref{bosholonomy} and \eqref{fermholonomy}. To see the precise form of the quantization condition on the momenta $k_3$, we consider the fourier transform of a function $f(x)\equiv f(\vec{x},x^3 )$ which is defined by \cite{Choudhury:2018iwf} 
\begin{equation}\label{ft}
f(\vec{x},x^3 ) = \int \frac{\mathcal{D}^3 k}{(2\pi)^3} \ e^{ \i \vec{k}\cdot \vec{x}+\i k_3x^3} \ f(\vec{k},k_3) \ . 
\end{equation} 
Depending upon the periodicity/antiperiodicity of the boundary conditions on the fields along the thermal circle, there are two possible cases. 
\subsubsection*{Bosonic Theory :} 
Bosonic fields are periodic along the thermal circle, i.e., $\phi(\vec{x},x^3+\beta )=\phi(\vec{x},x^3 ) $. 
This implies $e^{ \i k_3\beta} =e^{ 2n_k \pi \i }$, where $n_k \in \mathbb{Z}$; i.e., $k_3$ is quantized with the quantization condition $k_3=\frac{2n_k \pi }{\beta}$. 
It follows from \eqref{momint} and \eqref{mom3int} that the momentum integral with quantized $k_3$ in the bosonic theory for an arbitrary holonomy distribution $\rho_B(\alpha)$ is given by 
\begin{equation}\label{bosholonomy}
\int \frac{\mathcal{D}_B^3 k}{(2\pi)^3} f(\vec{k},k_3)= \int \frac{d^2 \vec{k}}{(2\pi)^2} \int \frac{\mathcal{D}_B k_3}{2\pi} f(\vec{k},k_3) =\frac{1}{\beta} \int \frac{d^2 \vec{k}}{(2\pi)^2} \int_{-\pi}^{\pi} \rho_{B}(\alpha)  ~d\alpha \sum_{n_k\in \mathbb{Z}} f\bigg(\vec{k},\frac{2n_k\pi + \alpha}{\beta} \bigg)  \ .
\end{equation} 
For later use, we here define the following function $\chi_B(z)$ of a real variable $z$
\begin{equation}\label{chiB}
\chi_B(z) =\frac{1}{2} \int_{-\pi}^{\pi} \rho_{B}(\alpha)  ~d\alpha ~\bigg[\coth\bigg(\frac{\beta z+\i \alpha}{2} \bigg)+\coth\bigg(\frac{\beta z-\i \alpha}{2} \bigg)\bigg] \ . 
\end{equation}
Following \cite{Choudhury:2018iwf}, we also define another function $\xi_B(z)$ as $\xi_B(z)=\int^{z}\chi_B(w)\  dw$, an (indefinite) integral over $\chi_B(z)$. The explicit form of the function $\xi_B(z)$ is given by \eqref{xiB} in the appendix \ref{appdef}. 

\subsubsection*{Fermionic Theory :} 
Fermion fields on the other hand satisfy the anti-periodicity boundary condition along the thermal cirlce, i.e., $\psi(\vec{x},x^3+\beta )=- \psi(\vec{x},x^3 ) $, which 
implies $e^{ \i k_3\beta} =e^{ \i (2n_k+1) \pi }$, where $n_k \in \mathbb{Z}$. This means $k_3$ in the fermionic theory is quantized with the quantization condition $k_3=\frac{(2n_k+1) \pi }{\beta}$.
So, the momentum integral with quantized $k_3$ in the fermionic theory for an arbitrary holonomy distribution $\rho_F(\alpha)$ is given by 
\begin{equation}\label{fermholonomy}
\int \frac{\mathcal{D}_F^3 k}{(2\pi)^3} f(\vec{k},k_3)=\int \frac{d^2 \vec{k}}{(2\pi)^2} \int \frac{\mathcal{D}_F k_3}{2\pi} f(\vec{k},k_3) =\frac{1}{\beta} \int \frac{d^2 \vec{k}}{(2\pi)^2} \int_{-\pi}^{\pi} \rho_{F}(\alpha)  ~d\alpha \sum_{n_{k}\in \mathbb{Z}} f\bigg(\vec{k},\frac{(2n_k+1)\pi +\alpha}{\beta} \bigg)  \ .
\end{equation} 
As above in bosonic theory, we define here the following function $\chi_F(z)$ 
\begin{equation}\label{chiF}
\chi_F(z) =\frac{1}{2} \int_{-\pi}^{\pi} \rho_{F}(\alpha)  ~d\alpha ~\bigg[\tanh\bigg(\frac{\beta z+\i \alpha}{2} \bigg)+\tanh\bigg(\frac{\beta z-\i \alpha}{2} \bigg)\bigg] , 
\end{equation}
which will be used later in the paper. As in the case of bosons, we define another function $\xi_F(z)$ as an integral over $\chi_F(z)$, i.e., as $\xi_F(z)=\int^{z}\chi_F(w)\  dw$. The explicit form of  $\xi_F(z)$ is given by \eqref{xiFz}. For other conventions and useful definitions see Appendix \ref{appdef}.

\section{Thermal correlators in the fermionic theory} \label{fermth}
\subsection{Brief review of the theory}
As discussed in the introduction, in this paper we study the massive, regular fermionic matter theory with a finite bare mass $m_F$ coupled to $U(N_F)$ Chern-Simons gauge fields at finite temperature in the large-$N$ limit. In the dimensional regulation scheme, the Euclidean action of this theory is given by 
\begin{align}\label{rflag} 
\mc{S}_{F}[A,\psi] = \frac{\i \kappa_F}{4\pi} \int \ud^3 x\ \epsilon^{\mu\nu\rho}\,\text{tr}\left( A_\mu \partial_\nu A_\rho - \frac{2\i}{3} A_\mu A_\nu A_\rho\right)\ +\int \ud^3 x \left(\bar\psi \gamma^\mu D_\mu \psi + m_F \bar\psi \psi\right) \ ,
\end{align}
where, covariant derivatives for fundamental and antifundamental fields are defined by, $D_{\mu} {\psi}=\partial_\mu {\psi}-\i A_{\mu} \psi$ and $D_{\mu} \bar{\psi}=\partial_\mu \bar{\psi}+\i \bar{\psi}A_{\mu}$, respectively \footnote{A note on the notations and conventions: we don't explicitly write the color indices, but the color contractions are easily understood from the context. $\psi$ and $\bar{\psi}$ can be thought of as $N_F$ component column and row vectors, respectively and $A_{\mu}$ can be thought of as $N_F\times N_F$ matrices in the color space. We don't show explicitly the spinor indices of $\psi$ and $\bar{\psi}$ which are two-component spinors in three dimensions. Also, we don't explicitly write the adjoint index of the gauge fields $A_{\mu}\equiv A_{\mu}^a T^a$, and color trace over the  gauge group generators $T^a$. One can easily restore these indices and explicitly use the normalizations $\text{Tr}(T^aT^b)=C(N_F)\delta^{ab}$. Conventionally, $C(N_F)=\frac{1}{2}$. The convention for the color contraction that is used here, is such that $\bar{\psi}\psi=\bar{\psi}_{i}\psi^{i}$; but $\psi\bar{\psi}$ does not necessarily have the contracted color indices. Also, $\bar{\psi}M\psi=\bar{\psi}_{i}M^{i}_{\ j} \psi^{j}$. The same convention is used for the bosonic theories where $N_F$ is replaced by $N_B$.}. The gauge field $A_\mu$  is in the adjoint representation of $U(N_F)$. The gamma matrices are chosen to be the ordinary $2\times 2$ Pauli matrices $\gamma^\mu = \sigma^\mu $, $\mu=1,2,3$.
Following the literature, we work in the gauge $A_{-}=0$. With this choice of gauge, the action \eqref{rflag} in momentum space takes the following form \footnote{Normalization for the Levi-Civita tensor used in this paper is such that $\epsilon^{123}=\epsilon_{123}=1 $ and $\epsilon^{+-3}=-\epsilon_{+-3}=-\i $.}
\begin{equation}\label{fermactinfourspace}
\begin{split}
\mathcal{S}_{F}[A,\psi]
 = & - \frac{\kappa_F \epsilon^{\mu -\nu} }{4\pi }\int \frac{\mc{D}_F^3k}{(2\pi)^3} \ A_{\mu }(-k) k_{-}A_{\nu }(k) \ + \ \int \frac{\mc{D}_F^3k}{(2\pi)^3} \ \bar{\psi}(-k)(\i \gamma^\mu k_\mu +m_F) \psi(k) \\
&-\i \int \frac{\mc{D}_F^3p}{(2\pi)^3} \frac{\mc{D}_F^3k}{(2\pi)^3}~\bar{\psi}(p)\g^{\mu} A_{\mu}(-p-k ) \psi(k) \ .
\end{split}
\end{equation}
In the momentum space, some of the Feynman rules for this theory are given below. 
\subsubsection*{Propagator for the gauge field :}
The propagator for the gauge field is defined as 
\begin{equation}
\langle A_{\mu}(p')A_{\nu}(p)\rangle = G^F_{\nu\mu}(p)~(2\pi)^3~\delta^{(3)}(p+p') \ , 
\end{equation}
where, 
\begin{equation}\label{gaugebospropforFermion} 
G^F_{\nu \mu}(p)=\frac{2\pi \epsilon_{\nu -\mu} }{\kappa_F \ p_{-}} \ .
\end{equation}
{As discussed in \cite{Aharony:2012ns}, the gauge field propagator is independent of $p_3$, so is not effected by the holonomy.}
\subsubsection*{Exact propagator for the Fermionic field :}
The exact fermionic propagator, in the strict large-$N$ limit, to all orders in the 't Hooft coupling parameter $\lambda_F$ \footnote{As discussed in the introduction, the 't Hooft coupling parameter for the fermionic theory is defined by $\lambda_F=\frac{N_F}{\kappa_F}$; similarly for the bosonic theory, $\lambda_B=\frac{N_B}{\kappa_B}$. The range of the couplings are $0< |\lambda|<1$, with $\lambda=0$ being the weak coupling limit and $\lambda=1$ being the strong-coupling limit of the corresponding theories.}, is given by
\begin{equation}
\langle \psi(p)\bar{\psi}(p')\rangle =\ S_F(p) \ (2\pi)^3~\delta^{(3)}(p+p') \ , 
\end{equation}
where \footnote{Alternatively, the exact propagator can be written as 
	\begin{equation}\label{exactpsiprop}
	S_F(p) = \frac{\tl{\Sigma}_{\mathbb{1} }(p) \mathbb{1} -\i \g^{\mu} \big(p_\mu +\Sigma_{\mu} \big)}{p^2+c_F^2} \equiv S^F_{\mu}(p)\g^\mu +S^F_{\mathbb{1}}(p)\mathbb{1} \ . 
	\end{equation} 
}, 
\begin{equation}
S_F(p) = \frac{1}{\i \gamma^\mu p_\mu +m_F\mathbb{1}+ \Sigma_F(p)} \ . 
\end{equation}
Here, $\Sigma_F(p)$ is the fermionic self energy. It is useful to expand $\Sigma_F(p)$ in the basis $\{\gamma^\mu, \mathbb{1}\}$ as $\Sigma_F=\i \gamma^{\mu}\Sigma_\mu +\Sigma_{\mathbb{1}}\mathbb{1} $, where, $\mathbb{1}$ is $2\times 2$ identity matrix. At non-zero temperature, the self energy was already computed in the literature (see, e.g. in \cite{Giombi:2011kc,Aharony:2012ns, Jain:2013py} for details) in the lightcone gauge, and is listed here for the purpose of later use \footnote{
The fermionic self energy is obtained by summing the $1PI$ graphs (see, e.g., Figure 5 in \cite{Aharony:2012ns}) and is given by \cite{Giombi:2011kc, Aharony:2012ns, Jain:2013py}
\begin{equation}
\Sigma_F(p)= - N_F \int \frac{\mc{D}_F^3\ell}{(2\pi)^3}  \ \mathcal{V}^{\mu}(\l,p)  S_F(\l) \ \mathcal{V}^{\nu}(p,\l) G^F_{\nu\mu}(\l-p)   \ .
\end{equation}}
\begin{equation}\label{fermionselfenrgy}
\begin{split}
\tl{\Sigma}_{\mathbb{1}}(p_s) =m_F +\lambda_{F}\xi_F(a(p_s)) \ ,  \  \Sigma_{+}(\vec{p} ) =\frac{p_{+}}{p_s^2}\big[c_F^2-\tl{\Sigma}^2_{\mathbb{1}}(p) \big] \ , \ \Sigma_{-}(p) \ = \ \Sigma_{3}(p)=0 \ ,
\end{split}
\end{equation}
where, $\tl{\Sigma}_{\mathbb{1}}=m_F +\Sigma_{\mathbb{1}}$, $a(p_s)=+\sqrt{p_s^2+c_F^2} \ $ and $p^2=2p_{+}p_{-}+p_3^2=p_s^2+p_3^2$. The function $\xi_F(z)$ is defined in \eqref{xiFz}. $c_F$ is the thermal mass of the fermion which is determined by the gap equation
\begin{equation}\label{masseqnforcF}
c_F =\sgn(h_F)  (m_F + \lambda_F \xi_F(c_F) )  \ ,
\end{equation}
where, $\sgn(h_F)= \sgn(m_F+\lambda_F \xi_{F}(c_F))$ (see e.g. in \cite{Choudhury:2018iwf}). We use the convention where, $c_F$ (also $c_B$ in the case of bosons) is always positive. There are two possibilities depending upon the two possible signs of $\sgn(h_F\lambda_F)$ \cite{Choudhury:2018iwf}) \footnote{The choice $\sgn(h_F\lambda_F)=+1$ corresponds to, under duality, the unHiggsed critical bosonic scalar theory, and the choice $\sgn(h_F\lambda_F)=-1$ corresponds to the Higgsed phases of critical bosons under duality.}. At zero temperature, $\sgn(h_F)$ is replaced by $\sgn(m_F)$.  
Diagramatically, the gauge field and the exact fermionic propagators are
\begin{displaymath}
\begin{tikzpicture}
\begin{feynman}
\vertex  (a) {\( \mu \)} ;
\vertex [right=2.5cm of a] (b) {\( \nu \)};
\diagram* {
	(a) -- [boson, momentum={[arrow shorten=0.35]\(p\)}] (b),
}; 
\draw (4.0,0) node {{$= \ G_{\nu\mu}^{F}(p)$}};
\end{feynman}
\end{tikzpicture} \\
\hspace{2cm}
\begin{tikzpicture}
\begin{feynman}
\coordinate (c) at (2.4,0) ; 
\vertex  (a) ;
\vertex [right=1.2cm of a] (b) ; 
\diagram* {
	(a) -- [fermion] (b),
}; 
\filldraw[black] (b) circle (3pt) ;
\draw (3.5,0) node {{$= \ S_{F}(p)$}};
\draw (b) -- (c) ; 
\draw (a)+(0.6, 0.3) node {{$p$}}; 
\end{feynman}
\end{tikzpicture}
\end{displaymath}
The vertex factor associated with the interaction term $\bar\psi(p) A_{\mu}(-p-k)  \psi(k)$ is given by 
\begin{equation}\label{vertexpsibarApsi}
\mathcal{V}^\mu(k,p)=\i \g^\mu  \ .
\end{equation}
We now compute various correlators of gauge invariant, single trace operators in this theory. 
\subsection{Case I: Spin zero}
The gauge invariant, single trace, spin $0$ operator in the regular fermionic theory is given by $J_0^{F}(x)=\bar{\psi}(x)\psi(x)$, which in the momentum space takes the form 
\begin{equation}\label{J0(-q)}
\begin{split}
J_0^{F}(-q)
& =\int \frac{\mathcal{D}_F^3 k}{(2\pi)^3}~\bar{\psi}(-(k+q))\psi(k) \\
\end{split} 
\end{equation}
In order to compute $\langle J_0^{F}J_0^{F} \rangle $ two-point correlator, we need to compute the exact $J_0^{F}$ insertion vertex, which we compute by solving the corresponding Schwinger-Dyson equations shown in Fig.\ref{sdexactvertx} in the subsection below. 
\subsubsection{Schwinger-Dyson equation for the exact vertex }
\begin{figure}[h!]
\begin{center}
\begin{tikzpicture}[scale=0.7, cross/.style={path picture={ 
		\draw[black]
		(path picture bounding box.south east) -- (path picture bounding box.north west) (path picture bounding box.south west) -- (path picture bounding box.north east);
}}]
\begin{feynman}
\coordinate (A) at (0,0);
\coordinate (D) at (2,2) ;
\coordinate (Dp) at (2,-2) ;
\coordinate (L) at (-1.1,0) ;
\diagram* {
	(Dp) -- [fermion] (A),  
}; 
\draw (A)+(2.5,0) node {{$= $}}; 
\node (c) [draw,circle,cross,minimum width=0.15 cm](b) at (A){};
\draw (A) -- (D) ; 
\draw [-] (A) -- (Dp) node [midway, below ] {$k$};
\draw[-latex] (L)--+(0.7,0);
\draw (L)+(0.3, 0.3) node {$q$}; 
\end{feynman}
\end{tikzpicture}  
\raisebox{7pt} {
\begin{tikzpicture}[scale=0.7, cross/.style={path picture={ \draw[black]
		(path picture bounding box.south east) -- (path picture bounding box.north west) (path picture bounding box.south west) -- (path picture bounding box.north east); }}]
\begin{feynman}
\coordinate (A) at (0,0);
\coordinate (D) at (2,1.7) ;
\coordinate (Dp) at (2,-1.7) ;
\coordinate (L) at (-1.1,0) ;
\diagram* {
	(Dp) -- [fermion] (A),  
}; 
\draw (A)+(2.5,0) node {{$+$}}; 
\draw (A) node {$\times$} ; 
\draw (A) -- (D) ; 
\draw [-] (A) -- (Dp) node [midway, below ] {$k$};
\draw[-latex] (L)--+(0.7,0);
\draw (L)+(0.3, 0.3) node {$q$}; 
\end{feynman}
\end{tikzpicture}  } 
\raisebox{-6pt}{ 
\begin{tikzpicture}[scale=0.7, cross/.style={path picture={ 
		\draw[black]
		(path picture bounding box.south east) -- (path picture bounding box.north west) (path picture bounding box.south west) -- (path picture bounding box.north east);
}}]
\begin{feynman}
\coordinate (A) at (0,0);
\coordinate (B) at (0.8,0.8) ;
\coordinate (C) at (1.5,1.5) ;
\coordinate (D) at (2,2) ;
\coordinate (Bp) at (0.8,-0.8) ;
\coordinate (Cp) at (1.5,-1.5) ;
\coordinate (Dp) at (2,-2) ;
\coordinate (L) at (-1.1,0) ;
\diagram* {
	(C) -- [boson, momentum={[arrow shorten=0.35]\(\ell -k \)}] (Cp),    }; 
\diagram* {
	(Dp) -- [fermion] (Cp),    }; 
\end{feynman}
\draw (A) -- (B) ;
\filldraw[black] (B) circle (3pt) ;
\draw (B) -- (C) ;
\draw (C) -- (D) ;
\draw (A) -- (Bp) ;
\filldraw[black] (Bp) circle (3pt) ;
\draw (Bp) -- (Cp) ;
\draw (Cp) -- (Dp) ;
\node [draw,circle,cross,minimum width=0.15 cm](b) at (A){};
\draw [-] (Cp) -- (Dp) node [midway, below ] {$k$};
\draw[-latex] (L)--+(0.7,0);
\draw (L)+(0.3, 0.3) node {$q$}; 
\end{tikzpicture} }
\end{center} 
\caption{Schwinger-Dyson equations for the exact vertices. The circled cross denote an insertion of an exact vertex and a filled circle denotes the exact fermion propagator. The bare cross denotes the `tree level' insertion.}
\label{sdexactvertx}
\end{figure}
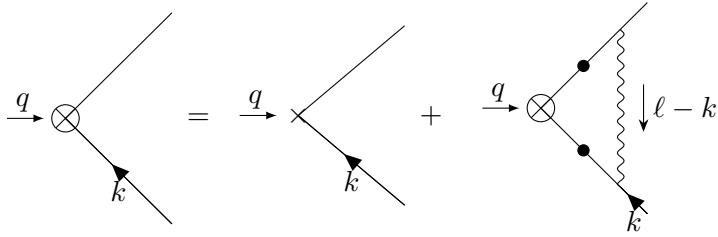

Diagrammatically, the Schwinger-Dyson equation for the exact $J_0^{F}$ vertex is shown in Fig.\ref{sdexactvertx}. The exact $J_0^{F}$ insertion vertex is defined by
\begin{equation}
\big\langle J_0^{F}(-q)\psi(k)\bar{\psi}(p) \big\rangle =V_{0}^F(k,q)~(2\pi)^3\delta^{(3)}(p+k+q) \ . 
\end{equation}
In the planar limit, the Schwinger-Dyson equation for $V_{0}^F(k,q)$ is given by 
\begin{equation}\label{SDforV0F} 
V_{0}^F(k,q)=\widetilde{V}_{0}^F(k,q)+\ N_F \int \frac{\mathcal{D}_F^3 \l }{(2\pi)^3} \big[  \mathcal{V}^\mu(\l+q,k+q) S_F(\l+q) V_{0}^F(\l,q) S_F(\l) \mathcal{V}^\nu(k,\ell)   \big] G^F_{\nu\mu}(\l-k) \ , 
\end{equation}
where, the factor of $N_F$ appearing in front of the second term is the contribution from the color factors. $\widetilde{V}_{0}^F$ is the tree-level insertion vertex corresponding to $J_{0}^F$. From the definition of $J_0^{F}$ operator in \eqref{J0(-q)}, it follows that $\widetilde{V}_{0}^F(k,q)=1$. We work in the `lightcone kinematics' $q_{\pm}=0$, in which case, the calculations simplfy considerably \footnote{The difficulty with working in the general case of $q_{\pm}\neq 0$ lies in the fact that the corresponding integrals over the spatial components $\ell_1$ and $\ell_2$ of the loop momenta $\ell$ are difficult to perform exactly.}. Using \eqref{vertexpsibarApsi} and \eqref{gaugebospropforFermion}, \eqref{SDforV0F} simplifies to
\begin{equation}\label{SDforV0Fsimplified} 
V_{0}^F(k,q)=1- 2\pi \i \lambda_{F} \int \frac{\mathcal{D}_F^3 \l }{(2\pi)^3} \ \Big[ \g^{[3|} S_F(\l+q) V_{0}^F(\l,q) S_F(\l) \g^{|+]}    \Big] \ \frac{1}{(\l-k)_{-}} \ .
\end{equation}
Here and in the rest of the paper, we use the notation 
\begin{equation}\label{gammaas}
\gamma^{[\rho|}A \g^{|\nu]} = \g^\rho A \g^\nu - \g^\nu A \g^\rho = 2\i \epsilon^{\rho \mu \nu} (A_{\mu}\mathbb{1}-A_{\mathbb{1}}\g_{\mu} ) \ ,
\end{equation}
 for any $2\times 2$ matrix $A= A_{\mu}\g^\mu + A_{\mathbb{1}}\mathbb{1}$. 
As $V_{0}^F(k,q)$ is a  $2\times 2$ matrix in the spinor space, it can be expanded in the complete basis $\{ \g^\mu, \mathbb{1} \}$ of $2\times 2$ matrices as $V_{0}^F(k,q)=V_{0,\mu}(k,q)\g^\mu +V_{0, \mathbb{1}}\mathbb{1} $. Comparing it with the RHS of \eqref{SDforV0Fsimplified}, we get $V_{0,-}(k,q)=V_{0,3}(k,q)=0$ and the following set of two non-trivial coupled integral equations 
\begin{equation}\label{componenteqnsforV0}
\begin{split}
V_{0,+}(k,q) & = - 4\pi \i \lambda_{F} \int \frac{\mathcal{D}_F^3 \l }{(2\pi)^3}  \  \big[S_F(\l+q) V_{0}^F(\l,q) S_F(\l)\big]_{\mathbb{1}} \  \frac{1}{(\l-k)_{-}}  \ , \\
V_{0,\mathbb{1} }(k,q) & = 1+ 4\pi \i \lambda_{F} \int \frac{\mathcal{D}_F^3 \l }{(2\pi)^3} \  \big[S_F(\l+q) V_{0}^F(\l,q) S_F(\l)\big]_{-} \  \frac{1}{(\l-k)_{-}} \ , 
\end{split}
\end{equation}
where, we have used the notation $[ABC]_{a}$ to denote the $a$-th component of the matrix product $ABC$ of three matrices $A$, $B$ and $C$, and $a\in (\mu,\mathbb{1})$. 
The RHS of the above equation \eqref{componenteqnsforV0} does not depend upon $k_3$. In the `lightcone kinematics' $q_{\pm} =0 $, the only non-zero component of $q$ is $q_3$. So, we use the $SO(2)$ rotational symmetry in the $1$-$2$ plane to write down 
\begin{equation}\label{definefgforV0}
V_{0,\mathbb{1} }(\vec{\l} ,q_3)\equiv f(\l_s,q_3) \ ,  \ \ \ \ \ 
V_{0,+}(\vec{\l} ,q_3)\equiv \frac{\l_{+}}{\ell_s^2}  g(\l_s,q_3) \ .
\end{equation}
The extra factor of $\ell_s^2$ in the denominator of the second expression above is kept for later convenience. Using \eqref{definefgforV0} and the explicit components of the exact fermion propagator \eqref{exactpsiprop}, the equations \eqref{componenteqnsforV0} can be simplified to give
\begin{equation}\label{eqnforfinV0}
f(k_s,q_3) = 1-4\pi \i \lambda_F \int\frac{\mathcal{D}_F^3\l }{(2\pi)^3} \frac{[-q_3+2\i \tl{\Sigma}_{\mathbb{1}}(\l_s)]f(\l_s,q_3)+ g(\l_s,q_3) } {(\l_3^2+a^2(\l_s))((\l_3+q_3)^2+a^2(\l_s)} \frac{\l_{-} }{(\l-k)_{-}} \ , 
\end{equation}
and
\begin{equation}\label{eqnforginV0}
\frac{k_{+}}{k_s^2} g(k_s,q_3) = 4\pi \i \lambda_F \int\frac{\mathcal{D}_F^3\l }{(2\pi)^3} \frac{[\l_3(\l_3+q_3)+a^2(\l_3)-2 \tl{\Sigma}^2_{\mathbb{1}}(\l_s)]f(\l_s,q_3)+\frac{1}{2}[q_3+2\i \tl{\Sigma}_{\mathbb{1}}(\l_s) ] g(\l_s,q_3) } {(\l_3^2+a^2(\l_s))((\l_3+q_3)^2+a^2(\l_s)) } \frac{1}{(\l-k)_{-}} \ . 
\end{equation}
Due to the $SO(2)$ rotational symmetry in the $\ell_1$-$\ell_2$ plane, it is useful to write the integration measure as 
\begin{equation}\label{DF3ell}
\mathcal{D}_F^3\l \equiv \ell_s d\ell_s d\theta_{\ell} \mathcal{D}_F \l_3 \ .
\end{equation} 
where, $\ell_s$ is the radial momenta in the $\ell_{+}$-$\ell_{+}$ plane (or equivalently, $\ell_{1}$-$\ell_{2}$ plane) and is given by $\ell_s^2=2\ell_{+}\ell_{-}=\ell_1^2+\ell_2^2$. $\theta_\ell$ is the angular variable in the lightcone plane. The measure $\mathcal{D}_F \l_3$ as before is given by \eqref{fermholonomy}. One can do the angular integration by using the result \eqref{angint} and also perform the integral over the momentum component $\l_3$ by using \eqref{GF2} and \eqref{GF3}; by doing so, we find
\begin{equation}\label{eqnforfinV0simpl}
f(k_s,q_3) = 1-2 \i \lambda_F \int_{k_s}^{\infty} \frac{\l_s d\l_s}{a(\l_s)} F_F(a(\l_s),q_3)\big[[-q_3+2\i \tl{\Sigma}_{\mathbb{1}}(\l_s)]f(\l_s,q_3)+ g(\l_s,q_3) \big] \ , 
\end{equation}
and
\begin{equation}\label{eqnforginV0simpl} 
 g(k_s,q_3) = -2 \i \lambda_F \int_{0}^{k_s}  \frac{\l_s d\l_s}{a(\l_s)} \ F_F(a(\l_s),q_3) \ \big[ 4[a^2(\l_s)- \tl{\Sigma}^2_{\mathbb{1}}(\l_s)]f(\l_s,q_3)+ [q_3+2\i \tl{\Sigma}_{\mathbb{1}}(\l_s) ] g(\l_s,q_3) \big]  \ .
\end{equation}
where, the function $F_F(z)$ is defined by $F_F(z)=\frac{\chi_F(z)}{q_3^2+4z^2}$, and $\chi_F(z)$ is given by \eqref{chiF}. In order to solve these two coupled integral equations \eqref{eqnforfinV0simpl} and \eqref{eqnforginV0simpl}, we first introduce a change of variable $a(\l_s) =+\sqrt{\l_s^2+c_F^2} =w$ and  $a(k_s) =+\sqrt{k_s^2+c_F^2} =z$. In terms of these reduced variables (as the functional dependence changes, so, we redefine $f(k_s,q_3)\equiv \tl{f}(z,q_3)$ and ${g}(k_s,q_3)\equiv \tl{{g} }(z,q_3)$ and $\tl{\Sigma}_{\mathbb{1}}(\l_s)\equiv h_F(w)$), 
\begin{equation}\label{eqnforfinV0reduced}
\tl{f}(z,q_3) = 1-2 \i \lambda_F \int_{z}^{\infty} dw \ F_F(w,q_3)\big[[-q_3+2\i h_F(w)]\tl{f}(z,q_3)+\tl{{g}}(w,q_3) \big] \ , 
\end{equation}
and
\begin{equation}\label{eqnforginV0reduced} 
\tl{{g}}(z,q_3) = -2 \i \lambda_F \int_{c_F}^{z}  dw  \ F_F(w,q_3) \ \big[ 4[w^2- h_F^2(w)]\tl{f}(w,q_3) + [q_3+2\i \tl{\Sigma}_{\mathbb{1}}(\l_s) ] \tl{{g}}(w,q_3) \big] \ .
\end{equation}
The above two coupled integral equations \eqref{eqnforfinV0reduced} and \eqref{eqnforginV0reduced} can be decoupled and solved easily, by converting them first to a set of two differential equations by taking derivatives w.r.t. the free parameter $z$. The corresponding differential equations take the following form 
\begin{equation}\label{eqnforfinV0diff}
\partial_{z} \tl{f}(z,q_3) = 2 \i \lambda_F \big[[-q_3+2\i h_F(z)] \tl{f}(z,q_3)+\tl{g}(z,q_3) \big] F_F(z,q_3) \ , 
\end{equation}
and
\begin{equation}\label{eqnforginV0diff} 
\partial_{z} \tl{{g}}(z,q_3) = -2 \i \lambda_F  \big[ 4[z^2-h_F^2(z) ]\tl{f}(z,q_3)+ [q_3+2\i h_F(z) ] \tl{{g}}(z,q_3) \big] \ F_F(z,q_3) \ .
\end{equation}
There is a particuar combination of the two equations \eqref{eqnforfinV0diff} and \eqref{eqnforginV0diff}, which can be written as a total derivative w.r.t. the variable $z$. This is easily done by multiplying \eqref{eqnforfinV0diff} by $(q_3+2\i h_F(z))$ and then adding that to \eqref{eqnforginV0diff}. From the definitions $h_F(z)=m_F+\lambda_F\xi_F(z)$, and $\xi_F(z)=\int^{z}\chi_F(w)dw$, it follows that $h_F'(z)=\lambda_F \xi_F'(z)=\lambda_F \chi_F(z)$. Constructing the particular combination mentioned in this paragraph, we find that it can be written as a total derivative as given below 
\begin{equation}\label{eqnfor(f+g)inV0diffsimpl}
\begin{split}
\partial_{z} \big[(q_3+2\i h_F(z)) \tl{f}(z,q_3)  \ + \  \tl{{g}}(z,q_3)\big] =0 \ . 
\end{split} 
\end{equation}
The general solution of \eqref{eqnfor(f+g)inV0diffsimpl} can be written as 
\begin{equation}\label{solnfor(f+g)inV0diff}
\begin{split}
(q_3+2\i h_F(z)) \tl{f}(z,q_3)  \ + \  \tl{{g}}(z,q_3)=\eta(q_3)  \ ,
\end{split} 
\end{equation}
where, $\eta(q_3) $ is an unknown function of $q_3$ to be determined from the boundary conditions. From \eqref{eqnforfinV0reduced} and \eqref{eqnforginV0reduced}, we see that the above differential equations \eqref{eqnforfinV0reduced} and \eqref{eqnforginV0reduced} must satisfy the following boundary conditions 
\begin{equation}\label{boundcondforfandginV0} 
\tl{f}(z=\Lambda,q_3) = 1, \ \ \ \ \  \tl{{g}}(z=c_F,q_3) = 0, \ \ \ \text{where}, \ \  \Lambda\rightarrow \infty  \ .
\end{equation} 
Here and also later in this paper, we have introduced a UV cutoff $\Lambda$ in the radial momentum integral, in the intermediate steps of calculations to keep track of the divergent terms in the integrals which we have eventually dropped away by regularization. Finally, $\Lambda$ is taken to infinity. Using \eqref{solnfor(f+g)inV0diff}  we substitute $\tl{{g}}(z,q_3)$ in \eqref{eqnforfinV0reduced} and get 
\begin{equation}\label{eqnforfinV0diffgreplaced}
\begin{split}
\partial_{z} \tl{f}(z,q_3) 
& = 2 \i \lambda_F \big[\eta(q_3)- 2q_3  \tl{f}(z,q_3) \ 
\big] F_F(z,q_3) \ .
\end{split}
\end{equation}
To solve the above equation \eqref{eqnforfinV0diffgreplaced}, we choose an ansatz of the form
\begin{equation}\label{ansatzf0}
\tl{f}(z,q_3) = A\bigg(1+B \exp\big[4\i \lambda_F q_3 \int_{z}^{\Lambda} dw \ F_F(w,q_3)\big]\bigg) \ . 
\end{equation}
Using the definition, $H_F(z,q_3)=\exp\big[4\i \lambda_F q_3 \int^{z} dw \ F_F(w,q_3)\big]$, \eqref{ansatzf0} can be written in the following form 
\begin{equation}\label{ansatzfoa}
\tl{f}(z,q_3) =A\bigg(1+B \frac{H_F(\Lambda,q_3)}{H_F(z,q_3)} \bigg) \ , 
\end{equation}
which is more useful in the intermediate steps of the calculations. 
There are three unknowns $A$, $B$ and $\eta(q_3)$ which need to be fixed to completely determine $\tl{f}$ and $\tl{g}$. Using the boundary condition \eqref{boundcondforfandginV0} on $\tl{f}$, we fix $A$ in terms of $B$. $\eta(q_3)$ is in terms of $B$ by substituting the ansatz \eqref{ansatzf0} in \eqref{eqnforfinV0diffgreplaced}. Finally, we use the boundary condition on $\tl{g}$ to determine $B$ from equation \eqref{solnfor(f+g)inV0diff}. The final result for the exact $J_0^F$ insertion vertex \eqref{definefgforV0} is given by
\begin{equation}\label{finalsolnforfandgwithB'} 
\begin{split}
\tl{f}(z,q_3) &=\frac{1+ B \frac{H_F(\Lambda,q_3)}{H_F(z,q_3)} }{1+B}  \ , \ \ \ 
\tl{g}(z,q_3)  = \frac{2q_3}{1+B} - \big(q_3+2\i (m_F+\lambda_F \xi_F(z))\big)\tl{f}(z,q_3) \ ,
\end{split}
\end{equation}
where, $B$ is given by 
\begin{equation}
\begin{split}
B =\frac{\big(q_3-2\i h_F(c_F)\big)}{\big(q_3+2\i h_F(c_F)\big)}\frac{H_F(c_F,q_3)}{H_F(\Lambda,q_3)} \ .
\end{split} 
\end{equation}
\subsubsection{Two-point function : }
\begin{figure}[h!]
	\begin{center}
		\begin{tikzpicture}[scale=0.7, cross/.style={path picture={ 
				\draw[black]
				(path picture bounding box.south east) -- (path picture bounding box.north west) (path picture bounding box.south west) -- (path picture bounding box.north east);
		}}]
		\begin{feynman}
		\coordinate (A) at (0,0) ;
		\coordinate (B) at (6,0) ;
		\coordinate (L) at (-1.1,0) ;
		\draw (A) node {$\otimes $} ; 
		\draw (B) node {$\times$} ; 
		\filldraw (3,1.5) circle (3pt) ;
		\filldraw (3,-1.5) circle (3pt) ;
		\end{feynman}
		\draw (0,0) .. controls (2,2) and (4,2) .. (6,0);
		\draw (0,0) .. controls (2,-2) and (4,-2) .. (6,0);
		\draw[-latex] (L)--+(0.7,0);
		\draw (L)+(0.3, 0.3) node {$q$}; 
		\draw (A)+(0.8,0) node {$\ J^{(s)}$};
		\draw (5,0) node {$\ J^{(s)}$};
		\end{tikzpicture}
		\caption{Diagram contributing to two-point function $\langle J^{(s)}(-q)J^{(s)}(q)\rangle$ for spin zero and spin one. Filled circle denotes the exact fermion propagator.}
		\label{jj2pt}
	\end{center} 
\end{figure}
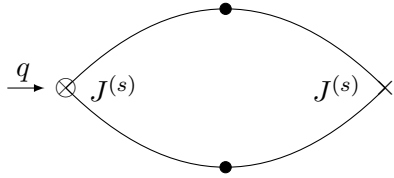

The two-point function is computed by computing the Feynman diagram shown in Fig.\ref{jj2pt}. To compute the two-point function $\langle J_0^{F}(q') J_0^{F}(q) \rangle $, only a single insertion of an exact vertex $J_0^{F}(q')$ is required to account for all the perturbative Feynman diagrams without any overcounting. 
For definiteness, we choose insertion at the left in Fig.\ref{jj2pt} as the exact vertex. Insertion on right side in Fig.\ref{jj2pt} is the `free' insertion vertex \footnote{We alternately use the terms `free', `tree level' and `bare' insertion vertex for the insertion on the RHS of the schematic diagrams Fig.\ref{jj2pt} or Fig.\ref{jj2ptBs1} of two-point correlators. What this exactly means is that it includes only the required diagrams (which may contain loops) to avoid any overcounting \cite{Aharony:2012nh, GurAri:2012is, Ghosh:2019sqf}. The required diagrams for the `bare' insertion vertex can be got from the definitions of the corresponding current operator.} which we define as follows :
\begin{equation}
\langle J_0^{F}(-q)\psi(k)\bar{\psi}(p)\rangle =U_{0}^F(k,q)~(2\pi)^3\delta^{(3)}(p+k+q) \ . 
\end{equation}
From the definition of $J_0^{F}(q)$ operator in \eqref{J0(-q)}, it follows that $U_{0}^F(k,q) =1$. 
We define the two-point correlation function of $J_0^{F}$ operator as 
\begin{equation}\label{J0(-q)J0(q)}
\langle J_0^{F}(q')J_0^{F}(q)\rangle =\mc{G}_0^F(q) \ (2\pi)^3 \delta^{(3)}(q'+q) \ . 
\end{equation}
From Fig.\ref{jj2pt}, the expression for $\mc{G}_0^F$ is given by 
\begin{equation}\label{integralforG0F(q)}
\mc{G}_0^F(q) =-N_F\int \frac{\mathcal{D}_F^3k}{(2\pi)^3} \ \text{Tr}_F \bigg[U_0^F(k+q,-q) \ S_F(k+q) \ V_0^F(k,q) \  S_F(k)\bigg] \ ,
\end{equation}
where, the extra $(-1)$ factor above in \eqref{integralforG0F(q)} is because of the integration over fermion loop. Here, $\text{Tr}_F$ indicates the trace in spinor space. Inserting the expressions of exact fermion propagator \eqref{exactpsiprop}, the exact insertion vertex $V_0^F$ computed in \eqref{finalsolnforfandgwithB'}, the expression of $U_0^F$, doing the gamma-matrix algebra and performing the momentum integral over the loop momenta $k$, the final result for $\langle J_0^{F}J_0^{F}\rangle$ that we find is given by
\begin{equation}\label{G0F(q)finalB'replaced}
\begin{split}
\mc{G}_0^F(q) 
& = \lim_{\Lambda \rightarrow \infty}  \bigg[
\frac{\i N_Fq_3 }{4\pi \lambda_{F} }\frac{ \frac{H_F(\Lambda,q_3)}{H_F(c_F,q_3)}e^{-2\i \sgn(h_F)\tan^{-1}\frac{q_3}{2c_F}}+ 1}{\frac{H_F(\Lambda,q_3)}{H_F(c_F,q_3)}e^{-2\i \sgn(h_F)\tan^{-1}\frac{q_3}{2c_F}} - 1} + \frac{N_F h_F(\Lambda)  }{2\pi \lambda_{F} } \ \bigg]  \ ,
\end{split} 
\end{equation}
where, $H_F(z,q_3)=\exp\big[4\i \lambda_F q_3 \int^{z} dw \ F_F(w,q_3)\big]$, $F_F(z,q_3)=\frac{\chi_F(z)}{q_3^2+4z^2}$ and $h_F(z)=m_F+\lambda_F \xi_F(z)$, and $\sgn(h_F)$ is defined in the gap equation \eqref{masseqnforcF}. From the expression of $\xi_F(w)$ in \eqref{xiFz}, it is clear that $\xi_F(\infty)$ diverges linearly. However, we regularize \cite{GurAri:2012is, Aharony:2012ns} the answer by subtracting the divergent term $ \xi_F(\infty)$ and take the $\Lambda\to \infty$ limit \footnote{One can use the dimensional regularization for the  radial momenta integrals as discussed in \cite{Choudhury:2018iwf} to remove the term $\xi_F(\infty)$. Alternatively, it can be done by adding a mass counterterm for the background source of $J^{F}_0$ as discussed in \cite{GurAri:2012is}.}. Thus, the renormalized two-point function $\langle J_0^{F}J_0^{F}\rangle$ is
\begin{equation}\label{G0F(q)thefinal}
\begin{split}
\mc{G}_0^F(q) 
=
\frac{\i N_Fq_3 }{4\pi \lambda_{F} }\frac{ \frac{H_F(\infty,q_3)}{H_F(c_F,q_3)}e^{-2\i \sgn(h_F)\tan^{-1}\frac{q_3}{2c_F}}+ 1}{\frac{H_F(\infty,q_3)}{H_F(c_F,q_3)}e^{-2\i \sgn(h_F)\tan^{-1}\frac{q_3}{2c_F}} - 1} + \frac{N_F m_F}{2\pi \lambda_{F} } \ . 
\end{split} 
\end{equation}
\eqref{G0F(q)thefinal} is valid at finite temperature and non-zero mass. In various limiting cases, this agrees with the existing results \cite{GurAri:2012is,Ghosh:2019sqf, amiyata1}.

\subsection{Case II: Spin one}
In this section we generalize the computaions of two-point correlators of spin one operator, done previously \cite{GurAri:2012is, Gur-Ari:2016xff, Ghosh:2019sqf, amiyata1}, to the general case of massive regular fermionic theory at finite temperature with arbitrary holonomy distribution. The regular fermionic theory \eqref{rflag} that we study in this paper has a gauge invariant, conserved $U(1)$ current, given by the single trace operator $J^{F}_\mu (x)=\i \ \bar{\psi}(x)\g_\mu \psi(x)$. 
In momentum space, this operator is given by
\begin{equation}\label{J1Fmu(-q)}
\begin{split}
J^{F}_{\mu}(-q)
& =\i \int \frac{\mathcal{D}_F^3 k}{(2\pi)^3}~\bar{\psi}(-(k+q)) \gamma_\mu \psi(k)  \ . \\
\end{split} 
\end{equation}
Our goal in this section is to compute the two-point correlator $\langle J^{F}_\mu J^{F}_\nu \rangle $, for which as in the case of $J_0^F$, we need the exact $J^{F}_\mu$ vertex. 
\subsubsection{Schwinger-Dyson equation for the exact insertion vertex}
Exact $J^{F}_\mu$ insertion vertex is computed by solving the Schwinger-Dyson equation given schematically in Fig.\ref{sdexactvertx}. The exact $J^{F}_\mu $ vertex is defined as
\begin{equation}
\langle J^{F}_\mu (-q)\psi(k)\bar{\psi}(p)\rangle =V_{(\mu) }^{F}(k,q)~(2\pi)^3\delta^{(3)}(p+k+q) \ .
\end{equation}
The corresponding Schwinger-Dyson equation for $V_{\mu}^{F}(k,q)$ is given by 
\begin{equation}\label{SDforV1Fmu} 
V_{(\mu)}^{F}(k,q)=\widetilde{V}_{(\mu)}^{F}(k,q)+\ N_F \int \frac{\mathcal{D}_F^3 \l }{(2\pi)^3} \big[  \mathcal{V}^\nu(\l+q,k+q) S_F(\l+q) V_{(\mu)}^{F}(\l,q) S_F(\l) \mathcal{V}^\rho(k,\l)   \big] G^F_{\rho \nu}(\l-k) \ ,
\end{equation}
where, $\widetilde{V}_{(\mu)}^{F}(k,q)$ denotes the tree-level insertion of $ J^{F}_\mu $. From \eqref{J1Fmu(-q)}, it follows that $\widetilde{V}_{(\mu)}^{F}(k,q)=\i \g_\mu $. Using \eqref{vertexpsibarApsi} and \eqref{gaugebospropforFermion} , the above equation \eqref{SDforV1Fmu} can be simplified to
\begin{equation}\label{SDforV1Fmusimplified} 
V_{(\mu)}^{F} (k,q)=\i\g_\mu - 2\pi \i \lambda_{F} \int \frac{\mathcal{D}_F^3 \l }{(2\pi)^3} \big[ \g^{[3|} S_F(\l+q) V_{(\mu)}^{F}(\l,q) S_F(\l) \g^{|+]}    \big] \frac{1}{(\l-k)_{-}}  \ .
\end{equation}
The only non-zero component of $\langle J_{\mu}^F(-q)J_{\nu}^F(q) \rangle $ in the  $x^1$-$x^2$ plane, in `lightcone kinematics' $q_{\pm}=0$, is the $(\mu\nu)\equiv (-+)$ (or equivalently $(+-)$) component \footnote{This folows from the $SO(2)$ rotational symmetry in the $1$-$2$ plane.}. Other components are zero \footnote{As the $U(1)$ current is classically conserved, it follows from the Ward identity, that in the momentum space, $q^{\mu}\langle J_{\mu}(-q)J_{\nu}(q)\rangle$ should vanish upto (at most) a contact term \cite{Ghosh:2019sqf}. With the external momentum choice $q_{\pm}=0$, it follows then that the component $\langle J_{3}J_{3}\rangle$ should vanish upto a contact term. We have also explicitly checked (though not provided here keeping in mind about the length of the paper) in the case of fermionic theory (with the $U(1)$ current given by \eqref{J1Fmu(-q)}), that this is indeed the case, i.e., $\langle J_{-}J_{-}\rangle $, $\langle J_{+}J_{+}\rangle$, $\langle J_{\pm}J_{3}\rangle$ components vanish in the choice $q_{\pm}=0$. Also we have explicitly checked that $\langle J_{3}J_{3}\rangle$ vanishes exactly in this momentum choice.}. So, from now on, we will be considering only $\langle J_{-}J_{+}\rangle$ (the same argument goes for the bosonic case as well).   Also as in the $J_0^F$ case, in order for the computation of two-point functions, only a single exact vertex is required. To compute $\langle J_{-}^F(-q)J_{+}^F(q) \rangle $, we will use the exact $J_{-}^F$ insertion vertex and the `tree level' insertion vertex for $J_{+}^F$. So, we need to compute the exact $J_{-}^F$ insertion vertex. It is clear from \eqref{SDforV1Fmusimplified} that $V_{(\mu)}^{F} (k,q)$ is independent of $k_3$. Also, in the lightcone kinematics $q_{\pm}=0$, the only non-zero component of the external momenta $q$ is $q_3$. So, the momentum dependence of $V_{(\mu)}^{F} (k,q)$ is basically $V_{(\mu)}^{F} (\vec{k},q_3)$. As in the $J_0^F$ case, $V_{(-)}^{F}(k,q)$ can be expanded as, 
\begin{equation}\label{Vmfgdef}
V_{(-)}^{F}(\vec{k},q_3)=V_{(-),\nu}^{F}(\vec{k},q_3) \g^\nu +V_{(-), \mathbb{1}}^{F}(\vec{k},q_3) \mathbb{1}  =  g_{m}(k_s,q_3) \g^+ + k_{-} f_{m}(k_s,q_3) \mathbb{1}  \ .
\end{equation} 
To solve for exact $V_{(-)}^{F}$, we plug \eqref{Vmfgdef} in the minus-component equation of \eqref{SDforV1Fmusimplified} and get a set of two coupled integral equations involving $f_m$ and $g_m$. As in the case of $J_0^F$, these equations for $f_m$ and $g_m$ get simplified a lot once we utilize the $SO(2)$ rotational symmetry in the lightcone plane and decompose the momentum integration measure as \eqref{DF3ell}. We perform the angular integration using \eqref{angint} and also perform the integral over momentum component $\l_3$ by using \eqref{GF2} and \eqref{GF3}. Performing a change of variable $a(\l_s) =+\sqrt{\l_s^2+c_F^2} =w$ and  $a(k_s) =+\sqrt{k_s^2+c_F^2} =z$,  and relabelling $f_m(k_s,q_3)\equiv \tl{f}_m(z,q_3)$ and $g_m(k_s,q_3)\equiv \tl{g}_m(z,q_3)$ and $\tl{\Sigma}_{\mathbb{1}}(\l_s)\equiv h_F(w)$, the equations for $\tl{f}_m$ and $\tl{g}_{m}$ can be written in a much simpler form, which takes the following form
\begin{equation}\label{eqnforfinV1Freduced}
\tl{f}_m(z,q_3) = 2 \i \lambda_F \int_{z}^{\infty} dw \ F_F(w,q_3)\bigg[\big(q_3-2\i h_F(w)\big) \tl{f}_m(z,q_3) \ - \ 2\tl{g}_m(w,q_3) \bigg] \ , 
\end{equation}
and
\begin{equation}\label{eqnforginV1Freduced} 
\tl{g}_m(z,q_3) =\i \ +\  2 \i \lambda_F \int_{z}^{\infty}  dw  \ F_F(w,q_3) \ \bigg[ 2\big(w^2- h_F^2(w)\big) \tl{f}_m(w,q_3) + \big(q_3+2\i \tl{\Sigma}_{\mathbb{1}}(\l_s) \big) \tl{g}_m(w,q_3) \bigg]  \ .
\end{equation}
To solve the above two equations \eqref{eqnforfinV1Freduced} and \eqref{eqnforginV1Freduced}, it is best to convert them into a set of differential equations. The boundary conditions that follow from \eqref{eqnforfinV1Freduced} and \eqref{eqnforginV1Freduced} are \footnote{The boundary conditions for the exact vertex $V_{(-)}^F$ is different from the the boundary conditions for that of the exact $V_{0}^F$ vertex.}
\begin{equation}\label{boundcondforfandginV1F} 
\tl{f}_m(z=\Lambda,q_3) = 0, \ \ \ \ \  \tl{g}_m(z=\Lambda ,q_3) = \i , \ \ \ \text{where}, \ \  \Lambda\rightarrow \infty  \ .
\end{equation} 
Solving the equations \eqref{eqnforfinV1Freduced} and \eqref{eqnforginV1Freduced}, by first converting them into differential equations with the boundary conditions \eqref{boundcondforfandginV1F}, we find the final solution for the exact $J_{-}^F$ vertex in terms of $f_m$ and $g_m$ as
\begin{equation}\label{finalsolnforfandginV1F} 
\begin{split}
\tl{f}_m(z,q_3) &=\frac{\i}{q_3}\bigg(1- \exp\big[4\i \lambda_F q_3 \int_{z}^{\Lambda} dw \ F_F(w,q_3)\big]\bigg) \ , \\ 
\tl{g}_m(z,q_3) & = \i  - \frac{1}{2}\big(q_3+2\i h_F(z)\big)\tl{f}_m(z,q_3) \ .
\end{split}
\end{equation}
\subsubsection{Two-point function}\label{2ptJ0F}
In this subsection, we compute $\langle J_{\mu}^{F}(-q)J_{\nu}^{F}(q)\rangle $. We define the two-point correlator of spin one operator $J_{\mu}^{F}$ as
\begin{equation}\label{J1Fmu(-q)J1Fnu(q)}
\langle J_{\mu}^{F}(-q)J_{\nu}^{F}(q)\rangle =\mc{G}_{\mu\nu}^{F}(q) \ (2\pi)^3 \delta^{(3)}(q'+q) \ .
\end{equation}
The corresponding Feynman diagram for this is given in Fig.\ref{jj2pt}. Similar to the $J_0^F$ case, the insertion on left side of the diagram is chosen to be the exact vertex, and the insertion on right side is chosen to be the `tree-level' insertion to avoid any overcounting of the diagrams. We define the `tree-level' insertion $J^{F}_{\nu}$ vertex as follows
\begin{equation}
\langle J_{\nu}^{F}(-q)\psi(k)\bar{\psi}(p)\rangle =U_{(\nu)}^{F}(k,q)~(2\pi)^3\delta^{(3)}(p+k+q) \ .
\end{equation}
From the definition of $J_{\mu}^{F}$ current in \eqref{J1Fmu(-q)}, it follows that $U_{(\nu)}^{F}(k,q) =\i \g_\nu $. 
From Fig.\ref{jj2pt}, we see that the two-point function of the spin one current is given by 
\begin{equation}\label{integralforG1F(q)}
\mc{G}_{\mu\nu}^F(q) =-N_F\int \frac{\mathcal{D}_F^3k}{(2\pi)^3} \ \tr_F \bigg[ \ S_F(k+q) \ V_{(\mu)}^{F}(k,q) \  S_F(k) \ U_{(\nu)}^{F}(k+q,-q)\ \bigg] \ , 
\end{equation}
where, again the extra $(-1)$ factor above in \eqref{integralforG1F(q)} is because of the integration over fermion loop. Using the expression of $U_\nu^F$ and performing the gamma matrix algebra \footnote{Use the fact that for a general matrix $M=M_\mu \g^\mu +M_{\mathbb{1}}\mathbb{1} $, $\tr_F(M\g_\nu )= M_{\mu} \tr_F(\g^\mu \g_\nu )=2M_{\nu } $. }, \eqref{integralforG1F(q)} reduces to 
\begin{equation}\label{integralforG1F(q)simpl2}
\mc{G}_{\mu\nu}^{F}(q) =-2\i N_F\int \frac{\mathcal{D}_F^3k}{(2\pi)^3} \ \big[ S_F(k+q) \ V_{(\mu)}^{F}(k,q) \  S_F(k)\big]_{\nu} \ .
\end{equation} 
Below we consider the $(\mu\nu)\equiv (-+)$ component of \eqref{integralforG1F(q)simpl2}. Using the expressions of $S_F$ and $V_{-}^F$ in \eqref{integralforG1F(q)simpl2} and performing the momentum integral, we find 
\begin{equation}\label{G1F(q)final4}
\begin{split}
\mc{G}_{-+}^{F}(q) 
&= \lim_{\Lambda \rightarrow \infty} \bigg[\frac{\i N_F q_3 }{16\pi \lambda_F}  \bigg(1+ \frac{2\i h_F(c_F) }{q_3 }\bigg)^2\bigg[\frac{H_F(\Lambda,q_3)}{H_F(c_F,q_3)} -1 \bigg] -\frac{N_F\xi_F(c_F)}{8 \pi}+ \frac{N_F\xi_F(\Lambda)}{8 \pi}\bigg] 
\end{split}
\end{equation}
As in the case of $J_0^F$,  \eqref{G1F(q)final4} is linearly divergent due to the appearance of $\xi_F(\infty)$. Regularizing the above answer by throwing away the term $\xi_F(\infty)$ \footnote{This divergence term can be removed by dimensional regularization as discussed in \cite{Choudhury:2018iwf}. Alternatively, one can also subtract this by adding the mass counterterm for the source field corresponding to $J_{\mu}^F$ \cite{GurAri:2012is}.}, we report below the final result for the renormalized current two-point correlator $\mc{G}_{-+}^F$ 
\begin{equation}\label{G1F(q)final}
\begin{split}
\mc{G}_{-+}^{F}(q) 
&= \frac{\i N_F q_3 }{16\pi \lambda_F}  \bigg(1+ \frac{2\i h_F(c_F) }{q_3 }\bigg)^2\bigg[\frac{H_F(\infty,q_3)}{H_F(c_F,q_3)} -1 \bigg] -\frac{N_F\xi_F(c_F)}{8 \pi}
\end{split}
\end{equation}
Alternatively, one could have considered the exact $J_{+}^F$ vertex and the `tree level' $J_{-}^F$ insertion vertex and perform the above excercise and would have got the same result \footnote{We also explicitly checked this (not provided here).}. This follows from the fact that $\mc{G}_{-+}^F(q)= \mc{G}_{+-}^F(-q)$.   

\section{Thermal correlators in the bosonic theory}\label{bosonth}
\subsection{Brief review of the theory} 
In this section, we study the mass deformed regular bosonic matter theory coupled to $SU(N_B)$ Chern-Simons gauge fields in the large $N_B$ limit, at finite temperature. The Euclidean action for this theory is given by 
\begin{equation}\label{RBlag} 
\begin{split}
\mc{S}_{\text{B}}[A, \phi]  &  =\frac{\i \kappa_B}{4\pi} \int \ud^3 x\ \epsilon^{\mu\nu\rho}\,\text{tr}\left( A_\mu \partial_\nu A_\rho - \frac{2\i}{3} A_\mu A_\nu A_\rho\right)\   \\
&  + \int d^3 x ~\bigg((D_\mu \bar{\phi})(D^\mu \phi)+m_B^2 \bar{\phi}\phi+\frac{b_4}{2N_B}(\bar{\phi}\phi)^2+\frac{b_6}{6N_B^2}(\bar{\phi}\phi)^3\bigg) \ . 
\end{split}
\end{equation}
One of our goal is to check the duality between the fermionic and the bosonic theories. We have studied the regular fermionic matter theory in the previous section which is dual to the critical boson theory. We will study critical boson theory by taking critical limit of the regular boson theory defined by the action \eqref{RBlag}, in the next section. We work in the lightcone gauge $A_{-}=0$. 
The Feynman rules in this theory include the following :
\subsubsection*{Propagator for the gauge field}
The gauge boson propagator in the lightcone gauge $A_{-}=0$, in Euclidean space, is given by 
\begin{equation}
\langle A_\mu (p') A_\nu(p)\rangle = G_{\nu\mu}^B(p) ~(2\pi)^{3}\delta^{(3)}(p+p') \ , 
\end{equation}
where, 
\begin{equation}\label{gaugebosonprop}
G^B_{\nu\mu} (p) =\frac{2\pi\epsilon_{\nu-\mu}}{\kappa_B p_{-}}  \ . 
\end{equation}
As also mentioned in the fermionic case, the gauge field propagator \eqref{gaugebosonprop} is independent of $p_3$, so is not effected by the holonomy \cite{Aharony:2012ns}. We label the gauge fields both in the case of fermionic theory and here in the bosonic theory by the same symbol $A_\mu$. However, the distinction should be obvious from the context whether we study fermionic or bosonic theory. Also we denote the gauge field propagator by the same feynman diagram as shown around equation \eqref{gaugebospropforFermion} but now it is equal to $G_{\nu\mu}^B(p)$. 

\subsubsection*{Propagator for the scalar field}
The exact propagator for the scalar fields in the Euclidean space is given by 
\begin{equation}\label{exactphipropdef}
\langle \phi(p)\bar{\phi}(p')\rangle = S_B(p) ~(2\pi)^{3}\delta^{(3)}(p+p') \ , 
\end{equation}
where,
\begin{equation}\label{exactphiprop}
S_{B}(p) =\frac{1}{p^2+c_B^2} \ .
\end{equation}
Here, $c_B$ is the thermal mass of the scalar field, which is related to the bare mass-squared $m_B^2$ as $c_B^2=m_B^2+\Sigma_{B}$, where $\Sigma_B$ is the bosonic self energy \footnote{As discussed in the literature in great details, one can compute the self energy either by summing the feynman diagrams or by intgerating out the matter fields and using the Hubbard-Stratonovich trick.}. The final result for the gap equation of the thermal is given by \footnote{As already mentioned before, we use the convention such that $c_B$ is always positive. In the zero temperature limit, $\xi_B(c_B)$ reduces to $c_B$. So, in this limit \eqref{thermalcB} reduces to $$c_B^2=m_B^2-\frac{b_4}{4\pi}c_B +\left(\frac{\lambda_{B}^2}{4}+\frac{b_6}{32\pi^2}\right)c_B^2 \ . $$}
\begin{equation}\label{thermalcB}
c_B^2=m_B^2-\frac{b_4}{4\pi}\xi_B(c_B)  +\left(\frac{\lambda_{B}^2}{4}+\frac{b_6}{32\pi^2}\right)\xi_B^2(c_B) \ , 
\end{equation}
where, the function $\xi_B(x)$ is defined in \eqref{xiB}. Solving the equation \eqref{thermalcB},  one finds the thermal mass $c_B$ of the regular boson theory. In Feynman diagram, the exact scalar propagator is denoted by 
\begin{displaymath}
\begin{tikzpicture}
\begin{feynman}
\coordinate (c) at (2.4,0) ; 
\vertex  (a) ;
\vertex [right=1.2cm of a] (b) ; 
\filldraw[black] (b) circle (3pt) ;
\draw (3.5,0) node {{$= \ S_{B}(p)$}};
\draw (a) -- (c) ; 
\draw (a)+(0.6, 0.3) node {{$p$}}; 
\end{feynman}
\end{tikzpicture}
\end{displaymath}

In the lightcone gauge $A_{-}=0$, 
the contribution to the vertex factor corresponding to the term $\bar\phi A_3 A^3 \phi $ is $\mathcal{V}_{\bar{\phi}A^2\phi}=-1$.
In momentum space of the action \eqref{RBlag}, 
the vertex contribution corresponding to the  interaction term $\bar{\phi}(p)A_{\mu}(-(p+k)) \phi(k)$, is given by $\mathcal{V}_{\bar{\phi}A\phi}^\mu(k,p)=(k-p)^{\mu}$, with explicit momentum conservation at the vertex. 

\subsection{Thermal four-point function of fundamental scalars}
In the zero-temperature theory, the connected scalar four-point function was computed previously in the literature (see, e.g., \cite{Aharony:2012nh, Jain:2014nza}). In \cite{Aharony:2012nh}, the authors computed the connected scalar four-point function in the massless regular boson theory. On the other hand, in the case of massive regular boson theory as given by the action \eqref{RBlag}, the off-shell connected scalar four-point function was computed in \cite{Jain:2014nza}, at zero temperature. In this section, we generalize these computations to the finite temperature for the massive theory. We closely follow the procedure given in \cite{Jain:2014nza} \footnote{For details of their method, see section 3.1 and appendix D of \cite{Jain:2014nza}.} to do the computation of four-point function of scalars. The corresponding Schwinger-Dyson equations that we need to solve are the same as given in \cite{Jain:2014nza} (for the relevant Schwinger-Dyson equations see \eqref{eqn1} and \eqref{eqn2} \footnote{Or, see e.g. equation 4.6 of \cite{Jain:2014nza}.}, and for the diagrams see Fig.\ref{4ptscalarsd},\ref{oneloop} and \ref{4ptvertex} \footnote{Or, see figure 5 and figure 4 of \cite{Jain:2014nza}.}). Here, we generalize these computations to the finite temperature by taking into account the effect of holonomy. In the Appendix \ref{4ptscalar}, we present the details of the computation and discuss about the relevant modifications we need to do, because of the holonomy, in the methods of computations given in \cite{Jain:2014nza}. In this subsection we present the finite temperature results of the off-shell four-point function of fundamental scalars. 

Following \cite{Aharony:2012nh, Jain:2014nza}, we define the exact, connected off-shell scalar four-point function by
\begin{equation}\label{4ptdef}
\langle \phi^i(p+q)\bar{\phi}_{j}(-(k+q)) \phi^m(k) \bar{\phi}_{n}(-p) \rangle  = \mathcal{A}^{im}_{jn}(p,k,q)  \ (2\pi)^3 \delta^{(3)}(0) \ . 
\end{equation}
Here, we have explicitly shown the color indices of the scalar fields. Without loss of generality, we choose the color contractions to be $\delta^{i}_{n} \delta^{m}_{j}$(terms with other possible color contractions are related to this by the permutation of momenta) \cite{Aharony:2012nh, Jain:2014nza}; so, we consider \footnote{$\mathcal{A}(p,k,q)$ here is the same as the quantity ${V}(p,k,q)$ in \cite{Jain:2014nza} but now at finite temperature. We use a different symbol to avoid the notational clash with the rest of the paper.} 
\begin{equation}
\mathcal{A}^{im}_{jn}(p,k,q)=\mathcal{A}(p,k,q) \ \delta^{i}_{n} \delta^{m}_{j} \ . 
\end{equation}
 Following \cite{Jain:2014nza} and including the finite temperature effects as discussed in the Appendix \ref{4ptscalar}, we compute $\mathcal{A}(p,k,q)$ in the choice of the overall external momenta $q_{\pm}=0$, by solving the Schwinger-Dyson equation given in Fig.\ref{4ptscalarsd}. The final result for $\mathcal{A}(p,k,q)$ is (see \eqref{exactscalar4ptapp} or equivalently \eqref{4ptfnexp})
 \begin{equation}\label{exactscalar4pt}
 \begin{split}
 N_B\mathcal{A}(\vec{p},\vec{k},q_3)
 & = \frac{H_B(a(p_s), q_3)}{H_B(a(k_s),q_3)}  \bigg\{(4\pi \i \lambda_{B} q_{3})\frac{(p+k)_{-}}{(p-k)_{-}} + j( q_3) \bigg\}   \ , 
 \end{split}
 \end{equation}
where, $H_B(z,q_3)$ is given by \eqref{Hbfunction} and $a(p_s)=+\sqrt{p_s^2+c_B^2}$. The function $j(q_3)$ is given by 
\begin{equation}\label{jq3}
\frac{j(q_3)}{ 4\pi \i \lambda_{B}q_3}=  \frac{4\pi \i \lambda_{B} q_3(H_B(c_B,q_3)-H_B(\infty,q_3)) +\tilde{b}_4 (H_B(c_B,q_3)+H_B(\infty,q_3)) }{4\pi \i \lambda_{B} q_3(H_B(c_B,q_3)+H_B(\infty,q_3)) +\tilde{b}_4 (H_B(c_B,q_3)-H_B(\infty,q_3)) } \ ,
\end{equation}
where, $\tl{b}_4$ is given by \eqref{b4tl}. An alternative and simplified form of $j(q_3)$ is given by \eqref{jq3simp}. In the zero temperature limit, this matches with the existing results given in \cite{Jain:2014nza}. We use the result \eqref{exactscalar4pt} to compute the correlation functions of gauge invariant, single trace operators of different spin. 

\subsection{Case I: Spin zero}
In the regular boson theory that we study here, there is a gauge invariant, single trace, spin 0 operator given by $J^B_{0}(x)=\bar{\phi}(x)\phi(x)$. In the momentum space, this takes the follwing form 
\begin{equation}\label{spin0current}
J^B_{0}(-q) =\int \frac{\mathcal{D}_B^3 k}{(2\pi)^3} \ \bar{\phi}(-(k+q)) \ \phi(k)  \ . 
\end{equation}
\subsubsection{Exact insertion vertex}
To compute the correlators involving $J^B_{0}$, we need the exact $J^B_{0}$ insertion vertex which is defined as
\begin{equation}\label{J0Bexactvertex}
\langle J^B_{0}(-q) \phi(k)\bar{\phi}(p) \rangle =V^B_{0}(k,q)  \ (2\pi)^3 \delta^{(3)} (p+k+q) \ .  
\end{equation}

\begin{figure}[h!]
	\begin{center}
		\begin{tikzpicture}[scale=0.8, cross/.style={path picture={ 
				\draw[black]
				(path picture bounding box.south east) -- (path picture bounding box.north west) (path picture bounding box.south west) -- (path picture bounding box.north east);
		}}]
		\begin{feynman}
		\coordinate (A) at (0,0);
		\coordinate (D) at (2.2,1.2) ;
		\coordinate (Dp) at (2.2,-1.2) ;
		\coordinate (L) at (-1.1,0) ;
		\diagram* {
			(Dp) -- [fermion] (A),  
		}; 
		\draw (A)+(2.5,0) node {{$= $}}; 
		\node (c) [draw,circle,cross,minimum width=0.15 cm](b) at (A){};
		\draw (A) -- (D) ; 
		\draw [-] (A) -- (Dp) node [midway, below ] {$k$};
		\draw[-latex] (L)--+(0.7,0);
		\draw (L)+(0.3, 0.3) node {$q$}; 
		\end{feynman}
		\end{tikzpicture}  
		\raisebox{1.5pt} {
			\begin{tikzpicture}[scale=0.8, cross/.style={path picture={ \draw[black]
					(path picture bounding box.south east) -- (path picture bounding box.north west) (path picture bounding box.south west) -- (path picture bounding box.north east); }}]
			\begin{feynman}
			\coordinate (A) at (0,0);
			\coordinate (D) at (2,1.2) ;
			\coordinate (Dp) at (2,-1.2) ;
			\coordinate (L) at (-1.1,0) ;
			\diagram* {
				(Dp) -- [fermion] (A),  
			}; 
			\draw (A)+(2.5,0) node {{$+$}}; 
			\draw (A) node {$\bigtimes$} ; 
			\draw (A) -- (D) ; 
			\draw [-] (A) -- (Dp) node [midway, below ] {$k$};
			\draw[-latex] (L)--+(0.7,0);
			\draw (L)+(0.3, 0.3) node {$q$}; 
			\end{feynman}
			\end{tikzpicture}  } 
		\raisebox{-20pt}{ 
			\begin{tikzpicture}[scale=0.7, cross/.style={path picture={ 
					\draw[black]
					(path picture bounding box.south east) -- (path picture bounding box.north west) (path picture bounding box.south west) -- (path picture bounding box.north east);
			}}]
			\begin{feynman}
			\def\m{2.0cm};
			\def\s{0.6};
			\coordinate (A) at (0,0);
			\coordinate (B) at (1,\s) ;
			\coordinate (C) at (2,1.2) ;
			\coordinate (D) at (3.5, 3.5*\s) ;
			\coordinate (Bp) at (1,-0.6) ;
			\coordinate (Cp) at (2,-1.2) ;
			\coordinate (Dp) at (3.5,-3.5*\s) ;
			\coordinate (L) at (-1.1,0) ;
			\diagram* { (Dp) -- [fermion] (Cp),    }; 
			\diagram* { (Cp) -- [fermion] (Bp),    }; 
			\end{feynman}
			\draw (A) -- (B) ;
			\filldraw[black] (B) circle (3pt) ;
			\draw (B) -- (C) ;
			\draw (C) -- (D) ;
			\draw (A) -- (Bp) ;
			\filldraw[black] (Bp) circle (3pt) ;
			\draw (Bp) -- (Cp) ;
			\draw (Cp) -- (Dp) ;
			\draw (A) node {$\bigtimes$} ;
			\draw (Bp)+(0.3,-0.7) node {$p$};
			\draw (Cp)+(0.7,-0.9) node {$k$};
			\draw[-latex] (L)--+(0.7,0);
			\draw (L)+(0.3, 0.3) node {$q$}; 
			\fill[gray] (2.2,0) ellipse [x radius=4.5mm, y radius=1.8cm, rotate=0];
			\end{tikzpicture} }
	\end{center} 
	\caption{Exact vertices $\langle J_{(s)}^B\phi \bar{\phi} \rangle$. The circled cross denotes an insertion of an exact vertex, the bare cross denotes the insertion vertex in the `free' theory. The filled circle denotes the exact scalar propagator. And the elliptic blob denotes the exact scalar four-point function.}
	\label{bosexactvertx}
\end{figure}
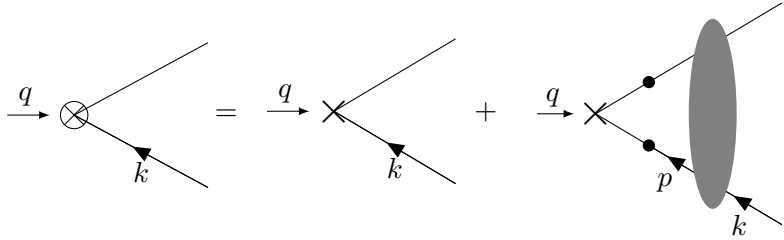
From \eqref{spin0current}, we see that the $J_0^B$ insertion in the `free' theory is $V^B_{0,\text{free}}(k,q)=1$. We work in the momentum $q_{\pm }=0$. The exact $J_0^B$ insertion vertex is shown in Fig.\ref{bosexactvertx}. In mathematical form,  this is given by 
\begin{equation}\label{bootstrapV0}
V^B_0(k,q) =1+N_B \int \frac{\mathcal{D}_B^3p}{(2\pi)^3} \ \frac{ \mathcal{A}(\vec{p}, \vec{k},q_3)}{(p_3^2+a^2(p_s))((p_3+q_3)^2+a^2(p_s))}  \ , 
\end{equation}
where, $a(p_s)=+\sqrt{p_s^2+c_B^2} \ $. The factor $N_B$ in the second term on the RHS of \eqref{bootstrapV0} comes from the color trace in the loop. It is useful to break the momentum integration measure as $\mathcal{D}_B^3p=p_sdp_s d\theta_p \mathcal{D}_Bp_3$ where, $p_s$ is the radial momentum in the lightcone plane and $\theta_p$ is the angular direction in that plane. $\mc{D}_Bp_3$ is the integration measure for the momentum component $p_3$ taking into account the effect of holonomy as given in \eqref{bosholonomy}. 
Doing the integration over the momenta $p_3$, we get 
\begin{equation}\label{}
V^B_0(k,q) =1+N_B \int \frac{d^2\vec{p}}{(2\pi)^2} \ \frac{ \chi_B(a(p_s))\mathcal{A}(\vec{p}, \vec{k},q_3)}{a(p_s)(p_3^2+a^2(p_s))}  \ . 
\end{equation}
Using the definition \eqref{Hbfunction} and the explicit expression of $\mathcal{A}(\vec{p}, \vec{k},q_3)$ given by \eqref{exactscalar4pt} and simplifying, we get, 
\begin{equation}\label{}
\begin{split}
V^B_0(k,q) =1+\bigg[\frac{1}{2 H(a(k_s),q_3)} & \int_0^{\infty} p_s( \partial_{p_s} H(a(p_s),q_3)) \\
& \int_0^{2\pi} \frac{d\theta_p}{2\pi} \bigg\{\frac{(p+k)_{-}}{(p-k)_{-}} +\frac{ j( q_3)}{4\pi \i \lambda_{B} q_{3}} \bigg\}  \bigg] \ .
\end{split}
\end{equation}
Performing the angular integral, we get 
\begin{equation}\label{}
\begin{split}
V^B_0(k,q) =1+\bigg[\frac{1}{2 H(a(k_s),q_3)} & \int_0^{\infty} p_s( \partial_{p_s} H(a(p_s),q_3))  \\
& \bigg\{2\Theta(p_s-k_s)-1+\frac{ j( q_3)}{4\pi \i \lambda_{B} q_{3}}  \bigg\}  \bigg]  \ .
\end{split}
\end{equation}
Finally, performing the radial integral and simplifying further, we find
\begin{equation}\label{V0Bfin}
\begin{split}
V^B_0(k,q) =\frac{ \frac{1}{2}\big(\frac{ j( q_3)}{4\pi \i \lambda_{B} q_{3}} +1\big)H(\infty,q_3)- \frac{1}{2}\big(\frac{ j( q_3)}{4\pi \i \lambda_{B} q_{3}} -1\big)H(c_B,q_3) }{H(a(k_s),q_3)}  \ .
\end{split}
\end{equation}
For later convenience, we label the numerator on the RHS of \eqref{V0Bfin} by the symbol $\tl{V}(q_3)$, in terms of which \eqref{V0Bfin} takes the following form
\begin{equation}\label{V0Bfina}
\begin{split}
V^B_0(k,q) =\frac{ \tilde{V}(q_3) }{H(a(k_s),q_3)}  \ .  
\end{split}
\end{equation}

\subsubsection{Two-point function}
\begin{figure}[h!]
	\begin{center}
		\begin{tikzpicture}[scale=0.85, cross/.style={path picture={ 
				\draw[black]
				(path picture bounding box.south east) -- (path picture bounding box.north west) (path picture bounding box.south west) -- (path picture bounding box.north east);
		}}]
		\begin{feynman}
		\coordinate (A) at (0,0) ;
		\coordinate (B) at (6,0) ;
		\coordinate (L) at (-1.1,0) ;
		\draw (A) node {$\otimes $} ; 
		\draw (B) node {$\times$} ; 
		\filldraw (3,1.05) circle (3pt) ;
		\filldraw (3,-1.05) circle (3pt) ;
		\end{feynman}
		\draw (0,0) .. controls (2,1.4) and (4,1.4) .. (6,0);
		\draw (0,0) .. controls (2,-1.4) and (4,-1.4) .. (6,0);
		\draw[-latex] (L)--+(0.7,0);
		\draw (L)+(0.3, 0.3) node {$q$}; 
		\draw (A)+(0.8,0) node {$\ J_0$};
		\draw (5,0) node {$\ J_0$};
		\end{tikzpicture}
		\caption{Diagram contributing to two-point function $\langle J_0(-q)J_0(q)\rangle$. The filled circle denotes the exact scalar propagator.}
		\label{jj2ptB}
	\end{center} 
\end{figure}
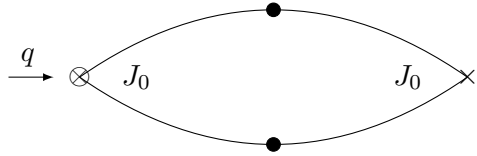

As in the case of fermions described in subsubsection \ref{2ptJ0F}, the two-point function for the spin zero operator $J_0^B$ is computed by evaluating the Feynman diagram shown in Fig.\ref{jj2ptB}. To compute the two-point function $\langle J_0^{B}(q') J_0^{B}(q) \rangle $, a single exact insertion vertex $J_0^{B}(q')$ is required to account for all the perturbative Feynman diagrams without any overcounting. 
We choose the insertion at the left in Fig.\ref{jj2ptB} as the exact vertex. Insertion on right side in Fig.\ref{jj2ptB} is the `free' insertion vertex which we define as follows :
\begin{equation}
\langle J_0^{B}(-q)\phi(k)\bar{\phi}(p)\rangle =U_{0}^B(k,q)~(2\pi)^3\delta^{(3)}(p+k+q) \ . 
\end{equation}
From the definition of $J_0^{B}(q)$ operator in \eqref{J0(-q)}, it follows that $U_{0}^B(k,q) =1$. 
We define the two-point correlation function of $J_0^{B}$ operator as 
\begin{equation}\label{J0B2pt}
\langle J_0^{B}(q')J_0^{B}(q)\rangle =\mc{G}_0^B(q) \ (2\pi)^3 \delta^{(3)}(q'+q) \ . 
\end{equation}
From the definition of the $J_0^B$, it follows that the other `free` insertion vertex is $U_0^B(k,q)=1$. 
We now have all the building blocks to compute the $\langle J_0^B J_0^B \rangle$ correlator. As discussed before, we compute the diagram shown in Fig.\ref{jj2ptB}, which translates to equation as 
\begin{equation}\label{spin02pt}
\mc{G}^B_{0}(q)= N_B\int \frac{\mathcal{D}_B^3k}{(2\pi)^3} \ \frac{ V^B_{0}(k,q)U^B_{0}(k+q,-q)}{(k^2+c_B^2)((k+q)^2+c_B^2)}  \ . 
\end{equation}
Working in the case $q_{\pm} = 0$, and inserting the expressions for $V_0^B$ and $U_0^B$, we find 
\begin{equation}\label{G0Bint}
\mc{G}^B_{0}(q)= N_B \tilde{V}(q_3) \int \frac{\mathcal{D}_B^3k}{(2\pi)^3} \ \frac{ 1 }{(k_3^2+a^2(p_s))((k_3+q_3)^2+a^2(p_s))H(a(k_s),q_3)}  \ . 
\end{equation}
The crucial fact is that the above momentum integral \eqref{G0Bint} can be carried out analytically. As before, we choose to separate the above integral as an integral over the radial momentum $k_s=\sqrt{k_1^2+k_2^2}=\sqrt{2k_{+}k_{-}} \ $, and an integral over an angular variable $\theta_k$. It is convenient to perform the $k_3$ integral first, which can be carried out using the integration result \eqref{FBfunction}. The angular integral in \eqref{G0Bint} contributes unity. Once these two integrals are performed, finally, the integration over radial momentum $k_s$ can also be performed. By changing the integration variable $k_s$ to another variable $a(k_s)=+\sqrt{k_s^2+c_B^2} \ $, the integrand in the radial integral can be written as a total derivative w.r.t. the variable $a(k_s)$. Thus the remaining integral can be carried out completely and the final result (inserting back the explicit form of $\tilde{V}(q_3)$) for the $\langle J_0^BJ_0^B\rangle $ two-point function in the regular boson theory at finite temperature is
\begin{equation}\label{spin02ptfinal}
\mc{G}^B_{0}(q)= -   \frac{N_B}{4\pi \i \lambda_{B} q_3\Big(\frac{H_B(c_B,q_3)+H_B(\infty,q_3)}{  H_B(c_B,q_3)-H_B(\infty,q_3)}\Big)+\tilde{b}_4} \ . 
\end{equation}

\subsection{Case II: Spin one}
The regular boson theory that we study here, has a global $U(1)$ symmetry and the corresponding conserved spin 1 current is
\begin{equation}\label{Jmu(x)} 
\begin{split}
J^{B}_\mu(x) & = \i \Big[(D_\mu \bar{\phi}) \phi -\bar{\phi}(D_\mu \phi)\Big] \ . \\
\end{split}
\end{equation}
In the momentum space, the $J_\mu^B$ can be written as 
\begin{equation}\label{J1Bmu(-q)final}
\begin{split}
J_\mu(-q) & 
= \int_{k} (2k+q)_{\mu} \ \bar{\phi}(-(k+q)) \ \phi(k) \  - 2 \int_{p,k}  \ \bar{\phi}(p) \ A_\mu(-(p+k+q) \ \phi(k) \ .
\end{split}
\end{equation}

\subsubsection{Exact insertion vertex}
In this subsection we compute the exact $J_{\mu}^{B}$ insertion vertex which is one of the building blocks for computing the corresponding correlators. We define the exact $J_\mu^B$ insertion vertex by
\begin{equation}\label{J1Bmu(-q)def}
\langle J_{\mu}^{B}(-q) \phi(k) \bar{\phi}(p) \rangle =V_{(\mu)}^{B}(k,q) \ (2\pi)^3 \ \delta^{(3)}(p+k+q)  \ .
\end{equation}
The exact insertion vertex is diagramatically shown in Fig.\ref{bosexactvertx}, by a circled cross. The insertion denoted by a bare cross is understood to be the insertion vertex in the `free' theory. In equation, $V_{(\mu)}^{B}(k,q)$ is given by 
\begin{equation}\label{bootstrapV1Bmu}
V_{(\mu)}^{B}(k,q) =V_{(\mu),\text{free} }^{B}(k,q) +N_B \int \frac{\mathcal{D}_B^3p}{(2\pi)^3} \ \bigg[S_B(p+k) V_{(\mu),\text{free} }^{B}(p,q) S_B(p)\mathcal{A}(p,k,q) \bigg] \ ,
\end{equation}
where, $S_B(k)$ is the exact scalar propagator given by \eqref{exactphiprop} and $\mathcal{A}(p,k,q)$ is the thermal scalar $4$-point function which in the `lightcone kinematics'  ($q_{\pm} =0 $), is given by \eqref{exactscalar4pt}. Substituting the exact scalar propagator we write this in a more convenient from as 
\begin{equation}\label{bootstrapV1Bmuexplicit}
V_{(\mu) }^{B}(k,q) =V_{(\mu),\text{free} }^{B}(k,q) +N_B \int \frac{\mathcal{D}_B^3p}{(2\pi)^3} \ \frac{ V_{(\mu),\text{free} }^{B}(p,q) \mathcal{A}(p,k,q)}{(p^2+c_B^2)((p+q)^2+c_B^2)}  \ , 
\end{equation}
To perform the integral in \eqref{bootstrapV1Bmuexplicit}, we need $V_{(\mu),\text{free} }^{B}(k,q)$, which can be easily read from the definition of $J_\mu^B$ current given in \eqref{J1Bmu(-q)final}. 

In this paper the explicit computations are performed in the lightcone gauge $A_{-}=0$ and with the external momenta $q_{\pm}=0$. For the moment, we will be interested in the computation of exact $J_{-}^B$ vertex. It follows from \eqref{J1Bmu(-q)final} that $V_{(-),\text{free} }^{B}(k,q) = 2k_{{-}} $. From \eqref{bootstrapV1Bmuexplicit}, it follows that
\begin{equation}\label{V1Bminusintegral}
V_{(-) }^{B}(k,q) =2k_{-} +N_B \int \frac{\mathcal{D}_B^3p}{(2\pi)^3} \ \frac{ (2p_{-} ) \mathcal{A}(\vec{p},\vec{k},q_3)}{(p^2+c_B^2)((p+q)^2+c_B^2)}  \ . 
\end{equation}
To perform the momentum space integral we follow the same procedure as in the case of $V_0^B$. We perform the intgeral over $p_3$ first and then carry out the angular integral by inserting the explicit form of $\mathcal{A}(\vec{p},\vec{k},q_3)$ to reduce into an one-dimensional integral over the radial momenta $p_s$. Interestingly enough, it is noted that the radial momenta integral can also be carried out analytically by writing the corresponding integrand as a total derivative w.r.t. the reduced integration variable $a(p_s)=+\sqrt{p_s^2+c_B^2} \ $. Performing the integral, we find the final result for the exact $J_{-}^B$ insertion vertex as
\begin{equation}\label{V1Bminusfinal}
V_{(-) }^{B}(k,q) =2k_{-} \ \frac{H_B(\infty,q_3)}{H_B(a(k_s),q_3)}  \ .
\end{equation}

\subsubsection{Two-point function}
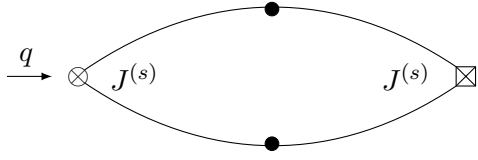
\begin{figure}[h!]
	\begin{center}
		\begin{tikzpicture}[scale=0.85, cross/.style={path picture={ 
				\draw[black]
				(path picture bounding box.south east) -- (path picture bounding box.north west) (path picture bounding box.south west) -- (path picture bounding box.north east);
		}}]
		\begin{feynman}
		\coordinate (A) at (0,0) ;
		\coordinate (B) at (6,0) ;
		\coordinate (L) at (-1.1,0) ;
		\draw (A) node {$\otimes $} ; 
		\node [draw,cross,minimum width=0.01 cm] at (B){};
		\filldraw (3,1.05) circle (3pt) ;
		\filldraw (3,-1.05) circle (3pt) ;
		\end{feynman}
		\draw (0.10,0.10) .. controls (2,1.4) and (4,1.4) .. (5.85,0.15);
		\draw (0.10,-0.10) .. controls (2,-1.4) and (4,-1.4) .. (5.85,-0.15);
		\draw[-latex] (L)--+(0.7,0);
		\draw (L)+(0.3, 0.3) node {$q$}; 
		\draw (A)+(0.8,0) node {$\ J^{(s)}$};
		\draw (5,0) node {$\ J^{(s)}$};
		\end{tikzpicture}
		\caption{Schematic diagram contributing to two-point function $\langle J^{(s)}(-q)J^{(s)}(q)\rangle $ for spin $s>0$. The filled circles denote the exact scalar propagator.}
		\label{jj2ptBs1}
	\end{center} 
\end{figure}

\begin{figure}[h!]
	\begin{center}
		\raisebox{1pt}{ 
			\begin{tikzpicture}[scale=0.55]
			\begin{feynman}
			\coordinate (a) at (-4,0) ;
			\coordinate (b) at (0,0) ;
			\coordinate (bp) at (2.5,0) ;
			\coordinate (c) at (4,0) ;
			\end{feynman}
			\draw (a) -- (b) ;
			\draw (b) -- (c) ;
			\draw[-Latex] (a) -- ++(2,0) ;
			\draw[-Latex] (0,-1.5) -- ++(0,1) ;
			\draw (0.5, -1.2) node {$q$} ; 
			\draw (-2, 0.5) node {$k$} ; 
			\draw (b) node {$\times$} ; 
			\end{tikzpicture} }
		\raisebox{1pt}{ 
			\begin{tikzpicture}[scale=0.55]
			\begin{feynman}
			\coordinate (a) at (-4,0) ;
			\coordinate (b) at (0,0) ;
			\coordinate (bp) at (2.5,0) ;
			\coordinate (c) at (4,0) ;
			\diagram* {
				(b) -- [boson, bend left] (bp)
			};
			\end{feynman}
			\draw (a) -- (b) ;
			\draw (b) -- (c) ;
			\filldraw (b)+(1.2,0) circle (5pt) ; 
			\draw[-Latex] (a) -- ++(2,0) ;
			\draw[-Latex] (0,-1.5) -- ++(0,1) ;
			\draw (0.5, -1.2) node {$q$} ; 
			\draw (-2, 0.5) node {$k$} ; 
			\draw (b) node {$\times$} ; 
			\end{tikzpicture} }
		\raisebox{1pt}{ 
			\begin{tikzpicture}[scale=0.55]
			\begin{feynman}
			\coordinate (a) at (-4,0) ;
			\coordinate (b) at (0,0) ;
			\coordinate (bp) at (2.5,0) ;
			\coordinate (c) at (4,0) ;
			\diagram* {
				(b) -- [boson, bend left] (bp)
			};
			\end{feynman}
			\draw (a) -- (b) ;
			\draw (b) -- (c) ;
			\filldraw (b)+(1.2,0) circle (5pt) ; 
			\draw[-Latex] (a) -- ++(2,0) ;
			\draw[-Latex] (2.5,-1.5) -- ++(0,1) ;
			\draw (3, -1.2) node {$q$} ; 
			\draw (-2, 0.5) node {$k$} ; 
			\draw (bp) node {$\times$} ; 
			\end{tikzpicture} }
	\end{center}
	\caption{Diagrams with nonzero contributions to the `bare' vertex $ U_{(+)}^{B}$ which is the boxed cross on the RHS in Fig.\ref{jj2ptBs1}. The filled circle denotes the exact scalar propagator. The loop momentum in the last two digrams is $\ell$.}
	\label{UpB}
\end{figure}
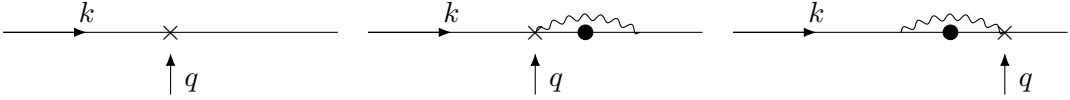

In this subsubsection, we compute the $\langle J_{-}J_{+} \rangle $ two-point correlator. The corresponding feynman diagram is shown in Fig.\ref{jj2ptBs1}. 
As mentioned before, we choose the insertion at the left of the digram to be the exact vertex which in our present case is the the exact $J_{-}^B$ vertex. The remaining thing is to compute the contribution from the insertion on the right side. To keep a distinction, we define the insertion on the right to be 
\begin{equation}\label{J1Bnu(-q)def}
\langle J_{\nu}^{B}(-q) \phi(k) \bar{\phi}(p) \rangle =U_{(\nu)}^{B}(k,q) \ (2\pi)^3 \ \delta^{(3)}(p+k+q)  \ . 
\end{equation}
To compute the two-point function $\langle J_{-}^BJ_{+}^B \rangle$, we need the `free' $J_{+}^B$ insertion vertex. From the definition \eqref{J1Bmu(-q)final}, it is clear that the `bare' insertion $ U_{(+)}^{B}(k,q) $ has one insertion without the gauge field i.e., the first term and the one involving the gauge field which is the second term; diagramatically, these are shown in Fig.\ref{UpB}.
Summing the diagrams in Fig.\ref{UpB}, total non-zero contribution to the `bare' insertion vertex $U_{(+)}^{B}$ is given by 
\begin{equation}\label{U1B(+)(k,q) }
\begin{split}
U_{(+)}^{B}(k,q)  
& = 2k_{+} - \ 4\pi  \lambda_B \epsilon_{+-\rho} q^{\rho}  \int \frac{\mathcal{D}_B^3 \l }{(2\pi)^3 }    \frac{S_B(\l)}{(\l-k)_{-}}    \ .     \\
\end{split}
\end{equation}
Here $S_B(\ell)$ is the exact scalar propagator given by \eqref{exactphiprop}. Performing the angular integral, the integral over $\l_3$ and finally performing the radial integral over $\l_s$ (also using the fact that $\epsilon_{+-\rho}=\i \delta_{\rho 3}$), we get
\begin{equation}\label{U1B(+)(k,q)final}
\begin{split}
U_{(+)}^{B}(k,q)  
& = 2k_{+} + \ \frac{\i \lambda_B q_3 }{k_{-}} \bigg[\xi_B(a(k_s))-\xi_B(c_B)\bigg]   \ ,
\end{split}
\end{equation}
where, the function $\xi_B(z)$ is given by \eqref{xiB}.  

We are now all set to compute the two-point function of the spin-one current operator. 
We define the two-point function $\langle J_{\mu}^{B}(-q) J_{\nu}^{B}(q)\rangle $ as
\begin{equation}\label{G1Bmunudefn} 
\langle J_{\mu}^{B}(q') J_{\nu}^{B}(q)\rangle  = \mc{G}^{B}_{\mu\nu} (q) \ (2\pi)^{3} \delta^{(3) }(q'+q)  \ . 
\end{equation}
Diagramatically, this is shown in Fig.\ref{jj2ptBs1}. In equation, it is given by 
\begin{equation}\label{G1Bmunu(q)}
\mc{G}^{B}_{\mu\nu}(q)= N_B\int \frac{\mathcal{D}_B^3k}{(2\pi)^3} \ \bigg[S_B(k+q) V^{B}_{(\mu)}(k,q)S_B(k)U^{B}_{(\nu)}(k+q,-q) \bigg]  \ , 
\end{equation}
where, $S_{B}(k)$ is the exact scalar field propagator given by \eqref{exactphiprop}. Explicitly, we write this as 
\begin{equation}\label{G1Bmunu(q)explicit}
\mc{G}^{B}_{\mu\nu}(q)= N_B\int \frac{\mathcal{D}_B^3k}{(2\pi)^3} \  \frac{ V^{B}_{(\mu)}(k,q)\ U^{B}_{(\nu)}(k+q,-q) }{(k^2+c_B^2)((k+q)^2+c_B^2)}  \ . 
\end{equation}
As we are working in the case of external momenta $q_{\pm}=0$, and in $A_{-}=0$ gauge, the non-trivial component in this case is $(\mu\nu)\equiv (-+)$ component. In this case, substituting the expressions \eqref{V1Bminusfinal} and \eqref{U1B(+)(k,q)final}, we find 
\begin{equation}\label{G1B-+(q)simpl}
\mc{G}^{B}_{-+}(q)= N_B\int \frac{\mathcal{D}_B^3k}{(2\pi)^3} \  (2k_{-})\frac{H_B(\infty,q_3) }{H_B(a(k_s),q_3)} \ \frac{ 2k_{+} -\frac{\i \lambda_B q_3}{k_{-} } \big[\xi_B(a(k_s))-\xi_B(c_B) \big]}{(k_3^2+a^2(k_s))((k_3+q_3)^2+a^2(k_s) )}   \ . 
\end{equation}
The interesting fact is that the momentum integral here in \eqref{G1B-+(q)simpl} can be carried out completely analytically. We first perform the intgeration over the momentum  component $k_3$ by using the result \eqref{GB2}. For the integrations in the lightcone plane, we use the polar coordinates to write the integration effectively as an integration over a radial momentum $k_s$ and an integration over an angular variable $\theta_k$. The angular integral in this case is trivial and contributes unity. Doing these, the integration in \eqref{G1B-+(q)simpl} is reduced to an integration over a single variable $k_s$. The $k_s$ intgeral can be performed analytically. The intgrand can be written as a total derivative of the reduced variable $a(k_s)=+\sqrt{k_s^2+c_B^2}\equiv z$, and so carrying out the radial integral, the final result for $\langle J_{-}J_{+}\rangle $ is given by 
\begin{equation}\label{G1B-+(q)finalanswer}
\begin{split}
\mc{G}^{B}_{-+}(q)
& = \lim_{\Lambda\to \infty} \bigg[ \frac{\i N_Bq_3 }{16\pi \lambda_B } \ \bigg(1+\frac{4c_B^2}{q_3^2}\bigg)  \bigg[ \frac{H_B(\Lambda,q_3)}{H_B(c_B, q_3)} -1\bigg] - \frac{N_B \xi_B(c_B)}{ 4 \pi} +\frac{N_B \xi_B(\Lambda)}{ 4 \pi}  \bigg] 
\end{split}
\end{equation}
where, $\Lambda$ is the UV cut-off in the radial momenta. The term $\xi_B(\Lambda\to \infty)$ is a pure divergent term. We regularize this answer by dropping this linearly divergent piece \footnote{This can be removed by using dimensional regularization in the integration over $k_s$ \cite{Choudhury:2018iwf}. This can also be done by turning on the mass counterterm for the background source which couples to the current $J_\mu$ \cite{Aharony:2012nh}.}, the renormalized two-point function $\langle J_{-}^BJ_{+}^B\rangle $, upto the momentum conserving delta function, is given by 
\begin{equation}\label{G1B-+(q)fin}
\begin{split}
\mc{G}^{B}_{-+}(q)
& = \frac{\i N_Bq_3 }{16\pi \lambda_B } \ \bigg(1+\frac{4c_B^2}{q_3^2}\bigg)  \bigg[ \frac{H_B(\infty,q_3)}{H_B(c_B, q_3)} -1\bigg] - \frac{N_B \xi_B(c_B)}{ 4 \pi} 
\end{split}
\end{equation}
As discussed in \cite{Gur-Ari:2016xff} (see e.g. around equation 80 of \cite{Gur-Ari:2016xff}), the two-point correlator of the $U(1)$ current \eqref{Jmu(x)} is $\langle J_{\mu}^BJ_{\nu}^B\rangle-2\delta_{\mu\nu}\langle\bar{\phi}\phi\rangle$. We report the relevant result by taking the contribution of $\langle\bar{\phi}\phi\rangle$ into account in subsubsection \ref{gis1b}.

\section{Analysis of results}\label{anlzresult}
In this section we analyze the results that we have obtained in this paper along with studying various limiting cases. 
\subsection{Fermionic result}
As discussed above in details, we have studied in this paper the massive regular fermions coupled to Chern-Simons theory at finite temperature with arbitrary holonomy distribution. Below we summarize the result and analyze the different relevant structures of the results. 
\subsubsection{Two-point correlator of spin 0 operator : }
The final result for $\langle J_0^F(-q)J_0^F(q) \rangle$ is given in \eqref{G0F(q)thefinal} and is rewritten below
\begin{equation}\label{G0Fan}
\begin{split}
\mc{G}_0^F(q) 
& = 
\frac{N_F|q_3|}{4\pi \lambda_{F} } \ \cot \bigg[ 2\lambda_F|q_3| \int_{c_F}^{\infty} \frac{dw \ \chi_F(w)}{q_3^2+4w^2} -\sgn(h_F)\tan^{-1}\frac{|q_3|}{2c_F} \bigg] + \frac{N_F m_F}{2\pi \lambda_{F} } \ , 
\end{split} 
\end{equation}
where, $\chi_F(w)$ is given by \eqref{chiF} and $\sgn(h_F)$ is defined around \eqref{masseqnforcF} (at zero temperature, $\sgn(h_F)$ effectively reduces to $\sgn(m_F)$). This result is general and is valid in the general case of massive theory at finite temperature. Interestingly, \eqref{G0Fan} is even in $q_3$. From the corresponding CFT results \cite{GurAri:2012is} \footnote{By corresponding CFT, we here mean the CFT from which one gets the massive theory that is being studied in this paper by deforming it with relevant deformations.}, we know that the two point function of spin zero scalar `current' operator is even in the general momentum $q$. The spin zero operator is even under parity in the corresponding CFT \cite{GurAri:2012is}, and it is unlikely that the finite temperature effects will break that structure. Assuming this to be true, we `covariantize' the result \eqref{G0Fan} by replacing $|q_3|$ with corresponding `$SO(3)$ rotation-invariant' generalization, i.e., with $|q|=+\sqrt{q_1^2+q_2^2+q_3^2}$ \footnote{Strictly speaking, the external momentum component $q_3$ is discrete at finite temperature. However, we formally treat this like a continuous variable for the purpose of `covariantization'. Also the true rotation symmetry is actually $SO(2)$ in the spatial plane. We however formally write the `covariantized' form by replacing $|q_3|$ with $+\sqrt{q_1^2+q_2^2+q_3^2}$ which reduces to $|q_3|$ in the choice $q_{\pm}=0$. The `covariantized' results are indeed the correct results obtained in this paper in the case $q_{\pm}=0$ in which case $|q|$ should be thought of as $|q_3|$. It will be nice to have a direct independent computation without setting $q_{\pm}$ to zero to check if this is indeed the case. \label{fncovs0f}}. The corresponding `covariant' generalization is given below
\begin{equation}\label{G0Fcov}
\begin{split}
\mc{G}_0^F(q) 
& = 
\frac{N_F|q|}{4\pi \lambda_{F} } \ \cot \bigg[ 2\lambda_F|q| \int_{c_F}^{\infty} \frac{dw \ \chi_F(w)}{q^2+4w^2} -\sgn(h_F)\tan^{-1}\frac{|q|}{2c_F} \bigg] + \frac{N_F m_F}{2\pi \lambda_{F} } \ , 
\end{split} 
\end{equation}

 We now study various limiting cases of this result and check that it matches with the existing previous results.
\subsubsection*{ Zero temperature and zero mass}
We start with the simplest possible limiting case, i.e., the case when both temperature and fermion bare mass parameter $m_F$ is zero. In this case, the effect of holonomy vanishes and it is clear from \eqref{chiF} that in the zero temperature limit, $\chi_F(w)=1$. Also from the gap equation \eqref{masseqnforcF} we see that in this case $c_F$ also vanishes; that is, there is no self energy correction to the pole mass $c_F$ in the corresponding CFT at zero temperature \cite{Giombi:2011kc}. As bare mass $m_F$ can have both possible signs, we choose to take the mass goes to zero limit from the side in which $\sgn(m_F)=\sgn(\lambda_F)$, which is known to be dual to the unhiggsed phase of bosonic theory. It follows from \eqref{G0Fan} or its `covariantized' form \eqref{G0Fcov}, that the two-point function of the scalar operator $J_0^F$ in this limiting case takes the form
\begin{equation}\label{G0FmF0T0}
\begin{split}
\mc{G}_0^F(q) 
& = -\frac{N_F |q| }{4\pi \lambda_{F} } \ \tan\bigg(\frac{\pi \lambda_F}{2} \bigg)
\end{split} 
\end{equation}
As expected, this result matches with the result in \cite{GurAri:2012is}. This provides a support to the correctness of the computation performed in this paper. 
\subsubsection*{Zero temperature and nonzero mass}
In this case, as before the effect of holonomy vanishes implying $\chi_F(w)=1$. The integral in the argument of cotangent in \eqref{G0Fcov} can be performed easily in this situation and the final result is given by 
\begin{equation}\label{G0FmFn0T0}
\begin{split}
\mc{G}_0^F(q) 
& = 
\frac{N_F|q|}{4\pi \lambda_{F} } \ \cot \bigg[ (\lambda_F-\sgn(m_F)) \tan^{-1}\frac{|q|}{2c_F} \bigg] + \frac{N_F m_F}{2\pi \lambda_{F} } \ .
\end{split} 
\end{equation}
\subsubsection*{Nonzero temperature and zero mass}
This is the case, when the effect of holonomy becomes non-trivial. The existing result in this case \cite{Ghosh:2019sqf} is only in the case when the holonomy takes the universal table-top distribution form, i.e., $\rho_F(\alpha)=\frac{\Theta(\pi|\lambda_F|-|\alpha|)}{2\pi|\lambda_F|}$. Performing the integral over the holonomy with this particular distribution, we get
\begin{equation}\label{chiFtabletop}
\chi_F(w) = \frac{\i}{\pi |\lambda_F|} \Big[\log \cosh\Big(\frac{\beta w-\i \pi |\lambda_F|}{2}\Big) - \log \cosh\Big(\frac{\beta w+\i \pi |\lambda_F|}{2}\Big) \Big] \ .
\end{equation}
Similar to the zero temperature case discussed above, when the bare mass of the fermion is zero, the theory is dual to the bosonic scalar theory for which  $\sgn(h_F)=\sgn(\lambda_F)$. The final result of the two-point function of spin zero operator $J_0^F$ in the thermal CFT in the `covariantized' form is given by 
\begin{equation}\label{G0FmF0Tn0}
\begin{split}
\mc{G}_0^F(q) 
& = 
\frac{N_F|q|}{4\pi \lambda_{F} } \ \cot \bigg[ 2\lambda_F|q| \int_{c_F}^{\infty} \frac{dw \ \chi_F(w)}{q^2+4w^2} -\sgn(\lambda_F)\tan^{-1}\frac{|q|}{2c_F} \bigg] .
\end{split} 
\end{equation}
where, $c_F$ is computed from the gap equation \eqref{masseqnforcF} by using the fact that $m_F=0$ and using \eqref{xiFtt}. $\chi_F(w)$ in this case is given by \eqref{chiFtabletop}. In the special choice of external momenta $q=(0,0,q_3)$, the result \eqref{G0FmF0Tn0} agrees with \cite{Ghosh:2019sqf}.

\subsubsection{Two-point correlator of spin 1 current : } 
The final result for spin 1 current two point function $\langle J_{-}^FJ_{+}^F\rangle$ is given by \eqref{G1F(q)final}. We rewrite this result below by using the fact that $h_F(c_F)=\sgn(h_F)c_F$ (this follows from the gap equation \eqref{masseqnforcF}) and the explicit form of $H_F(z,q_3)$.  It takes the following form \footnote{This does not have a definite even/odd property under $q_3\to -q_3$. However, this can formally be written as a sum of the even and odd parts, i.e., $\mc{G}_{-+}^{F}(q) =\mc{G}_{-+}^{F, \text{even}}(q) +\mc{G}_{-+}^{F,\text{odd}}(q)$, where, $\mc{G}_{-+}^{F, \text{even}}(-q)=\mc{G}_{-+}^{F, \text{even}}(q)$ and $\mc{G}_{-+}^{F,\text{odd}}(-q)=-\mc{G}_{-+}^{F,\text{odd}}(q)$. In this footnote, we propose to formally write the possible `covariantized form' of the correlator by replacing $|q_3|$ with $|q|=+\sqrt{q_1^2+q_2^2+q_3^2}$ as (also see footnote \ref{fncovs0f} for a related clarification)
\begin{equation}\label{GFevencov}
\mc{G}_{\mu \nu}^{F, \text{even}}(q) = \Big[ \frac{N_F}{16\pi \lambda_F}\Big(1-\frac{4c_F^2}{q^2}\Big)\sin\big(4\lambda_F|q|A_F\big) + \frac{N_F\sgn(h_F)c_F}{4\pi\lambda_F|q|} \Big(\cos(4\lambda_F|q|A_F)-1\Big)  +\frac{N_F\xi_{F}(c_F)}{4\pi|q|} \Big]  \frac{q_\mu q_\nu-\delta_{\mu\nu}q^2}{|q|} \ , 
\end{equation}
and
\begin{equation}\label{GFoddcov}
\mc{G}_{\mu \nu}^{F, \text{odd}}(q) =\frac{N_F}{8\pi \lambda_Fq^2} \Big[ (q^2-4c_F^2)\sin^2\big(2\lambda_F|q|A_F\big) + 2\sgn(h_F)c_F|q| \sin\big(4\lambda_F|q|A_F\big) \Big]  \epsilon_{\mu\nu\rho}q^{\rho} \ ,
\end{equation}
where, in the above two expressions $A_F=\int_{c_F}^{\infty}\frac{dw \ \chi_F(w)}{q^2+4w^2}$. It would be nice to have a direct independent check if this is indeed the correct form of $\mc{G}_{\mu\nu}^F(q)$ without choosing $q_{\pm}=0$. 
 Similar arguments apply for the correlator of spin one current in the bosonic case as given in \eqref{tlG1Ban}. 
 The author would like to thank S. Minwalla for asking a question about this. \label{fncovs1f}}
\begin{equation}\label{G1Fan}
\begin{split}
\mc{G}_{-+}^{F}(q) 
&= \frac{\i N_F q_3 }{16\pi \lambda_F}  \bigg(1+ \frac{2\i \sgn(h_F)c_F }{q_3 }\bigg)^2\bigg[\exp \bigg( 4\i \lambda_F q_3 \int_{c_F}^{\infty} \frac{dw \ \chi_F(w)}{q_3^2+4w^2}\bigg)  -1 \bigg] -\frac{N_F\xi_F(c_F)}{4 \pi} \ .
\end{split}
\end{equation}
Below we study limiting cases of this result and verify with the existing results. 
\subsubsection*{Zero temperature and zero mass}
As already discussed in the spin zero case, in the zero temperature limit, $\chi_F(w)=1$ and so, the integral appearing in the exponential in \eqref{G1Fan} can be performed exactly. Also as discussed before, the pole mass $c_F$ vanishes in this case. Using these and the fact that $m_F=0$ in this case, we get
\begin{equation}\label{G1FmF0T0}
\begin{split}
\mc{G}_{-+}^{F}(q) 
= \frac{\i N_F q_3}{16 \pi \lambda_F }   & \  \bigg[e^{\pi \i \lambda_{F} \sgn(q_3)}-1\bigg] \ .  
\end{split}
\end{equation}
This exactly matches with the corresponding CFT results at zero temperature reported in \cite{GurAri:2012is} \footnote{As discussed in \cite{GurAri:2012is}, this result can be separated into parity even and parity odd parts. And in this case, the covariantized form of the parity even part is given by $\frac{N_F}{16}\frac{\sin(\pi \lambda_F)}{\pi\lambda_F}\frac{q_\mu q_\nu-\delta_{\mu\nu}q^2}{|q|}$ as reported in \cite{GurAri:2012is}. The parity odd part can come from a contact term proportional to $\epsilon_{\mu\nu\rho}q^\rho$ \cite{GurAri:2012is} (see discussion around equation 35 of \cite{GurAri:2012is}).}.
\subsubsection*{Zero temperature and nonzero mass}
In this limiting case, the effect of holonomy goes away. As mentioned before, in the zero temperature limit $\chi_F(w)$ equals to unity and the integral appearing in the exponential in \eqref{G1Fan} can be performed exactly.  Keeping finite nonzero bare mass $m_F$, in this case the result \eqref{G1Fan} reduces to 
\begin{equation}\label{G1F(q)T0mFneq0final}
\begin{split}
\mc{G}_{-+}^{F}(q) 
= \frac{\i N_F q_3}{16 \pi \lambda_F }  \bigg(1+ \frac{2\i c_F\sgn(m_F) }{q_3} \bigg)^2  & \  \bigg[ e^{2\i \lambda_{F} \tan^{-1}\big(\frac{q_3}{2c_F}\big)}-1\bigg] -\frac{N_Fc_F}{4\pi} \ ,
\end{split}
\end{equation}
where, $c_F$ is the fermionic pole mass determined from the gap equation \eqref{masseqnforcF} at zero temperature which gives $c_F=\frac{m_F}{\sgn(m_F)-\lambda_F}$ . This exactly matches with the results of \cite{Gur-Ari:2016xff, amiyata1}.
\subsubsection*{Nonzero temperature and zero mass}
As discussed before, at finite temperature the effects of holonomy becomes non-trivial. The result in \eqref{G1Fan} is valid for any arbitrary holonomy distribution. In the specific case of table-top holonomy $\rho_F(\alpha)=\frac{\Theta(\pi|\lambda_F|-|\alpha|)}{2\pi|\lambda_F|}$, as used in \cite{Gur-Ari:2016xff, Ghosh:2019sqf}, the final result is given by \eqref{G1Fan}, where $\chi_F(w)$ is now given by \eqref{chiFtabletop}. $\xi_F(c_F)$ in this case is given by \eqref{xiFz} with the table-top holonomy distribution, i.e.,
\begin{equation}\label{xiFtt}
	\xi_{F}(c_F) = \frac{1}{2\pi|\lambda_F|\beta} \int_{-\pi|\lambda_F|}^{\pi|\lambda_F|} d\alpha \ \Big(\ln 2\cosh\big(\frac{\beta c_F+\i \alpha}{2}\big)+ \ln 2\cosh\big(\frac{\beta c_F-\i \alpha}{2}\big) \Big)  \ .
 \end{equation}
The results of the holonomy integral above are in terms of the dilogarithm functions as given, e.g., in \cite{Aharony:2012ns, Gur-Ari:2016xff, Ghosh:2019sqf}.



\subsection{Bosonic result} 
We have studied in this paper the bosonic theory as well. We have considered the massive regular bosonic matter theory coupled to $SU(N_B)$ Chern-Simons theory at finite temperature with arbitrary holonomy distribution. We have computed various correlators which we summarize below considering several limiting cases as well and check that it agrees with the existing results in limiting cases. 
\subsubsection{Two-point correlator spin 0 operator : }
The final result for the renormalized two point function of single trace, spin zero operator $J_0^B$ in the regular boson theory is given by \eqref{spin02ptfinal} which we rewrite below to analyze this result further.
\begin{equation}\label{G0Ban}
\mc{G}^B_{0}(q)=  \frac{N_B}{4\pi \lambda_{B} q_3 \ \cot \Big(2\lambda_B q_3 \int_{c_B}^{\infty} \frac{dw \ \chi_B(w)}{q_3^2+4w^2} \Big)-\tilde{b}_4} \ ,
\end{equation}
where, $\tl{b}_4$ is given by \eqref{b4tl} and thermal mass $c_B$ is given by \eqref{thermalcB}. First of all, it is easy to see that this result is even under $q_3\to -q_3$ (we have also seen the same property in the fermionic result in \eqref{G0Fan}). As described in the $J_0^F$ case (see footnote \ref{fncovs0f}), we `covariantize' this result by replacing $|q_3|$ with $|q|=+\sqrt{q_1^2+q_2^2+q_3^2}$. The `covariantized' form of this result is given by 
\begin{equation}\label{G0Bcov}
\mc{G}^B_{0}(q)=  \frac{N_B}{4\pi \lambda_{B} |q| \ \cot \Big(2\lambda_B |q| \int_{c_B}^{\infty} \frac{dw \ \chi_B(w)}{q^2+4w^2} \Big)-\tilde{b}_4} \ .
\end{equation}

\subsubsection*{Zero temperature and zero mass}
Let us consider the simple case of zero temperature and zero bare mass $m_B=0$. In the zero temperature limit $\chi_B(w)=1$ and $\xi_{B}(w)=w$. In this case, the pole mass of the scalar obtained by solving \eqref{thermalcB}. The integration appearing in the argument of \eqref{G0Bcov} can be performed explicitly in case of zero temperature. The final result in this case is given by 
 \begin{equation}\label{G0BT0mB0}
 \mc{G}^B_{0}(q)=  \frac{N_B}{4\pi \lambda_{B} |q| \ \cot \Big(\lambda_B\tan^{-1}\big(\frac{|q|}{2c_B} \big) \Big)-\tilde{b}_4} \ .
 \end{equation}
 Solving \eqref{thermalcB}, in this case, one finds the pole mass $c_B=0$. It follows that with nonzero quartic coupling $b_4$, the two-point function of the scalar operator in this case is given by 
  \begin{equation}\label{G0BT0mB0b4n0}
 \mc{G}^B_{0}(q)=  \frac{N_B}{4\pi \lambda_{B} |q| \ \cot \big(\frac{\pi \lambda_B}{2} \big)+{b}_4} \ .
 \end{equation}
This agrees with the results of \cite{Aharony:2012nh} (see e.g. equation 66 of \cite{Aharony:2012nh}). In the special case when $b_4$ is zero, \eqref{G0BT0mB0b4n0}  reduces to 
\begin{equation}\label{G0BT0mB0b40}
\mc{G}^B_{0}(q)=  \frac{N_B}{4\pi \lambda_{B} |q|} \ \tan \Big(\frac{\pi \lambda_B}{2} \Big) \ .
\end{equation}
This result exactly matches with the one given in \cite{Aharony:2012nh} (see e.g. equation 35 of \cite{Aharony:2012nh}). 
\subsubsection*{Zero temperature and nonzero mass}
As discussed above, at zero temperature the effect of holonomy disappears. As the bare mass $m_B$ is nonzero, the pole mass $c_B$ in this case is nonzero and is obtained by solving the gap equation \eqref{thermalcB} at zero temperature in which case $\xi_{B}(c_B)=c_B$. The two-point correlator of $J_0^B$ operator in this case is given by 
\begin{equation}\label{G0BT0mBn0}
\mc{G}^B_{0}(q)=  \frac{N_B}{4\pi \lambda_{B} |q| \ \cot \Big(\lambda_B\tan^{-1}\big(\frac{|q|}{2c_B} \big) \Big)-\tilde{b}_4} \ .
\end{equation}

\subsubsection*{Nonzero temperature and zero mass}
In this case, the non-trivial effects of holonomy becomes important. The result that we have given in \eqref{G0Bcov} is valid for any arbitrary holonomy distribution. In the case of specific table-top holonomy $\rho_B(\alpha)=\frac{\Theta(\pi|\lambda_B|-|\alpha|)}{2\pi|\lambda_B|}$,  $\rho_B(\alpha)$, as discussed in \cite{Ghosh:2019sqf}, the final result is given by \eqref{G0Bcov}, where $\chi_B(w)$ is now given by 
\begin{equation}\label{chiBtabletop}
\chi_B(w) = \frac{\i}{\pi |\lambda_B|} \Big[\log \sinh\Big(\frac{\beta w-\i \pi |\lambda_B|}{2}\Big) - \log \sinh\Big(\frac{\beta w+\i \pi |\lambda_B|}{2}\Big) \Big] \ .
\end{equation}
$\xi_B(c_B)$ in this case is given by \eqref{xiB} with the table-top holonomy distribution. This can be written as
\begin{equation}\label{xiBtt}
\xi_{B}(c_B) = \frac{1}{2\pi|\lambda_B|\beta} \int_{-\pi|\lambda_B|}^{\pi|\lambda_B|} d\alpha \ \Big(\ln 2\sinh\big(\frac{\beta c_B+\i \alpha}{2}\big)+ \ln 2\sinh\big(\frac{\beta c_B-\i \alpha}{2}\big) \Big) \ ,
\end{equation}
and the results of the holonomy integral are in terms of the dilogarithm functions as given, e.g., in \cite{Aharony:2012ns, Ghosh:2019sqf}. 

\subsubsection{Critical theory limit}
One of our goal in this paper is to check the conjectured bose-fermi duality. In this paper, we have studied the regular fermionic matter coupled to Chern-Simons theory which is conjectured to be dual to the critical bosons coupled to Chern-Simons theory. There is a calculational evidence of this conjectured duality from explicit computations of the thermal free energies on both sides of this duality. To check the duality of the two-point functions of gauge invariant operators, i.e., to match with the regular fermionic theory, we need to have the corresponding results in the critical bosonic theory. The `critical' limit of the theory is defined \cite{Jain:2014nza, Aharony:2012nh, GurAri:2012is} by taking $b_4\rightarrow \infty $ and $m_B^2\to \infty$ with $\frac{4\pi m_B^2}{b_4}=m_B^{\text{cri}}$ and $b_6$ kept fixed. The action for the critical theory can be obtained from the action \eqref{RBlag} of the regular boson theory, by introducing a Hubbard-Stratonovich field $\sigma_B$, and taking the critical limit (we set $b_6=0$) \cite{Jain:2014nza}. The action in this case takes the form
\begin{equation}\label{CBlag} 
\begin{split}
\mc{S}_{\text{CB}}[A, \phi, \sigma_B]  &  =\frac{\i \kappa_B}{4\pi} \int \ud^3 x\ \epsilon^{\mu\nu\rho}\,\text{tr}\left( A_\mu \partial_\nu A_\rho - \frac{2\i}{3} A_\mu A_\nu A_\rho\right)\   \\
&  + \int d^3 x ~\bigg((D_\mu \bar{\phi})(D^\mu \phi)+\sigma_B \Big( \bar{\phi}\phi+\frac{N_B}{4\pi}m_B^{\text{cri}} \Big) \bigg) \ . 
\end{split}
\end{equation}
We restrict our attention here to the case when $m_B^{\text{cri}}>0$. In the critical boson theory, the single trace scalar operator $J_0^B$ is simply the Lagrange multiplier field $\sigma_B$. The first term in the denominator of \eqref{G0Bcov} is finite (here thermal mass $c_B$ is assumed to be finite), however the second term is proportional to $b_4$ which grows without limit in the critical limit. However, we extract finite result by rescaling the $J_0^B$ operator with $b_4$ \footnote{The scaling is feasible, because in the corresponding CFT theory, the scaling dimension of the $J_0^B$ operator in the regular bosonic theory is $1$ in leading order in $N_B$ whereas, in the critical boson theory, the scaling dimension of the corresponding primary operator is $2$.} and define a new operator
\begin{equation}\label{scaledJ0B}
\tl{J}_0^B=\frac{b_4}{4\pi\lambda_B} \  J_0^B   \ . 
\end{equation}
Here, we have used the particlar normalization to exactly match its two-point function with the dual result in the regular fermionic theory. We define the two-point function of $\tl{J}_0^B$ operator by $\mc{\tl{G}}_0^B$ upto the momentum conserving delta function, and so, we see that in the critical boson theory, 
\begin{equation}\label{G0B(q)finalb4infty}
\mc{\tl{G}}^B_{0}(q)=  \lim_{b_4\rightarrow \infty}  \bigg(\frac{b_4}{4\pi\lambda_B}\bigg)^2 \mc{G}^B_{0}(q) \ . 
\end{equation}
Taking the critical limit, by sending $b_4$ and $m_B^2$ to infinity while keeping their ratio fixed, we find the two-point function of the operator $\tl{J}_0^B$ to be given by (we drop the contact term $\frac{N_Bb_4}{(4\pi\lambda_B)^2}$)
\begin{equation}\label{G0Bcri}
\mc{\tl{G}}^B_{0}(q)=   -\frac{N_B |q|}{4\pi\lambda_B} \ \cot \Big(2\lambda_B |q| \int_{c_B}^{\infty} \frac{dw \ \chi_B(w)}{q^2+4w^2} \Big)
\end{equation}
where, the thermal mass $c_B$ appearing in \eqref{G0Bcri} has to be computed from \eqref{thermalcB} in the critical limit which amounts to solving the gap equation 
 $\xi_{B}(c_B)=m_{B}^{\text{cri}}$ for a fixed $m_B^{\text{cri}}>0$. 
\paragraph*{Zero temperature and zero critical mass : }
As already mentioned above, in the zero temperature limit as the holonomy becomes trivial, it is easy see that $\chi_B(w)=1$ and $\xi_B(w)=w$. Thus, we can explicitly perform the integral appearing in the argument of \eqref{G0Bcri}.  Also in this case, as the critical mass is zero, i.e., $m_B^{\text{cri}}=0$, so, we see that the pole mass $c_B$ vanishes. In this case, \eqref{G0Bcri} reduces to 
\begin{equation}\label{tlG0BT0m0}
\mc{\tl{G}}^B_{0}(q)=   - \frac{N_B |q| }{4\pi\lambda_B} \ \cot\Big( \frac{\pi \lambda_B}{2} \Big)
\end{equation}
This result matches with the existing results \cite{Aharony:2012nh} in this limiting case.
\paragraph*{Zero temperature and nonzero critical mass : } 
This is similar to the above case with only difference being that in this case the pole mass is non zero, and is given by $c_B=m_B^{\text{cri}}$. So, from \eqref{G0Bcri} it follows that the final result in this case is 
\begin{equation}\label{tildeG0B(q)thefinalsimplifiedfurther}
\mc{\tl{G}}^B_{0}(q)=   - \frac{N_B |q| }{4\pi\lambda_B} \ \cot\Big[ \lambda_B \tan^{-1}\Big(\frac{|q|}{2c_B} \Big)\Big]
\end{equation}
\paragraph*{Nonzero temperature and zero critical mass : }  Taking into account the effect of holonomy, in this case, $\chi_B(w)$ is not simply unity but is given by \eqref{chiB}. The result \eqref{G0Bcri} is valid for arbitrary holonomy distribution. In the specific case of table-top holonomy, \eqref{chiB} reduces to \eqref{chiBtabletop}. So, in this case, the result is \eqref{G0Bcri} where, $\chi_{B}$ is now given by \eqref{chiBtabletop} and the thermal mass $c_B$ is determined from the gap equation $\xi_{B}(c_B)=0$, where, the function $\xi_B(c_B)$ is now given by \eqref{xiBtt}.

\subsubsection{Two-point correlator of spin 1 current : }\label{gis1b}
In \eqref{G1B-+(q)fin}, we have given the result of the two-point function $\langle J_{-}^BJ_{+}^B\rangle$ of the $U(1)$ current \eqref{Jmu(x)}. However, as discussed in \cite{Gur-Ari:2016xff} (see e.g. equation 80 of \cite{Gur-Ari:2016xff}), the gauge-invariant correlator of $U(1)$ currrent \eqref{Jmu(x)} is given by 
\begin{equation}\label{correlatorU1}
\langle J_{\mu}^BJ_{\nu}^B \rangle - 2\delta_{\mu\nu} \langle \bar{\phi}\phi\rangle 
\end{equation}
There is an extra $-2\langle \bar{\phi}\phi\rangle $ term which contributes to the correlator of the $U(1)$ current. The contribution corresponding to this part (apart from the overall momentum conserving delta function) is given by 
\begin{equation}\label{phibarphicont}
\begin{split}
& -2N_B\int \frac{\mathcal{D}_B^3 \l }{(2\pi)^3} \  \frac{1}{\l^2+c_B^2} = \frac{N_B \xi_B(c_B)}{2\pi}
\end{split}
\end{equation}
where, to obtain the RHS of \eqref{phibarphicont} we have used the intgration result \eqref{gBintegration}. We label the correlator \eqref{correlatorU1} of the $U(1)$ current by (apart from the overall delta function) $\mc{\tl{G}}_{\mu\nu}^B$, and so we find that 
\begin{equation}\label{tlG1Bdef}
\mc{\tl{G}}_{\mu\nu}^B(p) = \mc{{G}}_{\mu\nu}^B(p) + \delta_{\mu\nu} \ \frac{N_B \xi_B(c_B)}{2\pi}
\end{equation}
where, $\mc{{G}}_{\mu\nu}^B$ is defined by \eqref{G1Bmunudefn}. 
The final result of the correlator $\mc{\tl{G}}_{\mu\nu}^B$ of the $U(1)$ currrent (its $(-+)$ component) is \footnote{Following footnotes \ref{fncovs0f} and \ref{fncovs1f}, a possible `covariantized' form can be written formally as $\mc{\tl{G}}_{\mu\nu}^B(q)=\mc{\tl{G}}_{\mu\nu}^{B,\text{even}}(q)+\mc{\tl{G}}_{\mu\nu}^{B,\text{odd}}(q)$, where, 
\begin{equation}\label{GBevencov}
\mc{\tl{G}}_{\mu \nu}^{B, \text{even}}(q) = \Big[ \frac{N_B}{16\pi \lambda_B}\Big(1+\frac{4c_B^2}{q^2}\Big) \sin\big(4\lambda_B|q|A_B\big) -\frac{N_B\xi_{B}(c_B)}{4\pi|q|} \Big]  \frac{q_\mu q_\nu-\delta_{\mu\nu}q^2}{|q|} \ , 
\end{equation}
and
\begin{equation}\label{GBoddcov}
\mc{\tl{G}}_{\mu \nu}^{B, \text{odd}}(q) =\frac{N_B}{8\pi \lambda_B}\Big(1+\frac{4c_B^2}{q^2}\Big)\sin^2\big(2\lambda_B|q|A_B\big)  \epsilon_{\mu\nu\rho}q^{\rho} \ ,
\end{equation}
where, in the above two expressions, $A_B=\int_{c_B}^{\infty}\frac{dw \ \chi_B(w)}{q^2+4w^2}$. It would be nice to have a direct independent check if this is indeed the correct form of $\mc{\tl{G}}_{\mu\nu}^B(q)$.
\label{fncovs1b}}
\begin{equation}\label{tlG1Ban}
\begin{split}
\mc{\tl{G}}^{B}_{-+}(q)
& = \frac{\i N_Bq_3 }{16\pi \lambda_B } \ \bigg(1+\frac{4c_B^2}{q_3^2}\bigg)  \bigg[ \exp \bigg( 4\i \lambda_B q_3 \int_{c_B}^{\infty} \frac{dw \ \chi_B(w)}{q_3^2+4w^2}\bigg)  -1\bigg] + \frac{N_B \xi_B(c_B)}{ 4 \pi} 
\end{split}
\end{equation}
It is worth mentioning at this point that, unlike the spin zero case, there is no explicit appearance of the quartic and sextic coupling parameters $b_4$ and $b_6$ in the result \eqref{tlG1Ban} except implicitly through the thermal mass $c_B$ given by \eqref{thermalcB}. So, this result is unchanged also in the critical boson theory except in the critical boson theory the gap equation of the thermal mass $c_B$ is given by $\xi_B(c_B)=m_B^{\text{cri}}$. Below, we analyze various limiting cases of the above result \eqref{tlG1Ban}. 
\subsubsection*{Zero temperature and zero mass}
In the zero temperature and massless limit, $m_B=0$, the pole mass $c_B$ vanishes. So, in this limit, we have 
\begin{equation}\label{tlG1BT0m0}
\begin{split}
\mc{\tl{G}}_{-+}^{B}(q) = \frac{\i N_B q_3}{16 \pi \lambda_B}  \bigg[e^{\pi \i \lambda_{B} \sgn(q_3) }-1 \bigg] 
\end{split}
\end{equation}
This exactly matches with the zero temperature results at zero bare mass reported in \cite{Aharony:2012nh} (see equation 44 of \cite{Aharony:2012nh}) \footnote{As in the case of fermions, this result can be separated into parity even and parity odd parts as discussed in \cite{Aharony:2012nh}. The covariantized form of the parity even part in this case is given by $\frac{N_B}{16}\frac{\sin(\pi \lambda_B)}{\pi\lambda_B}\frac{q_\mu q_\nu-\delta_{\mu\nu}q^2}{|q|}$.}. 
\subsubsection*{Zero temperature and nonzero mass}
At zero temperature, the contribution of holonomy vanishes which implies $\chi_B(w)=1$ and $\xi_B(z)=z$. The integral appearing in \eqref{tlG1Ban} can be performed explicitly and the result is
\begin{equation}\label{G1B-+(q)T0mBneq0final}
\begin{split}
\mc{\tl{G}}^{B}_{-+}(q)
& =\frac{\i N_Bq_3 }{16\pi \lambda_B } \ \bigg(1+\frac{4c_B^2}{q_3^2}\bigg)  \bigg[ 
\exp \bigg(2\i \lambda_{B} \tan^{-1} \frac{q_3}{2c_B}\bigg)-1\bigg]  - \frac{N_B c_B}{ 4 \pi} 
\end{split}
\end{equation}
This exactly agrees with the results \cite{amiyata1}.  

\subsubsection*{Nonzero temperature and zero mass}
At nonzero temperature the holonomy contribution becomes non-trivial. The result \eqref{tlG1Ban} is valid for an arbitrary holonomy distribution function. In the specific case of tabletop holonomy $\rho_B(\alpha)=\frac{\Theta(\pi|\lambda_B|-|\alpha|)}{2\pi |\lambda_B|}$, the final result for $\tl{\mc{G}}^{B}_{-+}$ is given by \eqref{tlG1Ban}, where $\chi_B(w)$ is now given by \eqref{chiBtabletop} and the thermal mass $c_B$ is determined from \eqref{thermalcB} where $\xi_B(c_B)$ is now given by $\xi_{B}(c_B)$.


\section{Check of duality} \label{dualitycheck}
As already mentioned before, one of our goal in this paper is to check the bose-fermi duality at the level of correlation functions. In this section, we explicitly check the duality between the results that we have obtained in this paper both in the fermionic and in the bosonic theory. We check the dualities between the general results that we have got in the case of massive Chern-Simons matter theory at finite temperature with arbitrary holonomy distribution function. As mentioned in the introduction, the massive regular fermionic theory which is conjecturally dual to the massive critical bosonic theory. 
Below we list down the parameter map which is required to explicitly show that the results are dual to each other. 
\subsection{Duality map : }\label{dmap}
The large $N$ 't Hooft couplings $\lambda_F=\frac{N_F}{\kappa_F}$, $\lambda_B=\frac{N_B}{\kappa_B}$, the holonomies $\rho(\alpha)$ and the exact thermal masses $c_F/c_B$, under duality, are mapped via
\begin{equation}\label{dualitymap}
\begin{split}
\lambda_F & = \lambda_B-\sgn(\lambda_B) \ , \ \kappa_F=-\kappa_B \ , \ c_F=c_B  , \\
& \lambda_F \rho_{F}(\pi -\alpha ) = \lambda_B \rho_{B}(\alpha) - \frac{\sgn{(\lambda_B)}}{2\pi } \ .
\end{split} 
\end{equation}
In the present case $c_F$ is the thermal mass of the regular boson theory, determined by the gap equation \eqref{masseqnforcF}. The gap equation for the critical bosonic scalar theory is determined from $\xi_B(c_B)=m_B^{\text{cri}}$ \footnote{In the critical limit i.e., in the limit $b_4\to \infty, \ m_B^2 \to \infty $ with $\frac{4\pi m_B^2}{b_4}=m_B^{\text{cri}}$ and $b_6$ fixed, it follows from the gap equation \eqref{thermalcB} that $m_B^{\text{cri}}=\xi_B(c_B)$.}. The map between the UV parameters of the two theories in this case is given by $m_F=-\lambda_Bm_B^{\text{cri}}$. It follows from the definitions of $\
\chi_B(z)$ and $\chi_F(z)$ as given in \eqref{chiB} and \eqref{chiF}, \footnote{In the zero temperature limit ($\beta \rightarrow \infty $), this reduces to the familiar expression $\lambda_F = \lambda_B-\sgn(\lambda_B)$.}
\begin{equation}\label{chimap}
\lambda_F \chi_{F}(z) = \lambda_B \chi_{B}(z) - \sgn{(\lambda_B)}
\end{equation}
 Using the definition $\xi(z)=\int^{z} \chi(w) dw$, we can rewrite the above equation in terms of the duality map of $\xi(z)$ as below
\begin{equation}\label{ximap} 
\lambda_F \xi_{F}(z) = \lambda_B \xi_{B}(z) - \sgn{(\lambda_B)} z
\end{equation}
From the fermionic mass gap equation \eqref{masseqnforcF}, we can write the fermionic mass $m_F$ in terms of the bosonic variables as 
\begin{equation}
m_F = \big(1-\eta\big)\sgn(\lambda_B)c_B -\lambda_{B}\xi_B(c_B)
\end{equation}
where, we have used the definition $\eta = \sgn(h_F\lambda_F )$ \footnote{Also we use the fact that $\sgn(\lambda_F)=-\sgn(\lambda_B)$.}. The parameter $\eta$ can have two possible values \cite{Choudhury:2018iwf}. In the case of $\eta=+1$, the dual theory in the bosonic side is the scalar theory in the unhiggsed phase, in which case, we get
\begin{equation}
m_F = -\lambda_{B}\xi_B(c_B)
\end{equation}
However, we will not always write the maps between the bare parameters, instead use the map between the thermal masses, i.e., $c_F=c_B$. As we have considered the bosonic scalar theory \eqref{RBlag} in the paper, we will use the choice $\eta=+1$ to compare with the results in the bosonic theory.  On the other hand, we use $\eta=-1$ to predict the corresponding results for the critical boson theory in the Higgsed phase. 
\subsection{For spin 0 : }
We dualize the result of two-point correlator of spin zero operator $J_0^F$ given by \eqref{G0Fcov} in terms of the bosonic variables. 
Using duality map \eqref{dualitymap} and \eqref{chimap}, we rewrite \eqref{G0Fcov} in terms of bosonic variables as 
\begin{equation}\label{dG0Fcov}
\begin{split}
\mc{G}_0^F(q) 
& = 
-\frac{N_B|q|}{4\pi \lambda_{B} } \ \cot \bigg[ 2\lambda_B|q| \int_{c_B}^{\infty} \frac{dw \ \chi_B(w)}{q^2+4w^2} -(1-\eta)\sgn(\lambda_B)\tan^{-1}\frac{|q|}{2c_B} \bigg] \\  & +\frac{N_B\xi_{B}(c_B)}{2\pi} - (1-\eta) \frac{N_B c_B}{2\pi |\lambda_{B}| } \ .
\end{split} 
\end{equation}
Depending upon the two possible values $\eta=\pm 1$, there are two cases. The regular fermionic theory is dual to the critical boson theory in the unHiggsed phase when $\eta=+1$. On the other hand, in the case of $\eta=-1$, the regular fermionic theory is dual to the critical bosonic theory in the Higgsed phase (this is the case when $m_B^{\text{cri}}<0$) (for details about this see \cite{Choudhury:2018iwf}). 
\subsubsection{Duality check in the unhiggsed phase of critical bosons: }
As mentioned above, in the unhiggsed phase, $\eta =+1$. This implies that the fermionic result 
\begin{equation}\label{uHG0Fcov}
\begin{split}
\mc{G}_0^{uH}(q) 
& = 
-\frac{N_B|q|}{4\pi \lambda_{B} } \ \cot \bigg[ 2\lambda_B|q| \int_{c_B}^{\infty} \frac{dw \ \chi_B(w)}{q^2+4w^2} \bigg]  +\frac{N_B\xi_{B}(c_B)}{2\pi}  \ .
\end{split} 
\end{equation}
This matches with the result of two-point correlator of spin zero operator in the critical boson theory given by \eqref{G0Bcri} upto contact terms. Thus, we see that under duality, $J_{0}^F$ maps to $J_0^B$ (more precisely to $\tl{J}_{0}^B$ given in \eqref{scaledJ0B}) \footnote{As mentioned earlier, the single trace spin-zero scalar operator $J_0^B$ in the case of critical boson theory \eqref{CBlag} is simply the Lagrange multiplier field $\sigma_B$ appearing in the action \eqref{CBlag} \cite{Aharony:2012nh, GurAri:2012is}.}. 
\subsubsection{Prediction for the higgsed phase of critical bosons: }
In the higgsed phase, on the other hand, $\eta =-1$. So, the predicted answer of the two-point correlator of single trace, spin zero `scalar' current $J_0^H$ in the higgsed phases of critical bosons is given by \footnote{The single trace, spin-zero scalar operator in the Higgsed phase of critical bosons which is dual to corresponding spin zero operator $J_{0}^F=\bar{\psi}\psi$ in the massive regular fermionic theory, is given by $J_0^H=\overline{W}_\mu W^\mu +Z_\mu Z^\mu$ \cite{amiyata1}.}
\begin{equation}\label{HG0Fcov}
\begin{split}
\mc{G}_0^{H}(q) 
& = 
-\frac{N_B|q|}{4\pi \lambda_{B} } \ \cot \bigg[ 2\lambda_B|q| \int_{c_B}^{\infty} \frac{dw \ \chi_B(w)}{q^2+4w^2} -2\sgn(\lambda_B)\tan^{-1}\frac{|q|}{2c_B} \bigg] \\  & +\frac{N_B\xi_{B}(c_B)}{2\pi} - \frac{N_B c_B}{\pi |\lambda_{B}| } \ .
\end{split} 
\end{equation}
It is an interesting excercise to compute this (exactly in large $N$) directly from the higgsed phases of bosons, which we leave for future work. 
\subsection{For spin 1 : }
The final result that we have obtained for spin 1 current correlator in the fermionic theory, is given by \eqref{G1Fan}. 
Under duality maps \eqref{dualitymap}, \eqref{chimap} and \eqref{ximap}, $\mc{G}_{-+}^{F}(q)$ can be rewritten as 
\begin{equation}\label{dG1Fan}
\begin{split}
\mc{G}_{-+}^{F}(q) 
&= - \frac{\i N_B q_3 }{16\pi \lambda_B}  \bigg(1-\eta  \frac{2\i \sgn(\lambda_B)c_B}{q_3 }\bigg)^2\bigg[ - \bigg(\frac{ q_3+ 2 \i \sgn(\lambda_B)c_B}{q_3- 2 \i \sgn(\lambda_B)c_B}\bigg) \exp \bigg( 4\i \lambda_B q_3 \int_{c_B}^{\infty} \frac{dw \ \chi_B(w)}{q_3^2+4w^2}\bigg)  -1 \bigg] \\
& +\frac{N_B\xi_B(c_B)}{4 \pi } -\frac{N_Bc_B}{4 \pi |\lambda_B|}  \ .
\end{split}
\end{equation}
Depending upon the two possible values of $\eta$, we consider two phases separately below. 
\subsubsection{Duality check in the unhiggsed phase of critical bosons : }
As mentioned before, in the unhiggsed phase $\eta =+1$. So, it follows from \eqref{dG1Fan} that it dualizes to the following result in the unhiggsed phase of bosons, 
\begin{equation}\label{uHG1Fan}
\begin{split}
\mc{G}_{-+}^{uH}(q) 
&=  \frac{\i N_B q_3 }{16\pi \lambda_B}  \bigg(1+ \frac{4c_B^2}{q_3^2}\bigg)\bigg[ \exp \bigg( 4\i \lambda_B q_3 \int_{c_B}^{\infty} \frac{dw \ \chi_B(w)}{q_3^2+4w^2}\bigg)  -1 \bigg] +\frac{N_B\xi_B(c_B)}{4 \pi } +\frac{\i N_B q_3 }{8\pi \lambda_B}  \ .
\end{split}
\end{equation}
This exactly matches with the bosonic results \eqref{tlG1Ban} upto a contact term $\frac{\i N_B q_3}{8\pi \lambda_B }$. Thus, we see that $J_{\mu}^F$ maps to $J_{\mu}^B$ under duality. 

\subsubsection{Prediction for the higgsed phase of critical bosons : }
In the higgsed phase ($\eta = -1 $), the predicted result for the two-point correlator of the single trace spin one current is \footnote{A possible `covariant' form of $\langle J_{\mu}^H(q')J_{\nu}^H(q)\rangle=\mc{G}_{\mu\nu}^H\ (2\pi)^3\ \delta^{(3)}(q'+q)$, is obtained from \eqref{GFevencov} and \eqref{GFoddcov} with the substitution of $\sgn(h_F)=-\sgn(\lambda_F)$(which is equal to $\sgn(\lambda_B)$) and applying the duality map given in subsection \ref{dmap}. It would be nice to have a direct independent check if this is indeed the case.\label{fncovs1h}}
\begin{equation}\label{HG1Fan}
\begin{split}
\mc{G}_{-+}^{H}(q) 
&= \frac{\i N_B q_3 }{16\pi \lambda_B}  \bigg(1 +  \frac{2\i \sgn(\lambda_B)c_B}{q_3 }\bigg)^2\bigg[  \bigg(\frac{ q_3+ 2 \i \sgn(\lambda_B)c_B}{q_3- 2 \i \sgn(\lambda_B)c_B}\bigg) \exp \bigg( 4\i \lambda_B q_3 \int_{c_B}^{\infty} \frac{dw \ \chi_B(w)}{q_3^2+4w^2}\bigg)  +1 \bigg] \\
& +\frac{N_B\xi_B(c_B)}{4 \pi } -\frac{N_Bc_B}{4 \pi |\lambda_B|}  \ .
\end{split}
\end{equation}
We leave for the future work, the matching of this prediction with the exact large $N$ computation of $\langle J_{\mu}^{H}J_{\nu}^{H}\rangle $ directly in the Higgsed phases of critical bosons \footnote{In the Higgsed phase of critical bosons in the unitary gauge \cite{Choudhury:2018iwf}, the $U(1)$ current $J_{\mu}^H$ is proportional to the $Z$ bosons, i.e., $J_{\mu}^H\propto Z_{\mu}$ \cite{amiyata1}. So, in this case, $\mc{G}_{\mu\nu}^H\propto \langle Z_{\mu}Z_{\nu}\rangle$.}. 

\section{Discussions}\label{discussion}
In this work we have obtained several two-point momentum space correlators of gauge invariant, single trace operators in Chern-Simons coupled to massive fundamental matter theories in the large-$N$ 't Hooft limit, at finite temperature and considering arbitrary holonomy distribution for gauge holonomy. One of the challenging aspects of our work was to solve the correspoinding Schwinger-Dyson equations for the correlators because we chose to work with an arbitrary holonomy distribution, with massive matter at finite temperature. However, we have been able to overcome this technical difficulty and have solved these equations analytically by explicitly performing the loop momenta integrals. The results for the two-point correlators are indeed very simple in form. We have seen that in different limiting cases, the results obtained in this work agree with the existing previous results \cite{Aharony:2012nh, GurAri:2012is, Geracie:2015drf, Gur-Ari:2016xff, Ghosh:2019sqf}. We have also explicitly checked the bose-fermi duality between the results of the regular fermionic theory and the critical bosonic scalar theory. We now discuss about the applications of the results obtained and further possible generalizations of the analysis done in this paper. 

We computed the two-point correlators of the spin one conserved current of single trace operators both in the case of massive fermionic theory and in the case of massive bosonic scalar theory. We used the duality map to predict for the two-point function of the spin one current in the case of the higgsed phases of bosons. The analysis done in this paper is in the Euclidean signature and with the external momentum $q_{\mu}\equiv (0,0,q_3)$. By analytic continuations, the results in the Lorentzian signature can be obtained by Wick rotating back to Minkowski space $q_3\to \i \omega$. In \cite{Geracie:2015drf}, it was highlighted that the two-point correlator of the $U(1)$ current can be used to calculate the conductivity tensor by applying the Kubo formula. Their analysis was done at zero temperature in the fermionic theory. Also in \cite{Gur-Ari:2016xff}, the study of conductivity was continued in Chern-Simons fermionic matter theories. However in \cite{Gur-Ari:2016xff}, they considered the massless regular fermionic matter coupled to Chern-Simons gauge fields at finite temperature with table-top holonomy distribution. Moreover the results in \cite{Gur-Ari:2016xff} were given as an integral expressions. Our results of two-point correlators of spin-one current in the massive fermionic theory at finite temperature is a generalization of \cite{Geracie:2015drf, Gur-Ari:2016xff} and the final result is given in a much simplified form \eqref{G1Fan}. Following \cite{Geracie:2015drf, Gur-Ari:2016xff}, this can be used to study the conductivity of the massive regular fermionic theory at finite temperature by applying the Kubo formula. In section \ref{spins2pt}, we have commented on the computations of correlators of  arbitrary spin $s$ currents and we have seen a general structure of the results; following the same procedure used in the case of spin-zero and spin-one operators, one can get, e.g., the stress tensor two-point correlation function which can be used to study the viscosity of these theories \cite{Geracie:2015drf}.

As already discussed above, in the massive regular fermionic theory the results can be dual to either critical bosonic scalar theory in the unHiggsed phase or to the Higgsed phase of critical bosons depending upon the signs of $\sgn(h_F\lambda_F)$. We have explicitly checked the duality of the regular fermionic theory with the critical bosonic scalar theory in the unHiggsed phase, which is the case with $\sgn(h_F\lambda_F)=+1$. Also, in the case of $\sgn(h_F\lambda_F)=-1$, we have given the predictions for the two-point correlators of the corresponding current operators in the Higgsed phases of critical bosons. One may compute these correlators to all orders in 't Hooft coupling in the large-$N$ limit directly in the Higgsed phases of critical bosons coupled to Chern-Simons theory \cite{Choudhury:2018iwf, Dey:2018ykx} to explicitly check the duality in the higgsed phase. The technicalities of performing exact computations in the higgsed phases of bosons are rather involved, and we leave this excercise for future work. It is realized that the spin-one current of the massive fermionic theory is dual to the $Z$ bosons in the Higgsed phases \cite{amiyata1}. So, the dualized result of $\langle J_{\mu}^FJ_{\nu}^F\rangle $ in the higgsed phases is proportional to the $Z$ boson propagator $\langle Z_{\mu}Z_{\nu} \rangle$. One can try to explictly compute the exact $Z$ boson propagator directly in the Higgsed phases of bosonic theory \cite{Choudhury:2018iwf}, at least in large $N$ limit, and match with the results predicted here. 

The pair of theories that we have considered in this paper has global $U(1)$ symmetry. One can onsider the analysis presented here in the presence of a chemical potential $\mu$ by turning on a constant background gauge field $\mc{A}_{\nu}=\i \mu \delta_{\nu 3} $.  At least in the case when $|\mu|<c_B$ or $|\mu|<c_F$, the analysis seems to go through exactly the same way and the final result in presence of chemical potential is obtained by making the substitution $\alpha\to \alpha -\i \beta\mu$ in the integrand (except in the holonomy distribution $\rho(\alpha)$) of holonomy integral appearing in the results obtained in this paper at the zero chemical potential. It will be nice to explicitly check if the structure of the results remain the same even in the case when the chemical potential is larger than the thermal mass $c_B$ or $c_F$ \cite{Minwalla:2020ysu}. It seems like, even in the case when $|\mu|>c_B$ or $|\mu|>c_F$, the structure of the results of the correlators of this paper will remain the same; the only modifications will be through the functions $\chi_B(w)$ and $\chi_F(w)$ due to the contour deformations of the holonomy integrals away from the unit circle as prescribed in detail in \cite{Minwalla:2020ysu}. We leave the careful analysis about this as a future excercise.  

As discussed before, as part of the analysis in the bosonic scalar theory, following \cite{Jain:2014nza}, we have generalized the computations of four-point functions of fundamental scalars to finite temperature. As in \cite{Jain:2014nza}, this might be useful to study the scattering of fundamental scalars at finite temperature. 
In footnotes \ref{fncovs0f}, \ref{fncovs1f}, \ref{fncovs1b} and \ref{fncovs1h}, we have provided a possible `covariantized' form of the two-point correlators of the spin one currents of the fermionic and bosonic theories, valid for arbitrary values of the external momenta $q$. It would be nice to have an independent check of these by direct computations without assuming $q_{\pm}=0$.
One can also extend the computations of two-point current correlators to higher point correlators of single trace operators. We have already computed in this paper various exact insertion vertices. And one can use these to explicitly compute the higher-point current correlators by generalizing the existing results (e.g., three-point correlators \cite{Aharony:2012nh, GurAri:2012is}, four-point correlators \cite{Yacoby:2018yvy, Kalloor:2019xjb}) to the massive matter theories at finite temperature.  Presumably, one may also use the results of two-point correlators of single trace operators obtained in this paper to compute the thermal one-point functions using bootstrap approaches. In this paper we have not paid much attention to the contact terms appearing in the results of the two point-correlators. It would be interesting to understand if the contact terms appearing in this paper have any physical significance. Also, it would be interesting to extend the analysis of this paper by going beyond large-$N$ limit, taking into account the contributions from the non-planar diagrams. we leave all these problems for future work.

\subsection*{Acknowledgements}
The author would like to thank S. Chakraborty, S. Jain and N. Prabhakar for helpful conversations. The author would like to thank S. Minwalla for valuable discussions and comments. 
The author would also like to acknowledge support from the Infosys Endowment for the study of the Quantum Structure of Spacetime.

\appendix 

\section{Conventions and useful definitions}\label{appdef}
We work in the three dimensional spacetime with coordinates $(x^1,x^2,x^3)$, where $x^3$ is the Euclidean time coordinate. For a vector $a^{\mu}\equiv (a^1,a^2,a^3)$, the corresponding lightcone components are defined by $a^{\pm}\equiv a_{\mp}=\frac{a^1 \pm \i a^2}{\sqrt{2}}$ and the non-zero components of the metric in the lightcone coordinates are given by $\delta_{+-}=\delta_{-+}=\delta_{33}=1$. The totally antisymmetric Levi-Civita tensor $\epsilon^{\mu\nu\rho}$ that appears in the paper is normalized such that $\epsilon^{123}=\epsilon_{123}=1$. In the lightcone coordinates, $\epsilon_{+-3}=\epsilon^{-+3}=\i $. The momenta squared are given by $p^2=p_1^2+p_2^2+p_3^2=2p_{+}p_{-}+p_3^2=p_s^2+p_3^2$. Here, $p_s=+\sqrt{2p_{+}p_{-}}$ is the radial momentum in the spatial momentum plane. Throughout the paper, the overall external momentum flowing through the insertion vertices of the single trace current operators is denoted by $q$. We work in the momentum choice $q_{\pm}=0$. At finite temperature,
the component $q_3$ of the external momentum $q$ entering an insertion vertex is quantized \cite{Gur-Ari:2016xff};
however, we don't explicitly write the quantized version of it and write it simply as $q_3$.  

Below, we list down a few definitions of useful quantities that have appeared throughout the paper. Following \cite{Choudhury:2018iwf}, we use \eqref{chiB} to define the function $\xi_{B}(z)$ as an integral over $\chi_B(z)$, i.e., $\xi_B(z)=\int^{z} \chi_B(w) dw$. The final expression for $\xi_B(z)$ is given by 
\begin{equation}\label{xiB}
\xi_B(z) =\frac{1}{\beta} \int_{-\pi}^{\pi} \rho_{B}(\alpha)  ~d\alpha ~\bigg[\ln 2 \sinh\bigg(\frac{\beta z+\i \alpha}{2} \bigg)+\ln 2 \sinh\bigg(\frac{\beta z-\i \alpha}{2} \bigg)\bigg] \ .
\end{equation}
Following \cite{Choudhury:2018iwf}, we use the regularization where, $\xi_B(\infty)$ is a pure divergence term and is dropped away (for details, see discussion around equation 2.75 of \cite{Choudhury:2018iwf}). 

Similarly, for the fermionic theory, using \eqref{chiF} we define the function $\xi_F(z)=\int^{z} \chi_F(w) dw$. As before, we use the regularization where, $\xi_F(\infty)$ is purely a divergent term and is thrown away. Explicitly, $\xi_F(z)$ is given by 
\begin{equation}\label{xiFz}
\xi_F(z) =\frac{1}{\beta} \int_{-\pi}^{\pi} \rho_{F}(\alpha)  ~d\alpha ~\bigg[\ln 2 \cosh\bigg(\frac{\beta z+\i \alpha}{2} \bigg)+\ln 2 \cosh\bigg(\frac{\beta z-\i \alpha}{2} \bigg)\bigg] \ .
\end{equation}
We define another function which appears in the intermediate steps of the bosonic computations 
\begin{equation}\label{FBfunction} 
F_B(z,q_3)=\frac{\chi_B(z)}{\big(q_{3}^2+ 4z^2\big)} \ . 
\end{equation}
Similarly, in the case of fermions, we define 
\begin{equation}\label{FFfunction} 
F_F(z,q_3)=\frac{\chi_F(z)}{\big(q_{3}^2+ 4z^2\big)} \ . 
\end{equation}
Another set of useful definitions are 
\begin{equation}\label{Hbfunction} 
H_B(z,q_3)=\exp\Big(4\i \lambda_B q_3 \int^{z} F_B(w,q_3) dw  \Big)= \exp\Big(4\i \lambda_B q_3 \int^{z} \frac{dw \ \chi_B(w)}{q_{3}^2+ 4w^2} \Big)  \ ,
\end{equation}
and
\begin{equation}\label{HFfunction} 
H_F(z,q_3)=\exp\Big(4\i \lambda_F q_3 \int^{z} F_F(w,q_3) dw  \Big)= \exp\Big(4\i \lambda_F q_3 \int^{z} \frac{dw \ \chi_F(w)}{q_{3}^2+ 4w^2} \Big)  \ .
\end{equation}

Before concluding this appendix, we mention here that the multivalued function $\tan^{-1}(x)$ that appears in this paper is defined by choosing the branch $-\frac{\pi}{2}< \tan^{-1}(x)<\frac{\pi}{2}$. With this definition, it follows that $\tan^{-1}(-x)=\tan^{-1}(x)$. Also, we use the property that $\tan^{-1}(x)+\tan^{-1}(\frac{1}{x})=\sgn(x)\frac{\pi}{2} \ $. 
\section{Angular integrals :}
The angular integrals that are used in this paper, are given by
\begin{equation}\label{angint}
\begin{split}
\int_{0}^{2\pi} \frac{d\theta_l}{2\pi} \frac{(l_{-})^j }{(l-k)_-} & =(k_{-})^{j-1}\bigg[-\theta(-j) + \Theta(l_s- k_s) \bigg] \ .
\end{split}
\end{equation}
Here, on the right-hand side, $\theta$ and $\Theta$ both are unit step functions. The small $\theta$- function is used for the power of the momenta $\ell_{-}$ appearing in the numerator of the integrand and it is assumed to be defined such that $\theta(j=0)=1$.  This relation is true for $\forall j\in \mathbb{Z}$. The above integration result \eqref{angint} can be easily obtained by chosing $\ell_{-}=\frac{\ell_s}{\sqrt{2}}e^{\i \theta_{\ell}}$ and converting the resulting angular integral into a complex contour integral over a unit circle by substituting $z_{\ell}=e^{\i \theta_{\ell}}$. 
\section{Useful integrals}
In this appendix, we list down (non-vanishing) useful integrals which are repeatedly used in the main text. 
\subsection{Integrals in bosonic theory}
We list here a set of useful integrals and we present the integration results in terms of a quantity $a(k_s)=+\sqrt{k_s^2+c_B^2}$, and the function $\chi_B(z)$ given by \eqref{chiB}. We don't write the temperature dependence $\beta$ explicitly in the arguments of the functions that are being defined, but is implicitly assumed. The external momentum component $q_3$ appearing in the integrals is quantized but the explicit quantized form is not written for simplicity. 
\subsubsection*{Integration B1 :}
\begin{equation}\label{GB1}
\begin{split}
\int \frac{\mathcal{D}_B k_3}{2\pi} \frac{1}{k_{3}^2+ a^2(k_s)} & =\frac{1}{\beta} \int_{-\pi}^{\pi} \rho_{B}(\alpha)  ~d\alpha \sum_{n_k=-\infty}^{\infty} \frac{1}{\big(\frac{2n_k\pi + \alpha}{\beta} \big) ^2+ a^2(k_s)} \\ 
& = \frac{1}{2a(k_s)} \ \chi_B(a(k_s)) 
\end{split} 
\end{equation} 
where, $a(k_s)=+\sqrt{k_s^2+c_B^2}$ and $\chi_B(z)$ is given by \eqref{chiB}. 
In the zero-temperature limit, i.e., in $\beta\rightarrow \infty $ limit, it is clear from \eqref{chiB} that $\lim_{\beta\rightarrow \infty } \chi_B(z)=1$. In the zero temperature limit, the above integral \eqref{GB1} reduces to the usual integral 
\begin{equation}\label{IB1}
\int_{-\infty}^{\infty}  \frac{d k_3}{2\pi} \frac{1}{k_{3}^2+ a^2(k_s)}=\frac{1}{2a(k_s)}
\end{equation} 
We now consider the following integration
\begin{equation}\label{gBintegration}
\begin{split}
\int \frac{\mathcal{D}_B^3 k}{(2\pi)^3} \frac{1}{k_{3}^2+ a^2(k_s)}
& =\int_{0}^{\infty} \frac{k_sdk_s}{2\pi} \frac{\chi_B(a(k_s))}{2a(k_s)}\\
& = \frac{1}{4\pi}\int_{c_B}^{\infty} da(k_s) \ \chi_B(a(k_s)) \\
& = \frac{1}{4\pi} \bigg(\xi_B(\infty) -\xi_B(c_B) \bigg)\\
& =-\frac{\xi_B(c_B)}{4\pi} 
\end{split}
\end{equation} 
where, we have used the definition $\xi_B'(z)=\chi_B(z)\Rightarrow \xi_B(z)=\int^{z}\chi_B(w)dw$. We use the regularization scheme \cite{Choudhury:2018iwf} (see around equation 2.75 of \cite{Choudhury:2018iwf}), where, $\xi_B(\infty)$ is a pure divergence term and is dropped away. 

\subsubsection*{Integration B2 :}
There is another integration which is mostly used in the main text; this is given by
\begin{equation}\label{GB2}
\begin{split}
\int \frac{\mathcal{D}_B k_3}{2\pi} & \frac{1}{\big(k_{3}^2+ a^2(k_s)\big)\big((k_{3}+q_3)^2+ a^2(k_s)\big)} \\ 
&=\frac{1}{\beta} \int_{-\pi}^{\pi} \rho_{B}(\alpha)  ~d\alpha \sum_{n_k=-\infty}^{\infty} \frac{1}{\bigg(\big(\frac{2n_k\pi + \alpha}{\beta} \big) ^2+ a^2(k_s)\bigg)\bigg(\big(\frac{2n_k\pi + \alpha}{\beta} +q_3\big) ^2+ a^2(k_s)\bigg)} \\
& =\frac{\chi_B(a(k_s))}{a(k_s)\big(q_{3}^2+ 4a^2(k_s)\big)}
\end{split}
\end{equation} 
where, $\chi_B(z) $ is the same function defined in \eqref{chiB}. As expected, in the zero temperature limit, this reduces to the standard integral 
\begin{equation}\label{IB2}
\int_{-\infty}^{\infty}  \frac{d k_3}{2\pi} \frac{1}{\big(k_{3}^2+ a^2(k_s)\big)\big((k_{3}+q_3)^2+ a^2(k_s)\big)}=\frac{1}{a(k_s)\big(q_{3}^2+ 4a^2(k_s)\big)}
\end{equation} 

\subsubsection*{Integration B3 :}
Another useful integral is given by
\begin{equation}\label{GB3}
\begin{split}
\int \frac{\mathcal{D}_B k_3}{2\pi} & \frac{k_3(k_3+q_3)}{\big(k_{3}^2+ a^2(k_s)\big)\big((k_{3}+q_3)^2+ a^2(k_s)\big)}\\
&=\frac{1}{\beta} \int_{-\pi}^{\pi} \rho_{B}(\alpha)  ~d\alpha \sum_{n_k=-\infty}^{\infty} \frac{\big(\frac{2n_k\pi + \alpha}{\beta} \big)\big(\frac{2n_k\pi + \alpha}{\beta} +q_3\big)}{\bigg(\big(\frac{2n_k\pi + \alpha}{\beta} \big) ^2+ a^2(k_s)\bigg)\bigg(\big(\frac{2n_k\pi + \alpha}{\beta} +q_3\big) ^2+ a^2(k_s)\bigg)} \\
& =\frac{a(k_s) \ \chi_B(a(k_s))}{\big(q_{3}^2+ 4a^2(k_s)\big)}
\end{split}
\end{equation} 
where, $\chi_B(z) $ is defined by \eqref{chiB}. This reduces, in the zero temperature limit, to the usual integral 
\begin{equation}\label{IB3}
\int_{-\infty}^{\infty}  \frac{d k_3}{2\pi} \frac{k_3(k_3+q_3)}{\big(k_{3}^2+ a^2(k_s)\big)\big((k_{3}+q_3)^2+ a^2(k_s)\big)}=\frac{a(k_s)}{\big(q_{3}^2+ 4a^2(k_s)\big)}
\end{equation} 
\subsection{Integrals in fermionic Theory } 
The integrals that appear in the fermionic theory are of the same structure as that appear in the bosonic theory listed above. Below we provide a set of useful integrals 
\subsubsection*{Integration F1 :}
\begin{equation}\label{GF1}
\begin{split} 
\int \frac{\mathcal{D}_F k_3}{2\pi} \frac{1}{k_{3}^2+ a^2(k_s)} & =\frac{1}{\beta} \int_{-\pi}^{\pi} \rho_{F}(\alpha)  ~d\alpha \sum_{n_k=-\infty}^{\infty} \frac{1}{\big(\frac{(2n_k+1)\pi + \alpha}{\beta} \big) ^2+ a^2(k_s)} \\
 & =\frac{1}{2a(k_s)} \ \chi_F(a(k_s)) 
\end{split} 
\end{equation} 
where, $a(k_s)=+\sqrt{k_s^2+c_F^2}$ and $\chi_F(z)$ is given by \eqref{chiF}. We don't explictly write the temperature dependence $\beta$ in the arguments of the functions that are being defined, but is implicitly assumed. In the zero-temperature limit, i.e., in the $\beta\rightarrow \infty $ limit, it follows from \eqref{chiF}, $\lim_{\beta\rightarrow \infty } \chi_F(a(k_s))=1$. Thus this integral reduces, in the zero temperature limit, to the known integral
\begin{equation}\label{IF1}
\int_{-\infty}^{\infty}  \frac{d k_3}{2\pi} \frac{1}{k_{3}^2+ a^2(k_s)}=\frac{1}{2a(k_s)}
\end{equation} 
As in the case of bosons, we consider here the following integration 
\begin{equation}\label{gFintegration}
\begin{split}
\int \frac{\mathcal{D}_F^3 k}{(2\pi)^3} \frac{1}{k_{3}^2+ a^2(k_s)}
& =\int_{0}^{\infty} \frac{k_sdk_s}{2\pi} \frac{\chi_F(a(k_s))}{2a(k_s)}\\
& = \frac{1}{4\pi}\int_{c_F}^{\infty} da(k_s) \ \chi_F(a(k_s)) \\
& = \frac{1}{4\pi} \bigg(\xi_F(\infty) -\xi_F(c_F) \bigg)\\
& =-\frac{\xi_F(c_F)}{4\pi} 
\end{split}
\end{equation} 
where, we have used the definition $\xi_F'(z)=\chi_F(z)\Rightarrow \xi_F(z)=\int^{z}\chi_F(w)dw$. We use the regularization scheme, where, $\xi_F(\infty)$ is a pure divergent term and is dropped.

\subsubsection*{Integration F2 :}
As in the case of bosons, there is another useful integral 
\begin{equation}\label{GF2}
\begin{split}
\int \frac{\mathcal{D}_F k_3}{2\pi} & \frac{1}{\big(k_{3}^2+ a^2(k_s)\big)\big((k_{3}+q_3)^2+ a^2(k_s)\big)} \\ 
&=\frac{1}{\beta} \int_{-\pi}^{\pi} \rho_{F}(\alpha)  ~d\alpha \sum_{n_k=-\infty}^{\infty} \frac{1}{\bigg(\big(\frac{(2n_k+1)\pi + \alpha}{\beta} \big) ^2+ a^2(k_s)\bigg)\bigg(\big(\frac{(2n_k+1)\pi + \alpha}{\beta} +q_3\big) ^2+ a^2(k_s)\bigg)} \\
& =\frac{\chi_F(a(k_s))}{a(k_s)\big(q_{3}^2+ 4a^2(k_s)\big)}
\end{split}
\end{equation} 
where, $\chi_F(z)$ is given by \eqref{chiF}.  In the zero temperature limit, this integral reduces to the standard integral 
\begin{equation}\label{IF2}
\int_{-\infty}^{\infty}  \frac{d k_3}{2\pi} \frac{1}{\big(k_{3}^2+ a^2(k_s)\big)\big((k_{3}+q_3)^2+ a^2(k_s)\big)}=\frac{1}{a(k_s)\big(q_{3}^2+ 4a^2(k_s)\big)}
\end{equation} 
\subsubsection*{Integration F3 :}
The following integral is also useful for the computations in this paper
\begin{equation}\label{GF3}
\begin{split}
\int \frac{\mathcal{D}_F k_3}{2\pi} & \frac{k_3(k_3+q_3)}{\big(k_{3}^2+ a^2(k_s)\big)\big((k_{3}+q_3)^2+ a^2(k_s)\big)} \\
&=\frac{1}{\beta} \int_{-\pi}^{\pi} \rho_{B}(\alpha)  ~d\alpha \sum_{n_k=-\infty}^{\infty} \frac{\big(\frac{(2n_k+1)\pi + \alpha}{\beta} \big)\big(\frac{(2n_k+1)\pi + \alpha}{\beta} +q_3\big)}{\bigg(\big(\frac{(2n_k+1)\pi + \alpha}{\beta} \big) ^2+ a^2(k_s)\bigg)\bigg(\big(\frac{(2n_k+1)\pi + \alpha}{\beta} +q_3\big) ^2+ a^2(k_s)\bigg)} \\
& =\frac{a(k_s) \ \chi_F(a(k_s))}{\big(q_{3}^2+ 4a^2(k_s)\big)}
\end{split}
\end{equation} 
In the zero tempearature limit, this integral reduces to the familiar integral 
\begin{equation}\label{IF3}
\int_{-\infty}^{\infty}  \frac{d k_3}{2\pi} \frac{k_3(k_3+q_3)}{\big(k_{3}^2+ a^2(k_s)\big)\big((k_{3}+q_3)^2+ a^2(k_s)\big)}=\frac{a(k_s)}{\big(q_{3}^2+ 4a^2(k_s)\big)}
\end{equation} 

\section{Computation of thermal four-point function of fundamental scalars}\label{4ptscalar} 
In this appendix, we generalize the computation of four-point function of fundamental scalars in the regular boson theory presented in \cite{Jain:2014nza}(see e.g. section 4 and appendix D.3 of \cite{Jain:2014nza} for details of computation at zero temperature) to non-zero temperature. 
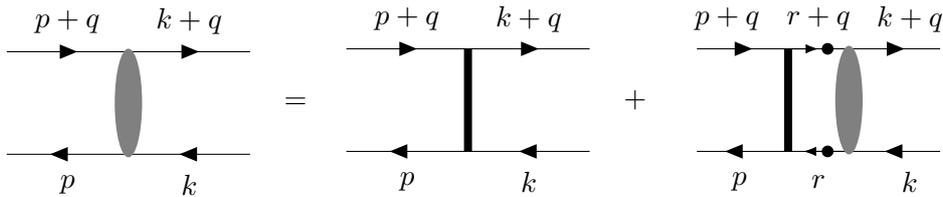
\begin{figure}[h!]
	\begin{center}
		\raisebox{1pt}{ 
			\begin{tikzpicture}[scale=0.4]
			\begin{feynman}
			\coordinate (a) at (-4,1.7) ;
			\coordinate (b) at (0,1.7) ;
			\coordinate (c) at (4,1.7) ;
			\coordinate (d) at (0,-1.7) ;
			\coordinate (e) at (-4,-1.7) ;
			\coordinate (f) at (4,-1.7) ; 
			\diagram* {
				(a) -- [fermion] (b),
				(b) -- [fermion] (c),
				(e) -- [anti fermion] (d),
				(f) -- [fermion] (d)
			}; 
			\end{feynman}
			\draw (a)+(2,1) node {$p+q$}; 
			\draw (b)+(2,1) node {$k+q$}; 
			\draw (e)+(2,-1) node {$p$}; 
			\draw (d)+(2,-1) node {$k$}; 
			\fill[gray] (0,0) ellipse [x radius=4.5mm, y radius=1.8cm, rotate=0];
			\draw (f)+(1.5,1.7) node {{$= $}}; 
			\end{tikzpicture} }
		\raisebox{2pt}{ 
			\begin{tikzpicture}[scale=0.4]
			\begin{feynman}
			\coordinate (a) at (-4,1.7) ;
			\coordinate (b) at (0,1.7) ;
			\coordinate (c) at (4,1.7) ;
			\coordinate (d) at (0,-1.7) ;
			\coordinate (e) at (-4,-1.7) ;
			\coordinate (f) at (4,-1.7) ; 
			\diagram* {
				(a) -- [fermion] (b),
				(b) -- [fermion] (c),
				(e) -- [anti fermion] (d),
				(f) -- [fermion] (d)
			}; 
			\end{feynman}
			\draw (a)+(2,1) node {$p+q$}; 
			\draw (b)+(2,1) node {$k+q$}; 
			\draw (e)+(2,-1) node {$p$}; 
			\draw (d)+(2,-1) node {$k$}; 
			\draw[line width=3pt] (0,1.7) -- (0,-1.7) ;
			\draw (f)+(1.5,1.7) node {{$+ $}}; 
			\end{tikzpicture} }
		\raisebox{2pt}{ 
			\begin{tikzpicture}[scale=0.4]
			\begin{feynman}
			\coordinate (a) at (-4,1.7) ;
			\coordinate (b1) at (-1,1.7) ;
			\coordinate (b2) at (1,1.7) ;
			\coordinate (c) at (4,1.7) ;
			\coordinate (d1) at (-1,-1.7) ;
			\coordinate (d2) at (1,-1.7) ;
			\coordinate (e) at (-4,-1.7) ;
			\coordinate (f) at (4,-1.7) ; 
			\diagram* {
				(a) -- [fermion] (b1),
				(b2) -- [fermion] (c),
				(e) -- [anti fermion] (d1),
				(f) -- [fermion] (d2)
			}; 
			\end{feynman}
			\draw (a)+(1,1) node {$p+q$}; 
			\draw (b2)+(2,1) node {$k+q$}; 
			\draw (e)+(1.4,-1) node {$p$}; 
			\draw (d2)+(2,-1) node {$k$}; 
			\draw (b1)+(1,1) node {$r+q$}; 
			\draw (d1)+(1,-1) node {$r$}; 
			\draw[-Latex] (b1) -- ++(1,0) ;
			\draw (b1) -- (b2) ;
			\draw[-Latex] (d2) -- ++(-1.5,0) ;
			\draw (d1) -- (d2) ;
			\filldraw (d1)+(1.3,0) circle (5pt) ; 
			\filldraw (b1)+(1.3,0) circle (5pt) ; 
			\draw[line width=3pt] (-1,1.7) -- (-1,-1.7) ;
			\fill[gray] (1,0) ellipse [x radius=4.5mm, y radius=1.8cm, rotate=0];
			\end{tikzpicture} }
	\end{center} 
	\caption{Schinger-Dyson equation for offshell four-point function of scalars. The elliptic blob denotes the exact scalar four-point function and is denoted by $\mc{A}(p,k,q)$. The thick vertical line corresponds to the `effective single particle exachange' four-point amplitude and is denoted by $\mc{A}_0(p,k,q)$; this amplitude is given by the sum of the diagrams of Fig.\ref{oneloop} and Fig.\ref{4ptvertex}. The filled circle denotes the exact scalar propagator.}
	\label{4ptscalarsd}
\end{figure}
The Schwinger-Dyson equations and the feynman diagrams are all the same as in \cite{Jain:2014nza}. We redo the computations here with the relevant modifications due to the finite temperature effect including holonomy of gauge fields. 
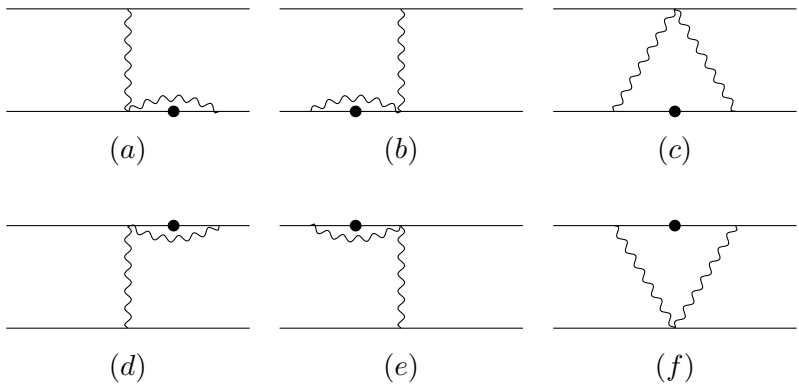
\begin{figure}[h!]
	\begin{center}
		\raisebox{1pt}{ 
			\begin{tikzpicture}[scale=0.4]
			\begin{feynman}
			\coordinate (a) at (-4,1.7) ;
			\coordinate (b) at (0,1.7) ;
			\coordinate (c) at (4,1.7) ;
			\coordinate (d) at (0,-1.7) ;
			\coordinate (e) at (-4,-1.7) ;
			\coordinate (f) at (4,-1.7) ; 
			\coordinate (dp) at (3,-1.7) ; 
			\diagram* {
				(b) -- [boson] (d) ,
				(d) -- [boson, bend left] (dp)
			}; 
			\end{feynman}
			\draw (a) -- (c) ;
			\draw (e) -- (f) ;
			\filldraw (d)+(1.5,0) circle (5pt) ; 
			\draw (0, -3) node {$(a)$}; 
			\end{tikzpicture} }
		\raisebox{1pt}{ 
			\begin{tikzpicture}[scale=0.4]
			\begin{feynman}
			\coordinate (a) at (-4,1.7) ;
			\coordinate (b) at (0,1.7) ;
			\coordinate (c) at (4,1.7) ;
			\coordinate (d) at (0,-1.7) ;
			\coordinate (e) at (-4,-1.7) ;
			\coordinate (f) at (4,-1.7) ; 
			\coordinate (dpp) at (-3,-1.7) ; 
			\diagram* {
				(b) -- [boson] (d) ,
				(dpp) -- [boson, bend left] (d)
			}; 
			\end{feynman}
			\draw (a) -- (c) ;
			\draw (e) -- (f) ;
			\filldraw (d)+(-1.5,0) circle (5pt) ; 
			\draw (0, -3) node {$(b)$}; 
			\end{tikzpicture} }
		\raisebox{1pt}{ 
			\begin{tikzpicture}[scale=0.4]
			\begin{feynman}
			\coordinate (a) at (-4,1.7) ;
			\coordinate (b) at (0,1.7) ;
			\coordinate (c) at (4,1.7) ;
			\coordinate (d) at (0,-1.7) ;
			\coordinate (e) at (-4,-1.7) ;
			\coordinate (f) at (4,-1.7) ; 
			\coordinate (dp) at (2,-1.7) ; 
			\coordinate (dpp) at (-2,-1.7) ; 
			\diagram* {
				(b) -- [boson] (dp) ,
				(b) -- [boson] (dpp) 
			}; 
			\end{feynman}
			\draw (a) -- (c) ;
			\draw (e) -- (f) ;
			\filldraw (d) circle (5pt) ; 
			\draw (0, -3) node {$(c)$}; 
			\end{tikzpicture} } \\
		\vspace{15pt}
		\raisebox{1pt}{ 
			\begin{tikzpicture}[scale=0.4]
			\begin{feynman}
			\coordinate (a) at (-4,1.7) ;
			\coordinate (b) at (0,1.7) ;
			\coordinate (c) at (4,1.7) ;
			\coordinate (d) at (0,-1.7) ;
			\coordinate (e) at (-4,-1.7) ;
			\coordinate (f) at (4,-1.7) ; 
			\coordinate (bp) at (3,1.7) ; 
			\diagram* {
				(b) -- [boson] (d) ,
				(b) -- [boson, bend right] (bp)
			}; 
			\end{feynman}
			\draw (a) -- (c) ;
			\draw (e) -- (f) ;
			\filldraw (b)+(1.5,0) circle (5pt) ; 
			\draw (0, -3) node {$(d)$}; 
			\end{tikzpicture} }
		\raisebox{1pt}{ 
			\begin{tikzpicture}[scale=0.4]
			\begin{feynman}
			\coordinate (a) at (-4,1.7) ;
			\coordinate (b) at (0,1.7) ;
			\coordinate (c) at (4,1.7) ;
			\coordinate (d) at (0,-1.7) ;
			\coordinate (e) at (-4,-1.7) ;
			\coordinate (f) at (4,-1.7) ; 
			\coordinate (bpp) at (-3,1.7) ; 
			\diagram* {
				(b) -- [boson] (d) ,
				(b) -- [boson, bend left] (bpp)
			}; 
			\end{feynman}
			\draw (a) -- (c) ;
			\draw (e) -- (f) ;
			\filldraw (b)+(-1.5,0) circle (5pt) ; 
			\draw (0, -3) node {$(e)$}; 
			\end{tikzpicture} }
		\raisebox{1pt}{ 
			\begin{tikzpicture}[scale=0.4]
			\begin{feynman}
			\coordinate (a) at (-4,1.7) ;
			\coordinate (b) at (0,1.7) ;
			\coordinate (c) at (4,1.7) ;
			\coordinate (d) at (0,-1.7) ;
			\coordinate (e) at (-4,-1.7) ;
			\coordinate (f) at (4,-1.7) ; 
			\coordinate (bp) at (2,1.7) ; 
			\coordinate (bpp) at (-2,1.7) ; 
			\diagram* {
				(d) -- [boson] (bp) ,
				(d) -- [boson] (bpp) 
			}; 
			\end{feynman}
			\draw (a) -- (c) ;
			\draw (e) -- (f) ;
			\filldraw (b) circle (5pt) ; 
			\draw (0, -3) node {$(f)$}; 
			\end{tikzpicture} } 
	\end{center} 
	\caption{Some of the diagrams contributing to one loop four-point amplitude $\mc{A}_0(p,k,q)$. The filled circle denotes the exact scalar propagator. The assignments of momenta in the external legs are the same as in Fig.\ref{4ptscalarsd}.}
	\label{oneloop}
\end{figure}
Summing all the diagrams in Fig.\ref{oneloop} (or see figure 19 of \cite{Jain:2014nza} (also including the diagram corresponding to the $(\bar{\phi}\phi)^3$ interaction vertex), the one-loop effective one-particle exchange can be written as 
\begin{equation}
\begin{split}
NA_{\text{one-loop}}& =\sum_{i=1}^{6}NA_{i} = -8\pi^2 \lambda_{B}^2 \int \frac{\mathcal{D}_B^3\ell }{(2\pi)^3}~\frac{1}{\ell^2+c_{B}^2} \\
& = -8\pi^2 \lambda_{B}^2\bigg(-\frac{\xi_B(c_{B}) }{4\pi}\bigg)\\
& = 2\pi \lambda_{B}^2 \xi_B(c_{B}) 
\end{split}
\end{equation}
where, the function $\xi_B(c_{B}) $ is given by \eqref{xiB}. 

\begin{figure}[h!]
	\begin{center}
		\raisebox{2pt} {
			\begin{tikzpicture}[scale=0.2]
		\coordinate (A) at (3,3) ;
		\coordinate (B) at (3,-3) ;
		\coordinate (C) at (-3,3) ;
		\coordinate (D) at (-3,-3) ;
		\draw (A) -- (D) ;
		\draw (B) -- (C) ; 
		\draw (0, -6) node {$(a)$}; 
		\end{tikzpicture}  }
		\hspace{2cm}
		\raisebox{2pt} {
			\begin{tikzpicture}[scale=0.5]
			\draw (-1.5,0) -- (1.5,0) ;
			\draw (0,0) -- (1.5,-1) ;
			\draw (0,0) -- (-1.5,-1) ;
			\draw (0,0.75) circle (0.75cm);
			\filldraw (0,1.5) circle (4pt) ; 
			\draw (0, -2) node {$(b)$}; 
			\end{tikzpicture} } 
			\hspace{2cm}
			\raisebox{1pt}{ 
			\begin{tikzpicture}[scale=0.3]
			\begin{feynman}
			\coordinate (a) at (-4,1.7) ;
			\coordinate (b) at (0,1.7) ;
			\coordinate (c) at (4,1.7) ;
			\coordinate (d) at (0,-1.7) ;
			\coordinate (e) at (-4,-1.7) ;
			\coordinate (f) at (4,-1.7) ; 
			\diagram* {
				(a) -- [fermion] (b),
				(b) -- [fermion] (c),
				(d) -- [boson,rmomentum'={[arrow shorten=0.3]\(p-k\)}] (b),
				(e) -- [anti fermion] (d),
				(f) -- [fermion] (d)
			}; 
			\end{feynman}
			\draw (a)+(2,1) node {$p+q$}; 
			\draw (b)+(2,1) node {$k+q$}; 
			\draw (e)+(2,-1) node {$p$}; 
			\draw (d)+(2,-1) node {$k$}; 
			\draw (0, -4) node {$(c)$}; 
			\end{tikzpicture} }
	\end{center} 
	\caption{Some of the diagrams contributing to the `effective single particle exchange' four-point amplitude $\mc{A}_0(p,k,q)$. Diagram $(a)$ corresponds to the four-point interaction vertex. Diagram $(b)$ corresponds to the contribution of the interaction vertex $(\bar{\phi}\phi)^3$ to the four-point amplitude; here, the filled circular dot denotes the exact scalar propagator. Diagram $(c)$ corresponds to the four-point amplitude due to the tree-level gauge boson exachange.}
	\label{4ptvertex}
\end{figure}
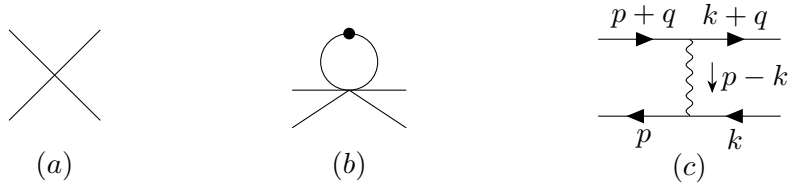

The contact contribution to the four-point function from the $b_4$ vertex shown in the Fig.\ref{4ptvertex}(a) is given by $-(\frac{b_{4}}{2N_B})\cdot 2=-\frac{b_{4}}{N_B} $.
As given in the action \eqref{RBlag}, contribution from the $b_6$ vertex to the four point function is shown in Fig.\ref{4ptvertex}(b) and is given by  $(-\frac{b_6}{N_B^2})\big(-\frac{\xi_B(c_B)}{4\pi}N_B\big)=\frac{\xi_B(c_B)b_{6}}{4\pi N_B}$. 
At the tree level, the contribution from the diagram shown in Fig.\ref{4ptvertex}(c) is given by 
\begin{equation}
\begin{split}
A_{\text{tree}}& =\mathcal{V}^{\mu}(p+q, -(k+q))G^B_{\nu\mu}(p-k)\mathcal{V}^\nu(k,-p)
=\frac{4\pi}{\kappa_B} \frac{q^\mu (k+p)^{\nu}\epsilon_{\nu-\mu}}{(p-k)_{-}}\\
&= -\frac{4\pi \i \lambda_{B}q_{3}}{N_B} \frac{(k+p)_{-}}{(k-p)_{-}}
\end{split}
\end{equation}
So, the total contribution to the `effective one-particle exchange' four-point amplitude $\mc{A}_0(p,k,q)$ 
is given by 
\begin{equation}\label{V0E}
\begin{split}
N_B\mathcal{A}_0(p,k,q_{3}) 
& = -4\pi \i \lambda_{B}q_{3} \frac{(k+p)_{-}}{(k-p)_{-}} + 2\pi \lambda_{B}^2 \xi_B(c_{B})-b_{4}+\frac{\xi(c_{B})b_{6}}{4\pi} \\ 
& = -4\pi \i \lambda_{B}q_{3} \frac{(k+p)_{-}}{(k-p)_{-}} +\tilde{b}_{4}
\end{split}
\end{equation}
where, in the last line $\tl{b}_4$ is defined by 
\begin{equation}\label{b4tl} 
\tilde{b}_{4} = -b_{4} +\frac{\xi_B(c_{B})b_{6}}{4\pi} + 2\pi \lambda_{B}^2 \xi_B(c_{B})
\end{equation}
\subsection{Schwinger-Dyson Equation}
The exact thermal scalar four-point function that we are interested in is defined by 
\begin{equation}
\langle \phi^i(p+q)\bar{\phi}_{j}(-(k+q)) \phi^m(k) \bar{\phi}_{n}(-p) \rangle  = \mathcal{A}^{im}_{jn}(p,k,q)  (2\pi)^3 \delta^3(0)
\end{equation}
For the purpose of calculation of the current correlators,  we are interested in the following contraction of the color indices
\begin{equation}
\mathcal{A}^{im}_{jn}(p,k,q)=\mathcal{A}(p,k,q) \ \delta^{i}_{n} \delta^{m}_{j}
\end{equation}
As explained in details in \cite{Jain:2014nza}, in order to calculate the exact four-point amplitude we solve the Schwinger-Dyson equation given in Figure 5 of \cite{Jain:2014nza}. 

In the mathematical form, the Schwinger Dyson equation can be written in the following two alternative but equivalent forms 
\begin{equation}\label{eqn1}
\mathcal{A}(p,k,q_3)=\mathcal{A}_{0}(p,k,q_3)+N_B\int \frac{\mathcal{D}_B^3 r}{(2\pi)^3}~\frac{\mathcal{A}_{0}(p,r,q_3)\mathcal{A}(r,k,q_3)}{(r^2+c_B^2)((r+q)^2+c_B^2)}
\end{equation}
\begin{equation}\label{eqn2}
\mathcal{A}(p,k,q_3)=\mathcal{A}_{0}(p,k,q_3)+N_B\int \frac{\mathcal{D}_B^3 r}{(2\pi)^3}~\frac{\mathcal{A}(p,r,q_3)\mathcal{A}_{0}(r,k,q_3)}{(r^2+c_B^2)((r+q)^2+c_B^2)}
\end{equation}
\\

These are the two equations that need to be solved for $\mathcal{A}(p,k,q_3)$. We work in the choice $q_{\pm} =0 $, and the expression (\ref{V0E}) for $\mathcal{A}_0(p,k,q_3)$ is independent of $p_3$. So, from the equation (\ref{eqn1}), we can write $\partial_{p_{3}}\mathcal{A}(p,k,q_3)=0$. In the same way, we can write from the equation (\ref{eqn2}), $\partial_{k_{3}}\mathcal{A}(p,k,q_3)=0$. 

This implies that $\mathcal{A}(p,k,q_3)$ is independent of $p_{3}$ and $k_{3}$. So, using the integration result \eqref{GB2}, we integrate $r_3$ momenta. 
And so, from (\ref{eqn1}) after integrating over $r_{3}$, and using the definition \eqref{FBfunction} we get
\begin{equation}\label{eqn1.1}
\mathcal{A}(\vec{p},\vec{k},q_3)=\mathcal{A}_0(\vec{p},\vec{k},q_3)+N_B\int \frac{d^2 \vec{r}}{(2\pi)^2}~\frac{F_B(a(r_{s}),q_3)}{a(r_{s})}\mathcal{A}_0(\vec{p},\vec{r},q_3)\mathcal{A}(\vec{r},\vec{k},q_3)
\end{equation}
where, $a(r_s)=+\sqrt{r_s^2+c_B^2} \ $. Similarly from (\ref{eqn2}) after integrating over $r_{3}$, we get 
\begin{equation}\label{eqn2.1}
\mathcal{A}(\vec{p},\vec{k},q_3)=\mathcal{A}_0(\vec{p},\vec{k},q_3)+N_B \int \frac{d^2 \vec{r}}{(2\pi)^2}~\frac{F_B(a(r_{s}),q_3)}{a(r_{s})} \mathcal{A}(\vec{p},\vec{r},q_3,\lambda_{B})\mathcal{A}_0(\vec{r},\vec{k},q_3)
\end{equation}
Treating the lightcone momenta as the complex coordinates, 
$ p_{-} \rightarrow z, \ p_{+} \rightarrow \bar{z} $, 
and using the standard complex integration formula 
$\frac{\partial }{\partial \bar{z}}\frac{1}{(z-a)}=2\pi~\delta^{(2)}(\vec{r}-\vec{a})$,
from (\ref{V0E}) we calculate
\begin{equation}\label{derA0delta} 
\begin{split}
N_B \partial_{p_{+}}\mathcal{A}_0(\vec{p},\vec{k},q_3)
& = (2\pi)^2 4\i \lambda_{B} q_{3} p_{-}\delta^{(2)}(\vec{p}-\vec{k})\\
\end{split}
\end{equation}
From equation (\ref{eqn1.1}) and the integration result \eqref{FBfunction}, one can easily find 
\begin{equation}\label{eqn1.2}
\begin{split}
\partial_{p_{+}}(\mathcal{A}(\vec{p},\vec{k},q_3)-\mathcal{A}_0(\vec{p},\vec{k},q_3))
& =\frac{4\i \lambda_{B}q_{3} p_{-}}{a(p_{s})}F_B(a(k_s),q_3) \mathcal{A}(\vec{p},\vec{k},q_3)
\end{split}
\end{equation}
Similarly from (\ref{eqn2.1}), one gets
\begin{equation}\label{eqn2.2}
\begin{split}
\partial_{k_{+}}(\mathcal{A}(\vec{p},\vec{k},q_3)-\mathcal{A}_0(\vec{p},\vec{k},q_3))
& =-\frac{4\i \lambda_{B}q_{3} k_{-}}{a(r_{s})} F_B(a(k_s),q_3)  \mathcal{A}(\vec{p},\vec{k},q_3)
\end{split}
\end{equation}
Equations (\ref{eqn1.2}) and (\ref{eqn2.2}) may be regarded as the first order ordinary differential equations in the variables $p_{+}$ and $k_{+}$ respectively. These equations are easily solved. For convenience, we define the following function 
\begin{equation}\label{HBfunction}
H_B(z,q_3) =\exp\big( {4\i \lambda_{B} q_3 \int^{z} F_B(w,q_3) dw}  \big) 
\end{equation}
The inverse of this function is given by $H_B^{-1}(z,q_3) = \frac{1}{H_B(z,q_3)} $. 
From (\ref{eqn1.2}), we get 
\begin{equation}\label{eqn1.3}
\begin{split}
\partial_{p_{+}}\bigg(H_B^{-1}(a(p_s),q_3) \mathcal{A}(\vec{p},\vec{k},q_3)\bigg) 
& = H_B^{-1}(a(p_s),q_3) \partial_{p_{+}}\mathcal{A}_0(\vec{p},\vec{k},q_3)
\end{split}
\end{equation}
In the same way, the equation (\ref{eqn2.2}) can be recast into the following form 
\begin{equation}\label{eqn2.3}
\begin{split}
\partial_{k_{+}}\bigg(H_B(a(k_s),q_3) \mathcal{A}(\vec{p},\vec{k},q_3)\bigg) 
& = H_B(a(p_s),q_3) \partial_{k_{+}}\mathcal{A}_0(\vec{p},\vec{k},q_3)
\end{split}
\end{equation}
The equations \eqref{eqn1.3} and \eqref{eqn2.3} can be easily solved by integrations.
We have already found in \eqref{derA0delta} that $\partial_{p_{+}}\mathcal{A}_0(\vec{p},\vec{k},q_3)\propto \delta^{(2)}(\vec{p}-\vec{k})$; so, we use the formula $$f(\vec{r})\delta^{(2)}(\vec{r}-\vec{a})=f(\vec{a})\delta^{(2)}(\vec{r}-\vec{a})$$
Thus, we can replace the prefactors on the RHS of equation \eqref{eqn1.3} by replacing $\vec{p}$ by $\vec{k}$.
By integrating both sides of \eqref{eqn1.3} and multiplying by $N_B$ and introducing the integration constant $h(\vec{k},\vec{p},q_{3})$, we find
\begin{equation}\label{eqn1.4}
\begin{split}
H_B^{-1}(a(p_s),q_3) N_B\mathcal{A}(\vec{p},\vec{k},q_3)
& = H_B^{-1}(a(k_s),q_3) \bigg\{(4\pi \i \lambda_{B} q_{3})\frac{(p+k)_{-}}{(p-k)_{-}}\bigg\} + h(\vec{k}, p_{-},q_3)
\end{split}
\end{equation}
Similarly, integrating (\ref{eqn2.3}) and introducing another integration constant $\tl{h}(\vec{k},\vec{p},q_{3})$, we get 
\begin{equation}\label{eqn2.4}
\begin{split}
H_B(a(k_s),q_3) N_B\mathcal{A}(\vec{p},\vec{k},q_3)
& = H_B(a(p_s), q_3)  \bigg\{(4\pi \i \lambda_{B} q_{3})\frac{(p+k)_{-}}{(p-k)_{-}}\bigg\}  + \tilde{h}(k_{-}, \vec{p}, q_3)
\end{split}
\end{equation}
Comparing (\ref{eqn1.4}) and (\ref{eqn2.4}), we see that the $k_{+}$ dependence of $h$ and $p_{+}$ dependence of $\tilde{h}$ are determined. So, we conclude
\begin{equation}\label{A}
\begin{split}
N_B\mathcal{A}(\vec{p},\vec{k},q_3)
& = \frac{H_B(a(p_s), q_3)}{H_B(a(k_s),q_3)}  \bigg\{(4\pi \i \lambda_{B} q_{3})\frac{(p+k)_{-}}{(p-k)_{-}} + j(k_{-}, p_{-}, q_3) \bigg\}  
\end{split}
\end{equation}
where, $j$ is now the integration constant to be determined shortly. Following the arguments of \cite{Jain:2014nza}, we see that in the above equation \eqref{A}, the function $j(k_{-},p_{-},q_{3})$ must be a function of charge zero, and so, must be a function of $\frac{k_{-}}{p_{-}}$. It must also be a singularity free, i.e., the derivative w.r.t. both $k_{+}$ and $p_{+}$ must vanish. This seems impossible unless the function $j$ is a constant w.r.t. $k$ and $p$. And so, we conclude 
\begin{equation}\label{exactscalar4ptapp}
\begin{split}
N_B\mathcal{A}(\vec{p},\vec{k},q_3)
& = \frac{H_B(a(p_s), q_3)}{H_B(a(k_s),q_3)}  \bigg\{(4\pi \i \lambda_{B} q_{3})\frac{(p+k)_{-}}{(p-k)_{-}} + j( q_3) \bigg\}  
\end{split}
\end{equation}
\\

In order to evaluate $j(q_3)$, we now replace equation \eqref{exactscalar4pt} back into equations (\ref{eqn1.1}), and get 
\eqref{FBfunction} we get
\begin{equation}\label{insertbackA}
\begin{split}
N_B (\mathcal{A}(\vec{p},\vec{k},q_3)&-\mathcal{A}_0(\vec{p},\vec{k},q_3)) 
 = \frac{1}{(2\pi)H_B(a(k_s),q_3)} \int_{0}^{\infty} \frac{r_s dr_s}{a(r_{s})}\  F_B(a(r_{s}),q_3) \ H_B(a(r_s), q_3)\ I(r_s) 
\end{split}
\end{equation}
where, 
\begin{equation}\label{integralIrs}
\begin{split}I(r_s)& 
= \int_{0}^{2\pi} \frac{d\theta_{r}}{2\pi}~\bigg\{-(4\pi \i \lambda_{B} q_{3})\frac{(r+p)_{-}}{(r-p)_{-}}+\tilde{b}_4 \bigg\}\bigg\{(4\pi \i \lambda_{B} q_{3})\frac{(r+k)_{-}}{(r-k)_{-}}+j(q_{3})\bigg\}
\end{split}
\end{equation}
One can easily perform the angular integrals by using \eqref{angint},
and the final result of the integration is 
\begin{equation}{\label{I(r)}}
\begin{split}
I(r_s) 
& = [4\pi \i \lambda_{B} q_{3} +\tilde{b}_4][-4\pi \i \lambda_{B} q_{3}+j(q_3)] \\
& - \Theta(r_s-p_s)(8\pi \i \lambda_{B} q_{3})\bigg\{(4\pi \i \lambda_{B} q_{3})\frac{p_{-}+k_{-}}{p_{-}-k_{-}}+j(q_{3})\bigg\} \\
& +  \Theta(r_s-k_s)(8\pi \i \lambda_{B} q_{3})\bigg\{-(4\pi \i \lambda_{B} q_{3})\frac{k_{-}+p_{-}}{k_{-}-p_{-}}+\tilde{b}_4\bigg\}
\end{split}
\end{equation}
From \eqref{insertbackA}, we define 
\begin{equation}\label{LHSdef}
\text{LHS}=(8\pi \i \lambda_{B} q_{3}) H_B(a(k_s),q_3) N_B \bigg(\mathcal{A}(\vec{p},\vec{k},q_3)-\mathcal{A}_{0}(\vec{p},\vec{k},q_3)\bigg) 
\end{equation}
Also, we get from \eqref{insertbackA}, 
\begin{equation}\label{RHS}
\begin{split}
\text{RHS} & =  \int \frac{r_s dr_s}{a(r_{s})}\  (4\i \lambda_B q_3 )\ F_B(a(r_{s}),q_3) \ H_B(a(r_s), q_3)\ I(r_s) \\
& = [4\pi \i \lambda_{B} q_{3} +\tilde{b}_4][-4\pi \i \lambda_{B} q_{3}+j(q_3)][H_B(\infty,q_3)-H_B(c_B,q_3)] \\
& - (8\pi \i \lambda_{B} q_{3})\bigg\{(4\pi \i \lambda_{B} q_{3})\frac{(p+k)_{-}}{(p-k)_{-}}+j(q_{3})\bigg\}[H_B(\infty,q_3)-H_B(a(p_s),q_3)] \\
& + (8\pi \i \lambda_{B} q_{3})\bigg\{-(4\pi \i \lambda_{B} q_{3})\frac{(k+p)_{-}}{(k-p)_{-}}+\tilde{b}_4\bigg\}[H_B(\infty,q_3)-H_B(a(p_s),q_3)]
\end{split}
\end{equation}
From \eqref{LHSdef}, by using \eqref{exactscalar4ptapp} and the expression obtained for $N\mathcal{A}_0(\vec{p},\vec{k},q_3)$, we can write 
\begin{equation}\label{LHS}
\begin{split}
\text{LHS}
& = (8\pi \i \lambda_{B} q_{3})\bigg[H_B(a(p_s),q_3)\bigg\{4\pi \i \lambda_{B}q_3 \frac{(p+k)_{-}}{(p-k)_{-}}+j(q_3)\bigg\} \\
& - H_B(a(k_s),q_3)\bigg\{-4\pi \i \lambda_{B}q_3 \frac{(p+k)_{-}}{(k-p)_{-}}+\tilde{b}_4 \bigg\}\bigg]
\end{split}
\end{equation}
Comparing \text{LHS} and \text{RHS}, 
we get
\begin{equation}\label{jq3app}
\frac{j(q_3)}{ 4\pi \i \lambda_{B}q_3}=  \frac{4\pi \i \lambda_{B} q_3(H_B(c_B,q_3)-H_B(\infty,q_3)) +\tilde{b}_4 (H_B(c_B,q_3)+H_B(\infty,q_3)) }{4\pi \i \lambda_{B} q_3(H_B(c_B,q_3)+H_B(\infty,q_3)) +\tilde{b}_4 (H_B(c_B,q_3)-H_B(\infty,q_3)) }
\end{equation}



So, the final result for the thermal four-point function of fundamanetal scalars is given by \eqref{exactscalar4ptapp} with $j(q_3)$ given by \eqref{jq3app}. 

The expression for $j(q_3)$ given in \eqref{jq3app} can be written as an equivalent but simplified form as below
\begin{equation}\label{jq3simp}
j(q_3) = 4\pi \lambda_{B} q_3 \tan\Big( \tan^{-1}\Big(\frac{\tl{b}_4}{4\pi\lambda_Bq_3}\Big)+2\lambda_{B}q_3 \int_{c_B}^{\infty} \frac{dw \ \chi_{B}(w)}{q_3^2+4w^2} \ \Big) \ . 
\end{equation}
 It is clear from \eqref{jq3simp} that $j(q_3)$ is an even function of $q_3$, i.e., $j(-q_3)=j(q_3)=j(|q_3|)$.  It is also clear from \eqref{jq3simp} that $j(q_3)$ or rather $j(q_3,\lambda_B)$ is also an even function of $\lambda_B$ separately, i.e., $j(q_3,\lambda_B)=j(q_3,-\lambda_B)=j(q_3, |\lambda_B|)$. The large $N$ exact four-point function \eqref{exactscalar4ptapp} can be explictly written as 
 \begin{equation}\label{4ptfnexp}
 \begin{split}
 N_B\mathcal{A}(\vec{p},\vec{k},q_3)
 & = \bigg\{(4\pi \i \lambda_{B} q_{3})\frac{(p+k)_{-}}{(p-k)_{-}} + j( q_3) \bigg\}  \exp\Big(4\i \lambda_B q_3 \int_{a(k_s)}^{a(p_s)} \frac{dw \ \chi_{B}(w)}{q_3^2+4w^2} \Big)
 \end{split}
 \end{equation}
 where, $j(q_3)$ is given by \eqref{jq3app} or equivalently by \eqref{jq3simp}. 

\section{Comment on arbitrary spin $s$ current two-point function}\label{spins2pt}
Fundamental matter coupled to Chern-Simons theories contain a spectrum of higher spin currents \cite{Giombi:2011kc, Maldacena:2011jn, Maldacena:2012sf, Frishman:2013dvg}. The relevant modifications in the expressions of the currents $J_{\mu_1\cdots \mu_s}$ in the massive matter theory is discussed in detail in \cite{Frishman:2013dvg}. However, these modifications will not be important to what we are going to discuss below.
In this subsection, we outline the generalization of the computations correlators of the previous sections to the case of correlator of arbitrary spin $s$ single trace operator which we schematically label as $J_{(s)}$.  
\begin{figure}[h!]
	\begin{center}
		\begin{tikzpicture}[scale=0.85, cross/.style={path picture={ 
				\draw[black]
				(path picture bounding box.south east) -- (path picture bounding box.north west) (path picture bounding box.south west) -- (path picture bounding box.north east);
		}}]
		\begin{feynman}
		\coordinate (A) at (0,0) ;
		\coordinate (B) at (6,0) ;
		\coordinate (L) at (-1.1,0) ;
		\draw (A) node {$\otimes $} ; 
		\node [draw,cross,minimum width=0.01 cm] at (B){};
		\filldraw (3,1.05) circle (3pt) ;
		\filldraw (3,-1.05) circle (3pt) ;
		\end{feynman}
		\draw (0.10,0.10) .. controls (2,1.4) and (4,1.4) .. (5.85,0.15);
		\draw (0.10,-0.10) .. controls (2,-1.4) and (4,-1.4) .. (5.85,-0.15);
		\draw[-latex] (L)--+(0.7,0);
		\draw (L)+(0.3, 0.3) node {$q$}; 
		\draw (A)+(0.8,0) node {$\ J_{(s)}$};
		\draw (5,0) node {$\ J_{(s)}$};
		\end{tikzpicture}
		\caption{two-point correlator $\langle J_{(s)}(-q)J_{(s)}(q)\rangle $. The circled cross at the left is the exact insertion vertex and the boxed cross at the right is the `bare' insertion vertex. The circular dot denotes the exact propagator.}
		\label{JsJs}
	\end{center} 
\end{figure}
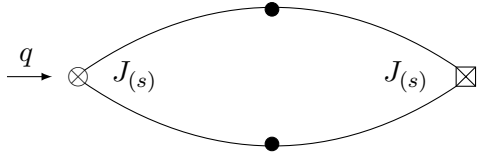
The schematic diagram for the two-point correlator $\langle J_{(s)}J_{(s)}\rangle$ is shown in Fig.\ref{JsJs}. In the lightcone gauge $A_{-}=0$, the explicit computation of the correlators (specifically, solving the corresponding Schwinger-Dyson equations) becomes simple in the kinematical regime in which the external momentum $q$ is chosen to be such that $q_{\pm}=0$, i.e., in the explicit component form $q\equiv (0,0,q_3)$. With these choices, we will specifically talk about the component of the correlator $\langle J_{(s)}J_{(s)}\rangle$ in which one of the $J_{(s)}$ insertion vertex has all the minus signs ($s$ times) and the other $J_{(s)}$ insertion vertex which has all the plus signs ($s$ times). As already mentioned in previous cases, to compute the two-point correlators of current operator $J_{(s)}$, only a single exact insertion vertex is required and the other insertion vertex is the `bare' insertion which is computed from the explicit definition of the corresponding current operator. For definiteness, we choose the exact $J_{(s)}$ insertion vertex to be the one in which the spacetime indices are all minuses, and the `bare' insertion vertex $J_{(s)}$ to be the one in which all spacetime indices are pluses. 
The definition of the gauge invariant, single-trace, higher spin spin current operator is obtained from the corresponding generating functions (see e.g. \cite{Giombi:2011kc}). Below, we consider fermionic and bosonic theories separately and discuss about the specific details regarding those theories there. 
\subsection{Fermionic theory}
In the fermionic theory coupled to Chern-Simons theory, there is a spectrum of gauge invariant, single trace, higher spin currents. In the lightcone gauge, and in the case of external momenta $q_{\pm}=0$, the expression for the current operator - with all component signs being minuses - is given by 
\begin{equation}
J_{-\cdots -}^F(-q) \equiv J_{(s_m)}^F(-q) = \i  \alpha_{s} \int \frac{\mc{D}_F^3k}{(2\pi)^3} \ (k_{-})^{s-1} \ \bar{\psi}(-(q+k)) \gamma_{-} \ \psi(k) \ , 
\end{equation}
for $s\geq 1$. Here, we have kept an explicit factor of $\i$ for later convenience (comparing it with the case $s=1$) and the factor $\alpha_{s}$ is the normalization factor. Here, a schematic symbol $s_{m}$ denotes that the spin $s$ current operator carry $s$ minus signs. 
\subsubsection*{Schwinger-Dyson equation for exact vertex }
The corresponding Schwinger-Dyson equation for the exact $J_{(s_m)}^F$ vertex is schematically given by the Fig.\ref{sdexactvertx} where, the `tree level' insertion in this case is given by $\widetilde{V}^F_{(s_m)}=\i \alpha_s (k_{-})^{s-1} \gamma_{-}$.  

The exact $J^{F}_{(s_m)}$ insertion vertex is defined as
\begin{equation}
\langle J^{F}_{(s_m)} (-q)\psi(k)\bar{\psi}(p)\rangle =V_{(s_m) }^{F}(k,q)~(2\pi)^3\delta^{(3)}(p+k+q) \ .
\end{equation}
From Fig.\ref{sdexactvertx}, we find that the corresponding Schwinger-Dyson equation for $V_{(s_m)}^{F}(k,q)$ is given by 
\begin{equation}\label{SDforVsmF} 
V_{(s_m)}^{F}(k,q)=\widetilde{V}_{(s_m)}^{F}(k,q)+\ N_F \int \frac{\mathcal{D}_F^3 \l }{(2\pi)^3} \big[  \mathcal{V}^\nu(\l+q,k+q) S_F(\l+q) V_{(s_m)}^{F}(\l,q) S_F(\l) \mathcal{V}^\rho(k,\l)   \big] G^F_{\rho \nu}(\l-k) \ ,
\end{equation}
Using \eqref{vertexpsibarApsi} and \eqref{gaugebospropforFermion} , the above equation \eqref{SDforV1Fmu} can be simplified to
\begin{equation}\label{SDforVsmFxp}
V_{(s_m)}^{F}(k,q)=\i\alpha_{s} (k_{-})^{s-1} \g_{-} - 2\pi \i \lambda_{F} \int \frac{\mathcal{D}_F^3 \l }{(2\pi)^3} \big[ \g^{[3|} S_F(\l+q) V_{(s_m)}^{F}(\l,q) S_F(\l) \g^{|+]}    \big] \frac{1}{(\l-k)_{-}}  \ .
\end{equation}
Following exactly the same procedure that is used to compute the exact $J_{-}^{F}$ vertex in the case of spin one two-point correlator, one can solve the above Schwinger-Dyson equation \eqref{SDforVsmFxp}. Here, briefly outline the steps and present the final result. In `lightcone kinematics' $q_{\pm}=0$, it is clear from \eqref{SDforVsmFxp} that $V_{(s_m)}^{F} (k,q)$ is independent of $k_3$. Also, in the lightcone kinematics $q_{\pm}=0$, the only non-zero component of the external momenta $q$ is $q_3$. So, the momentum dependence of $V_{(s_m)}^{F} (k,q)$ is basically $V_{(s_m)}^{F} (\vec{k},q_3)$. As in the $J_{-}^F$ case, $V_{(s_m)}^{F}k,q)$ can be expanded as, 
\begin{equation}\label{Vsmfgdef}
V_{(s_m)}^{F}(\vec{k},q_3)=V_{(s_m),\nu}^{F}(\vec{k},q_3) \g^\nu +V_{(s_m), \mathbb{1}}^{F}(\vec{k},q_3) \mathbb{1}  = (k_{-})^{s-1} g_{m}^{(s)}(k_s,q_3) \g^+ + (k_{-})^{s} f_{m}^{(s)}(k_s,q_3) \mathbb{1}  \ .
\end{equation} 
Plugging \eqref{Vmfgdef} in \eqref{SDforVsmFxp}, we get a set of two coupled integral equations involving $f_m^{(s)}$ and $g_m^{(s)}$. Simiar to the case of $J_{-}^F$, these equations for $f_m^{(s)}$ and $g_m^{(s)}$ simplifies a lot once we utilize the $SO(2)$ rotational symmetry in the lightcone plane and decompose the momentum integration measure as \eqref{DF3ell}, perform the angular integration using \eqref{angint} and also perform the integral over momentum component $\l_3$ by using \eqref{GF2} and \eqref{GF3}. Performing a change of variable $a(\l_s) =+\sqrt{\l_s^2+c_F^2} =w$ and  $a(k_s) =+\sqrt{k_s^2+c_F^2} =z$,  and relabelling $f_m^{(s)}(k_s,q_3)\equiv \tl{f}_m^{(s)}(z,q_3)$ and $g_m(k_s,q_3)^{(s)}\equiv \tl{g}_m^{(s)}(z,q_3)$ and $\tl{\Sigma}_{\mathbb{1}}(\l_s)\equiv h_F(w)$, the equations for $\tl{f}_m^{(s)}$ and $\tl{g}^{(s)}$ can be written in a much simpler form, which takes the following form
\begin{equation}\label{fsminteqn}
\tl{f}_m^{(s)}(z,q_3) = 2 \i \lambda_F \int_{z}^{\infty} dw \ F_F(w,q_3)\bigg[\big(q_3-2\i h_F(w)\big) \tl{f}_m^{(s)}(z,q_3) \ - \ 2\tl{g}_m^{(s)}(w,q_3) \bigg] \ , 
\end{equation}
and
\begin{equation}\label{gsminteqn} 
\tl{g}_m^{(s)}(z,q_3) =\i \alpha_{s} \ +\  2 \i \lambda_F \int_{z}^{\infty}  dw  \ F_F(w,q_3) \ \bigg[ 2\big(w^2- h_F^2(w)\big) \tl{f}_m^{(s)}(w,q_3) + \big(q_3+2\i \tl{\Sigma}_{\mathbb{1}}(\l_s) \big) \tl{g}_m^{(s)}(w,q_3) \bigg]  \ .
\end{equation}
To solve the above two equations \eqref{fsminteqn} and \eqref{gsminteqn}, it is best to convert them to a set of differential equations. The boundary conditions that follow from \eqref{fsminteqn} and \eqref{gsminteqn} are 
\begin{equation}\label{bcfsmgsm} 
\tl{f}_m^{(s)}(z=\Lambda,q_3) = 0, \ \ \ \ \  \tl{g}_m^{(s)}(z=\Lambda ,q_3) = \i \alpha_s , \ \ \ \text{where}, \ \  \Lambda\rightarrow \infty  \ .
\end{equation} 
Solving the equations \eqref{fsminteqn} and \eqref{fsminteqn}, by first converting them into differential equations with the boundary conditions \eqref{bcfsmgsm} , we find the final solution for the exact $J_{(s_m)}^F$ vertex in terms of $f_m^{(s)}$ and $g_m^{(s)}$ as
\begin{equation}\label{finsolfsmgsm} 
\begin{split}
\tl{f}_m(z,q_3) &=\frac{\i \alpha_s }{q_3}\bigg(1- \exp\big[4\i \lambda_F q_3 \int_{z}^{\Lambda} dw \ F_F(w,q_3)\big]\bigg) \ , \\ 
\tl{g}_m(z,q_3) & = \i \alpha_s  - \frac{1}{2}\big(q_3+2\i h_F(z)\big)\tl{f}_m(z,q_3) \ .
\end{split}
\end{equation}
\subsubsection*{Two-point function}\label{2ptJsF}
Schematic diagram for the two-point function $\langle J_{s_m}^FJ_{s_p}^F\rangle$ is shown in Fig.\ref{JsJs}. One needs the `bare' insertion vertex $J_{s_p}^F$ as shown in the right of Fig.\ref{JsJs}. Let us define the `bare' insertion $J^F_{(s_{p})}(q)$ vertex as 
\begin{equation}\label{spinspFvertex}
\langle J^F_{(s_p)}(-q) \psi(k)\bar{\psi}(p) \rangle =U^F_{(s_p)}(k,q)  \ (2\pi)^3 \delta^{(3)} (p+k+q) 
\end{equation}
$U^F_{(s_p)}$ can be computed from the explicit expressions of the corresponding currents and the number of digrams included in this insertion should be such that there is no overcounting. The two-point function $\langle J_{s_m}^FJ_{s_p}^F\rangle$ is then given by 
\begin{equation}\label{GsmspF(q)}
G_{s_ms_p}^F(q) =-N_F\int \frac{\mathcal{D}_F^3k}{(2\pi)^3} \ \tr_F \bigg[ \ S_F(k+q) \ V_{(s_m)}^{F}(k,q) \  S_F(k) \ U_{(s_p)}^{F}(k+q,-q)\ \bigg]
\end{equation}
where, the extra $(-1)$ factor above in \eqref{integralforG0F(q)} is because of the integration over fermion loop. 
For various spins, one needs to compute \eqref{spinspFvertex} and then perform the integral \eqref{GsmspF(q)}.

\subsection{Bosonic theory}
Similar to the fermionic sase, we consider the two point function of spin $s$ current operator, in which one of the operator carries all the minus signs and the other operator carries all the plus signs. We define this two-point function of general spin $s$ current $J^B_{(s)}$ in the bosonic theory as 
\begin{equation}\label{spins2ptdef}
\langle J^B_{(s_m)}(q') J^B_{(s_p)}(q) \rangle =G^B_{s_ms_p}(q)  \ (2\pi)^3 \delta^{(3)} (q'+q) \ , 
\end{equation}
where, $s_m$ denotes that the operator $J^B_{(s_m)}(q')$ carries $s$ minus signs and similarly $s_p$ denotes that the operator $J^B_{(s_p)}(q)$ carries $s$ plus signs. To compute this, we first compute the exact $J^B_{(s_m)}(q')$ vertex. The other insertion $J^B_{(s_p)}(q)$ is computed upto the leading non-trivial orders to account for all the diagrams that contribute to \eqref{spins2pt} without overcounting. 
We define the exact $J^B_{(s_m)}(q')$ insertion vertex by separating the overall momentum conserving $\delta$-function as follows 
\begin{equation}\label{spinsexactvertex}
\langle J^B_{(s_m)}(-q) \phi(k)\bar{\phi}(p) \rangle =V^B_{(s_m)}(k,q)  \ (2\pi)^3 \delta^{(3)} (p+k+q) 
\end{equation}

Following the same procedure as spin one current, we compute the exact $J^B_{(s_m)}(q')$ insertion vertex by 
\begin{equation}\label{bootstrapVsexplicit}
V^B_{(s_m)}(k,q) =V^B_{(s_m),\text{free}}(k,q) +N_B \int \frac{\mathcal{D}_B^3p}{(2\pi)^3} \ \frac{ V^B_{(s_m),\text{free}}(k,q)\mathcal{A}(p,k,q)}{(k^2+c_B^2)((k+q)^2+c_B^2)} 
\end{equation}
In the case of lightcone gauge $A_{-}=0$ and in the case $q_{\pm}=0$, it follows from the rotational symmetry, that $V^B_{(s_m),\text{free}}(k,q)=\tl{\alpha}_{s} (k_{-})^s$ for $s\geq 1$, where, $\tl{\alpha}_{s}$ depends upon the normalization of the current operator. Following the same procedure as in the case of spin one current, we perform the momentum integral and get
\begin{equation}\label{finans}
V^B_{(s_m)}(k,q) =\tl{\alpha}_{s} (k_{-})^{s} \ \frac{H_B(\infty,q_3)}{H_B(a(k_s),q_3)} 
\end{equation}
And we also define the `tree level' insertion $J^B_{(s_{p})}(q)$ vertex as 
\begin{equation}\label{spinspvertex}
\langle J^B_{(s_p)}(-q) \phi(k)\bar{\phi}(p) \rangle =U^B_{(s_p)}(k,q)  \ (2\pi)^3 \delta^{(3)} (p+k+q) 
\end{equation}
To compute \eqref{spins2ptdef} we need \eqref{spinspvertex} upto leading non-trivial orders to account for all the diagrams that contribute to \eqref{spins2ptdef} without overcounting. 
Once this is computed, it is easy to compute the spin $s$ current two-point function  $G^B_{s_ms_p}(q)$ as 
\begin{equation}\label{spins2ptcompute}
G^B_{s_ms_p}(q)= N_B\int \frac{\mathcal{D}_B^3k}{(2\pi)^3} \ \bigg[S_B(k+q) V^B_{(s_m)}(k,q)S_B(k)U^B_{(s_p)}(k+q,-q) \bigg]
\end{equation}
where, $S_{B}(k)$ is the exact scalar field propagator given by \eqref{exactphiprop}. Explicitly, this can be written as
\begin{equation}\label{spins2ptexplicit}
G^B_{s_ms_p}(q)= N_B H_B(\infty,q_3) \ \int \frac{\mathcal{D}_B^3k}{(2\pi)^3} \ \frac{(k_{-})^{s} \ U^B_{s_{p}} (k+q,-q)}{(k^2+c_B^2)((k+q)^2+c_B^2)} 
\end{equation}
As in the case of fermions, for various spins, one needs to compute \eqref{spinspvertex} for varrious spins and then perform the above integral \eqref{spins2ptexplicit}.

\bibliography{ref}\bibliographystyle{JHEP}

\end{document}